\documentclass[preprintnumbers,article,amsmath,amssymb,floatfix,10pt,prd,onecolumn,superscriptaddress,nofootinbib]{revtex4}
\usepackage[colorlinks=true, pdfstartview=FitV, linkcolor=blue, citecolor=red, urlcolor=magenta]{hyperref}
\usepackage{bbm}
\usepackage{amsfonts}
\usepackage{amsfonts,graphicx,amsmath,amsthm,amssymb,epsfig}
\usepackage{float}
\allowdisplaybreaks
\usepackage{mathrsfs}
\usepackage{latexsym}
\usepackage{epsfig}
\usepackage{epstopdf}
\usepackage{epstopdf}
\usepackage{graphicx}
\usepackage{amssymb}
\usepackage{amsmath}
\usepackage{dcolumn}
\usepackage{bm}
\usepackage{color}
\usepackage{comment}
\begin{document}
\title{\bf Existence of Non-singular Stellar Solutions within the context of Electromagnetic Field:
A Comparison between Minimal and Non-minimal Gravity Models}

\author{Tayyab Naseer}
\email{tayyabnaseer48@yahoo.com; tayyab.naseer@math.uol.edu.pk}\affiliation{Department of Mathematics and Statistics, The University of Lahore,\\
1-KM Defence Road Lahore-54000, Pakistan}

\author{Jackson Levi Said}
\email{jackson.said@um.edu.mt}\affiliation{Institute of Space Sciences and Astronomy, University of Malta, Malta, MSD 2080}
\affiliation{Department of Physics, University of Malta, Malta, MSD 2080}

\begin{abstract}
In this paper, we explore the existence of various non-singular
compact stellar solutions influenced by the Maxwell field within the
matter-geometry coupling based modified gravity. We start this
analysis by considering a static spherically symmetric spacetime
which is associated with the isotropic matter distribution. We then
determine the field equations corresponding to two specific
functions of this modified theory. Along with these models, we also
adopt different forms of the matter Lagrangian. We observe several
unknowns in these equations such as the metric potentials, charge
and fluid parameters. Thus, the embedding class-one condition and a
particular realistic equation of state is used to construct their
corresponding solutions. The former condition provides the metric
components possessing three constants, and we calculate them through
junction conditions. Further, four developed models are graphically
analyzed under different parametric values. Finally, we find all our
developed solutions well-agreeing with the physical requirements,
offering valuable insights for future explorations of
the stellar compositions in this theory.  
\end{abstract}

\maketitle

\date{\today}

\section{Introduction}

Cosmologists have recently revealed revolutionary discoveries that
defy traditional beliefs regarding the spatial organization of
celestial structures in our cosmos. Rather than presenting a random
dispersion, these formations exhibit a discernible order, sparking
significant curiosity among researchers. The meticulous study of
these entities has become a central topic of exploration for
scientists committed to unraveling the mystery surrounding the
accelerated expansion of the universe. Empirical testimonies
strongly indicate the existence of an extensive counter-force to
gravitational attraction, driving the observed rapid expansion.
Referred to as dark energy, this enigmatic force presents a
significant puzzle for scientists. While Einstein's general
relativity (GR) provides some insights into this expansion, it
encounters difficulties in fully explaining dark energy,
particularly in relation to the cosmological constant $\Lambda$.
Therefore, it has been necessary to introduce modifications to the
existing theory to better comprehend and enhance our knowledge
regarding fundamental dynamics of the cosmos.

Einstein's GR has straightforwardly been modified to
$f(\mathcal{R})$ gravity, representing a substantial advancement in
experimental physics. This theory alters the action function by
interchanging the curvature scalar $\mathcal{R}$ and its general
functional. Notable progress has been made within this gravity
theory, with implications reaching into the study of celestial
structures \cite{2}-\cite{4}. Astashenok with his collaborators
\cite{1n} investigated the upper mass limit for massive objects in
the current framework. Their research produced an intriguing result
that as a second object in the binary GW190814, there must be either
a rapidly rotating neutron star or a black hole. A significant body
of literature underscores the remarkable contributions made by
various researchers \cite{1o}-\cite{9}. One notable contribution
comes from Bertolami and his colleagues \cite{10}, who were
instrumental in put forwarding the coupling between matter and
spacetime geometry in $f(\mathcal{R})$ gravity. Their methodology
involved integrating the matter Lagrangian and $\mathcal{R}$ into a
unified functional form, known as
$f(\mathcal{R},\mathcal{L}_\mathrm{m})$ theory. This novel concept
prompted astronomers to focus on discussions related to the rapid
universe' expansion \cite{11}.

Following these developments, Harko et al. \cite{20} introduced a
ground-breaking gravitational theory, called
$f(\mathcal{R},\mathcal{T})$ gravity at the action level. This
theory utilizes a generalized function that leads to a non-conserved
phenomenon, resulting in the emergence of an extra force, causing
moving particles to follow non-geodesic path \cite{22}. Houndjo
\cite{22a} employed a particular model based on the minimal
interaction to explain the shift from one cosmic era to the other
phase in which we are living right now. Among the various functional
forms of $f(\mathcal{R},\mathcal{T})$ theory, the
$\mathcal{R}+2\beta\mathcal{T}$ candidate has attracted considerable
attention in scientific literature due to its ability to generate
physically existing internal structures. Different researchers,
including Das et al. \cite{22b}, utilized a similar model to develop
a three-layer gravastar geometry. Various methodologies were
implemented to explore diverse geometrical structures in this
context \cite{25af}-\cite{25ae}. An essential facet of the
$f(\mathcal{R},\mathcal{T})$ gravity is its incorporation of some
effects at quantum level, which introduces the potential for
particle creation. This characteristic is of great significance in
astronomical investigations as it sets out a connection between the
extended theory and quantum mechanics. Notable findings in this area
have been produced and can be seen in \cite{1j,1k}. In a recent
research endeavor, Zaregonbadi et al. \cite{1l} have examined the
feasibility of this modification to GR to study the impact of dark
matter on clusters of galaxies.

The $f(\mathcal{R},\mathcal{T})$ theory has indeed presented an
intriguing extension to GR, showing a diverse range of phenomenology
in modern research. However, researchers \cite{25ag,25aga} delved
into the challenges associated with constructing a viable and
realistic cosmology within this theory. Their study demonstrated
that the currently discussed models of this theory do not yield an
expandable cosmic background. In response to these challenges,
Haghani and Harko \cite{25ah} undertook a considerable effort by
simultaneously unify two categories of gravitational theories, and
call it the $f(\mathcal{R},\mathcal{L}_{m},\mathcal{T})$ gravity.
This strategic approach aims to address the limitations encountered
in the previously discussed gravity models and offers a more
comprehensive understanding of the intricate dynamics governing the
universe. They explored the Newtonian limit of the field equations
and provided some terms representing an extra-acceleration,
particularly focusing on scenarios involving small velocities of
particles and weak fields of gravity. This exploration enlightens
how different choices of Lagrangian influence the description of the
cosmic expansion. Zubair et al. \cite{25ai} reconstructed some
cosmological solutions such as de Sitter and $\Lambda$CDM models in
this theory and found them to be cosmologically stable through
suitable perturbations.

The study of celestial entities characterized by the field equations
possessing high non-linearity, either in the framework of GR or
extended theories, has prompted astronomers to actively seek their
numerical or exact solutions. The significance of compact interiors
lies in the physical interest they hold, contingent upon the
satisfaction of specific conditions by the developed model. Various
methodologies have been engaged in the scientific literature to
derive such solutions, including the utilization of a specific
ansatz or the implementation of particular equations of state, among
other techniques. One approach to solving this challenge is through
the implementation of the embedding class-one phenomenon, which
posits that one can embed any space in another having at least one
higher dimension. Bhar et al. \cite{36} employed the same method,
coupled with particular metric potentials and derived physically
existing anisotropic solutions. Maurya et al. \cite{37,37a} used the
same approach to construct a new solution, delving into its
stability and exploring the impact of anisotropic pressure on
relativistic systems. Singh with his collaborators \cite{37b}
devised a singularity-free solution for spherical geometry by
proposing a specific metric function within the framework of this
technique. Exploring this condition into a matter-geometry coupled
theory, several works have yielded stable as well as viable
solutions \cite{38c}-\cite{38a}.

In this paper, we explore various isotropic solutions in conjunction
with the Maxwell field within the framework of
$f(\mathcal{R},\mathcal{L}_{m},\mathcal{T})$ theory. The paper's
structure is organized as follows. The following section establishes
some basics of this extended theory and derives the generalized
field equations. Section \textbf{III} presents the Karmarkar
condition, which aids in determining the metric potentials.
Additionally, we utilize the Reissner-Nordstr\"{o}m vacuum solution
and compute the constants associated with the overhead condition. We
outline particular criteria that, once fulfilled, guarantee the
model's physical validity in section \textbf{IV}. Advancing further,
section \textbf{V} reveals the newly formulated solutions and offers
a visual representation to aid in understanding the physical
relevance of the obtained results. Conclusively, in the final
section, our findings are encapsulated, summarizing the main
outcomes and insights acquired in this study.

\section{Fundamentals of Modified Theory}

The action of the modified
$f(\mathcal{R},\mathcal{L}_{m},\mathcal{T})$ theory is obtained
after replacing the Ricci scalar with this functional \cite{25ah}.
This has the form
\begin{equation}\label{g1}
S=\int
\sqrt{-g}\left[\frac{f(\mathcal{R},\mathcal{L}_{m},\mathcal{T})}{16\pi}
+\mathcal{L}_{m}+\mathcal{L}_{\mathcal{E}}\right]d^{4}x,
\end{equation}
where the electric charge and ordinary matter have Lagrangian
densities, denoted by $\mathcal{L}_{\mathcal{E}}$ and
$\mathcal{L}_{m}$, respectively. Also, $g=|g_{\epsilon\omega}|$ with
$g_{\epsilon\omega}$ being the metric tensor and the two lines
enclosing it symbolize the determinant. Varying the action
\eqref{g1} w.r.t. $g_{\epsilon\omega}$, the tensorial form of the
modified field equations become
\begin{equation}\label{g2}
\mathcal{G}_{\epsilon\omega}=8\pi
\mathcal{T}_{\epsilon\omega}^{(eff)},
\end{equation}
where the entity $\mathcal{G}_{\epsilon\omega}$, namely the Einstein
tensor, expresses the geometry of the considered fluid distribution
and $\mathcal{T}_{\epsilon\omega}^{(eff)}$ refers to the matter
enclosed by that geometry. This effective term is further classified
into three different energy-momentum tensors as
\begin{equation}\label{g3}
\mathcal{T}_{\epsilon\omega}^{(eff)}=\frac{1}{f_{\mathcal{R}}}\left(\mathcal{T}^{(m)}_{\epsilon\omega}
+\mathcal{E}_{\epsilon\omega}\right)+\mathcal{T}_{\epsilon\omega}^{(cr)},
\end{equation}
where
\begin{itemize}
\item $\mathcal{T}^{(m)}_{\epsilon\omega}$ correspond to the ordinary
matter configuration,
\item $\mathcal{E}_{\epsilon\omega}$ indicates the presence of charge
in the self-gravitating system,
\item $\mathcal{T}^{(cr)}_{\epsilon\omega}$ are modified correction terms.
\end{itemize}
We express $\mathcal{T}^{(m)}_{\epsilon\omega}$ as follows
\begin{equation}\nonumber
\mathcal{T}_{\epsilon\omega}^{(m)}=-\frac{2}{\sqrt{-g}}\left\{\frac{\delta\left(\sqrt{-g}\mathcal{L}_{m}\right)}
{\delta g^{\epsilon\omega}}\right\} \quad \Rightarrow \quad
\mathcal{T}_{\epsilon\omega}^{(m)}=g_{\epsilon\omega}\mathcal{L}_{m}
-\frac{\partial\mathcal{L}_{m}}{\partial g_{\epsilon\omega}}.
\end{equation}

On the other hand, the last term on the right side of Eq.\eqref{g3}
have the value given by
\begin{align}\nonumber
\mathcal{T}_{\epsilon\omega}^{(cr)}&=\frac{1}{8\pi
f_{\mathcal{R}}}\bigg[\frac{1}{2}\big(2f_{\mathcal{T}}+f_{\mathcal{L}_{m}}\big)\mathcal{T}_{\epsilon\omega}^{(m)}
-(g_{\epsilon\omega}\Box-\nabla_{\epsilon}\nabla_{\omega})f_{\mathcal{R}}\\\label{g4}
&+\frac{1}{2}\big(f-\mathcal{R}f_{\mathcal{R}}\big)g_{\epsilon\omega}
-\big(2f_{\mathcal{T}}+f_{\mathcal{L}_{m}}\big)\mathcal{L}_{m}g_{\epsilon\omega}
+2f_{\mathcal{T}}g^{\zeta\beta}\frac{\partial^2
\mathcal{L}_{m}}{\partial g^{\epsilon\omega}\partial
g^{\zeta\beta}}\bigg],
\end{align}
where $f_{\mathcal{T}}=\frac{\partial
f(\mathcal{R},\mathcal{L}_{m},\mathcal{T})}{\partial \mathcal{T}}$,
$f_{\mathcal{L}_{m}}=\frac{\partial
f(\mathcal{R},\mathcal{L}_{m},\mathcal{T})}{\partial
\mathcal{L}_{m}}$ and $f_{\mathcal{R}}=\frac{\partial
f(\mathcal{R},\mathcal{L}_{m},\mathcal{T})}{\partial \mathcal{R}}$.
The mathematical definitions of the D'Alembertian operator and
covariant derivative are
$\Box\equiv(-g)^{\frac{-1}{2}}\partial_\epsilon\big(\sqrt{-g}g^{\epsilon\omega}\partial_{\omega}\big)$
and $\nabla_\epsilon
f_{\mathcal{R}}=f_{\mathcal{R},\epsilon}-\Gamma^{\omega}_{\epsilon\omega}f_{\mathcal{R}}$,
respectively. Equations \eqref{g2}-\eqref{g4} provides after
combining as
\begin{align}\nonumber
\mathcal{G}_{\epsilon\omega}&=\frac{1}{f_{\mathcal{R}}}\bigg[\big\{8\pi+\frac{1}{2}\big(2f_{\mathcal{T}}
+f_{\mathcal{L}_{m}}\big)\big\}\mathcal{T}_{\epsilon\omega}^{(m)}+8\pi\mathcal{E}_{\epsilon\omega}
-(g_{\epsilon\omega}\Box-\nabla_{\epsilon}\nabla_{\omega})f_{\mathcal{R}}\\\label{g4a}
&+\frac{1}{2}\big(f-\mathcal{R}f_{\mathcal{R}}\big)g_{\epsilon\omega}
-\big(2f_{\mathcal{T}}+f_{\mathcal{L}_{m}}\big)\mathcal{L}_{m}g_{\epsilon\omega}
+2f_{\mathcal{T}}g^{\zeta\beta}\frac{\partial^2
\mathcal{L}_{m}}{\partial g^{\epsilon\omega}\partial
g^{\zeta\beta}}\bigg].
\end{align}

The energy-momentum tensor plays a pivotal role in formulating the
gravitational field equations, enabling a precise representation of
the interaction between matter and spacetime curvature. This tensor
proves indispensable in understanding a wide array of physical
phenomena, from celestial bodies' gravitational influences to the
dynamics of fluid systems. Its incorporation not only facilitates
the development of accurate models for diverse astrophysical
scenarios but also contributes to the exploration of fundamental
principles in the broader context. The models possessing the
isotropic fluid among all existing in the literature holds
significance. Its application proves instrumental in various
scientific disciplines, contributing to the development of accurate
and tractable models for the study of diverse physical processes.
Such matter distributions can be defined in the following way
\cite{1i}
\begin{equation}\label{g5}
\mathcal{T}_{\epsilon\omega}^{(m)}=\rho\mathcal{V}_{\epsilon}\mathcal{V}_{\omega}
+\big(\mathcal{V}_{\epsilon}\mathcal{V}_{\omega}+g_{\epsilon\omega}\big)P,
\end{equation}
where $P$ being the pressure, $\rho$ symbolizes the energy density
and $\mathcal{V}_{\epsilon}$ indicates the four-velocity. The
stress-energy tensor expressing the electromagnetic field is defined
by \cite{38b}
\begin{equation*}
\mathcal{E}_{\epsilon\omega}=\frac{1}{4\pi}\left[\frac{1}{4}g_{\epsilon\omega}\mathcal{W}^{\alpha\eta}\mathcal{W}_{\alpha\eta}
-\mathcal{W}^{\eta}_{\epsilon}\mathcal{W}_{\eta\omega}\right],
\end{equation*}
whereas we can write Maxwell equations in concise (or tensorial)
form as
\begin{equation}\label{g5b}
\mathcal{W}^{\epsilon\omega}_{;\omega}=4\pi \varepsilon^{\epsilon},
\quad \mathcal{W}_{[\epsilon\omega;\eta]}=0.
\end{equation}
Here,
$\mathcal{W}_{\epsilon\omega}=\varphi_{\omega;\epsilon}-\varphi_{\epsilon;\omega}$
is written in terms of the four potential defined by
$\psi_{\omega}=\psi(r)\delta^{0}_{\omega}$. Also, the current
$\varepsilon^{\epsilon}$ and charge density $\varpi$ are combined
with each other through the relation $\varepsilon^{\epsilon}=\varpi
\mathcal{V}^{\epsilon}$.

Determining the trace of Eq.\eqref{g4a}, we have the following
\begin{align}\nonumber
&2\big\{f-\big(2f_{\mathcal{T}}+f_{\mathcal{L}_{m}}\big)\mathcal{L}_{m}\big\}
+\mathcal{T}\left(f_\mathcal{T}+8\pi+\frac{1}{2}f_{\mathcal{L}_{m}}\right)\\\nonumber
&-3\Box f_{\mathcal{R}}-\mathcal{R}f_\mathcal{R}
+2f_\mathcal{T}g^{\zeta\beta}g^{\epsilon\omega}
\frac{\partial^2\mathcal{L}_{m}}{\partial g^{\zeta\beta}\partial
g^{\epsilon\omega}}=0.
\end{align}
As functional of this theory is generalized in terms of the geometry
and matter terms, the divergence of the stress-energy tensor becomes
non-null. As a result, a supplementary force emerges within the
gravitational field of a massive object, leading to modifications in
the geodesic trajectory of moving test particles. This force is
mathematically expressed as follows
\begin{align}\nonumber
\nabla^\epsilon\mathcal{T}_{\epsilon\omega}^{(m)}&=\frac{1}{16\pi+2f_{\mathcal{T}}+f_{\mathcal{L}_{m}}}
\bigg[\nabla_\omega\big\{\big(2f_\mathcal{T}+f_{\mathcal{L}_{m}}\big)\mathcal{L}_{m}\big\}
-\mathcal{T}_{\epsilon\omega}^{(m)}\nabla^\epsilon\big(2f_\mathcal{T}+f_{\mathcal{L}_{m}}\big)\\\label{g5a}
&-\big(f_{\mathcal{T}}\nabla_\omega\mathcal{T}+f_{\mathcal{L}_{m}}\nabla_\omega\mathcal{L}_{m}\big)
-8\pi\nabla^\epsilon\mathcal{E}_{\epsilon\omega}-4g^{\zeta\beta}\nabla^\epsilon\left(f_{\mathcal{T}}
\frac{\partial^2\mathcal{L}_{m}}{\partial g^{\epsilon\omega}\partial
g^{\zeta\beta}}\right)\bigg].
\end{align}

Considering a spherical spacetime as an interior geometry is a
significant starting point as its investigation involves
understanding the curvature dynamics and gravitational interactions
specific to a spherical space. The following metric represents such
geometry as
\begin{equation}\label{g6}
ds^2=-e^{\varrho_1(r)} dt^2+e^{\varrho_2(r)}
dr^2+r^2\big(d\theta^2+\sin^2\theta d\phi^2\big),
\end{equation}
where radial/temporal components depend only on the radial
coordinate, showing that the geometry under consideration is static.
We observe the presence of the four-vector in Eq.\eqref{g5} which
now becomes
\begin{equation}\label{g7}
\mathcal{V}_\epsilon=-\delta^0_\epsilon
e^{\frac{\varrho_1}{2}}=(-e^{\frac{\varrho_1}{2}},0,0,0).
\end{equation}
Equation \eqref{g5b} (left) along with the metric \eqref{g6} yields
\begin{equation}\nonumber
\psi''+\frac{1}{2r}\big[4-r(\varrho_1'+\varrho_2')\big]\psi'=4\pi\varpi
e^{\frac{\varrho_1}{2}+\varrho_2},
\end{equation}
where $'=\frac{\partial}{\partial r}$. Implementing an integration
on the above second-order equation results in the following
expression
\begin{equation}\nonumber
\psi'=\frac{s}{r^2}e^{\frac{\varrho_1+\varrho_2}{2}},
\end{equation}
where the total interior charge is defined as $s\equiv
s(r)=\int_{0}^{r}\varpi e^{\frac{\varrho_2}{2}}\bar{r}^2d\bar{r}$.

The isotropic modified field equations representing spherical
structure are now formulated by combining Eqs.\eqref{g4a},
\eqref{g5} and \eqref{g6}. The non-vanishing components are given by
\begin{align}\nonumber
&e^{-\varrho_2}\left(\frac{\varrho_2'}{r}-\frac{1}{r^2}\right)
+\frac{1}{r^2}=\frac{1}{f_{\mathcal{R}}}\bigg[\big\{8\pi+\frac{1}{2}\big(2f_{\mathcal{T}}
+f_{\mathcal{L}_{m}}\big)\big\}\rho+(\Box-\nabla^{0}\nabla_{0})f_{\mathcal{R}}\\\label{g8}
&\quad\quad\quad\quad\quad\quad\quad\quad\quad~
+\frac{s^2}{r^4}-\frac{1}{2}\big(f-\mathcal{R}f_{\mathcal{R}}\big)
+\big(2f_{\mathcal{T}}+f_{\mathcal{L}_{m}}\big)\mathcal{L}_{m}\bigg],\\\nonumber
&e^{-\varrho_2}\left(\frac{1}{r^2}+\frac{\varrho_1'}{r}\right)
-\frac{1}{r^2}=\frac{1}{f_{\mathcal{R}}}\bigg[\big\{8\pi+\frac{1}{2}\big(2f_{\mathcal{T}}
+f_{\mathcal{L}_{m}}\big)\big\}P-(\Box-\nabla^{1}\nabla_{1})f_{\mathcal{R}}\\\label{g9}
&\quad\quad\quad\quad\quad\quad\quad\quad\quad~
-\frac{s^2}{r^4}+\frac{1}{2}\big(f-\mathcal{R}f_{\mathcal{R}}\big)
-\big(2f_{\mathcal{T}}+f_{\mathcal{L}_{m}}\big)\mathcal{L}_{m}\bigg],
\\\nonumber
&\frac{e^{-\varrho_2}}{4}\left[\varrho_1'^2-\varrho_2'\varrho_1'+2\varrho_1''-\frac{2\varrho_2'}{r}+\frac{2\varrho_1'}{r}\right]
=\frac{1}{f_{\mathcal{R}}}\bigg[\big\{8\pi+\frac{1}{2}\big(2f_{\mathcal{T}}
+f_{\mathcal{L}_{m}}\big)\big\}P+\frac{s^2}{r^4}\\\label{g10}
&\quad\quad\quad\quad\quad\quad\quad\quad\quad~
-(\Box-\nabla^{1}\nabla_{1})f_{\mathcal{R}}+\frac{1}{2}\big(f-\mathcal{R}f_{\mathcal{R}}\big)
-\big(2f_{\mathcal{T}}+f_{\mathcal{L}_{m}}\big)\mathcal{L}_{m}\bigg].
\end{align}
Also, the terms $\mathcal{T}$ and $\mathcal{R}$ are defined as
\begin{align}\nonumber
\mathcal{T}&=-\rho+3P,\\\nonumber
\mathcal{R}&=e^{-\varrho_2}\left[\varrho_1''+2(1-e^{\varrho_2})+\frac{\varrho_1'^2}{2}-\frac{\varrho_1'\varrho_2'}{2}
+\frac{2(\varrho_1'-\varrho_2')}{r}\right].
\end{align}
Solving Eqs.\eqref{g8}-\eqref{g10} presents a complex challenge due
to the intricate relationships among multiple quantities, including
$(\varrho_1, \varrho_2, \rho, P, q)$. To address this complexity and
arrive at a definitive solution, it is essential to introduce
specific constraints. Without these constraints, obtaining a unique
solution proves to be an insurmountable task.

\subsection{Embedding Class-one Condition and Smooth Matching of Interior and Exterior Spacetimes}

The incorporation of embedding class-one condition is crucial in
discussing compact stars as they provide essential constraints and
insights into the equilibrium and stability of these astrophysical
objects. This mathematical condition contribute to a more
comprehensive understanding of the physical properties governing
celestial systems, aiding researchers in formulating accurate models
and predictions for their behavior in extreme environments.
According to this, if a tensor $\mathcal{Q}_{\epsilon\omega}$
possessing the property of being symmetry fulfills the Gauss-Codazzi
equations given in the following
\begin{equation}
\mathcal{R}_{\epsilon\omega\alpha\eta}=2\mathbf{p}\mathcal{Q}_{\epsilon[\alpha}\mathcal{Q}_{\eta]\omega},
\quad
\mathcal{Q}_{\epsilon[\omega;\alpha]}-\Gamma^\eta_{\omega\alpha}\mathcal{Q}_{\epsilon\eta}
+\Gamma^\eta_{\epsilon[\omega}\mathcal{Q}_{\alpha]\eta}=0,
\end{equation}
then an $(n-2)$-dimensional space can be embedded into the space of
$(n-1)$-dimension. Here, $\mathcal{Q}_{\omega\epsilon}$ and
$\mathcal{R}_{\epsilon\omega\alpha\eta}$ symbolize the coefficients
of second differential form and the curvature tensor, respectively,
and $\mathbf{p}=\pm1$. The above left equation, known as the Gauss
equation, characterizes the intrinsic geometry of the surface by
relating its curvature to that of the ambient space. This equation
is crucial for understanding how the surface curves within the space
it is embedded. On the other hand, the Codazzi-Mainardi (or Codazzi)
equation given on the right side, expresses the compatibility
between the intrinsic and extrinsic geometry of the surface.
Mathematically, this condition can be written as follows \cite{41i}
\begin{equation}
\mathcal{R}_{2323}\mathcal{R}_{0101}-\mathcal{R}_{0303}\mathcal{R}_{1212}-\mathcal{R}_{1303}\mathcal{R}_{1202}=0,
\end{equation}
resulting in the second-order differential equation after merging
with the metric \eqref{g6} as
\begin{equation}
\big(\varrho_2'-\varrho_1'\big)\varrho_1'e^{\varrho_2}+2\big(1-e^{\varrho_2}\big)\varrho_1''+\varrho_1'^2=0,
\end{equation}
that provides one component, say radial, in terms of the temporal
coefficient. This takes the form
\begin{equation}\label{g14h}
\varrho_2(r)=\ln\big(1+b_1\varrho_1'^2e^{\varrho_1}\big),
\end{equation}
involving $b_1$ as an integration constant. To calculate the
$g_{rr}$ component accurately, it is essential to adopt the temporal
coefficient. For this, we refer to widely recognized $g_{tt}$
component within the astrophysics research community \cite{37,37a}.
This is taken by
\begin{equation}\label{g14i}
\varrho_1(r)=2b_2r^2+\ln b_3,
\end{equation}
possessing two positive constants, denoted as $b_2$ and $b_3$, the
values of which remain unspecified yet. Lake \cite{41j} introduced a
criterion to assess the physical relevance of the metric potentials
under consideration. By applying this evaluation to the specific
component \eqref{g14i}, it is determined whether such component
holds significance in the context of the study. Therefore, we have
\begin{equation}\nonumber
\varrho_1'(r)=4b_2r, \quad \varrho_1''(r)=4b_2.
\end{equation}
We notice that $\varrho_1(r)=\ln b_3,~\varrho_1'(r)=0$ and
$\varrho_1''(r)>0$ at $r=0$, representing the star's center. This
validates the suitability of Eq.\eqref{g14i}. Upon insertion into
Eq.\eqref{g14h}, the function $\varrho_2(r)$ assumes the following
value
\begin{equation}\label{g15}
\varrho_2(r)=\ln\big(1+b_2b_4r^2e^{2b_2r^2}\big),
\end{equation}
where $b_4=16b_1b_2b_3$.

By enforcing consistency at the boundary of the object, junction
conditions enable a smooth transition between different regions of
spacetime, preserving the physical integrity of the model. This is
crucial for accurately representing the gravitational field both
inside and outside the compact object, contributing to a more
realistic understanding of its structure and gravitational effects.
Since a charged interior sphere \eqref{g6} is considered, it must be
adopted the Reissner-Nordstr\"{o}m metric as an exterior spacetime.
With $\mathcal{S}$ and $\mathcal{M}$ as the total charge and mass,
this metric is given as follows
\begin{equation}\label{g20}
ds^2=-\left(1-\frac{2\mathcal{M}}{\mathrm{R}}+\frac{\mathcal{S}^2}{\mathrm{R}^2}\right)dt^2
+\left(1-\frac{2\mathcal{M}}{\mathrm{R}}+\frac{\mathcal{S}^2}{\mathrm{R}^2}\right)^{-1}dr^2
+r^2\big(d\theta^2+\sin^2\theta d\phi^2\big).
\end{equation}
It must be stressed here that the first fundamental forms equals the
radial as well as temporal components of both of the exterior and
interior spacetime at the surface boundary, say mathematically
$\Sigma: r=\mathrm{R}$. This is also true for the term $g_{tt,r}$.
Following this, we have
\begin{eqnarray}\label{g21}
g_{tt}&{_=^\Sigma}&e^{\varrho_1(\mathrm{R})}=b_3e^{2b_2\mathrm{R}^2}=1-\frac{2\mathcal{M}}{\mathrm{R}}
+\frac{\mathcal{S}^2}{\mathrm{R}^2},\\\label{g21a}
g_{rr}&{_=^\Sigma}&e^{\varrho_2(\mathrm{R})}=1+b_2b_4\mathrm{R}^2e^{2b_2\mathrm{R}^2}
=\bigg(1-\frac{2\mathcal{M}}{\mathrm{R}}+\frac{\mathcal{S}^2}{\mathrm{R}^2}\bigg)^{-1},\\\label{g22}
\frac{\partial g_{tt}}{\partial
r}&{_=^\Sigma}&\varrho_1'(\mathrm{R})=4b_2\mathrm{R}=\frac{2\mathcal{M}\mathrm{R}-2\mathcal{S}^2}
{\mathrm{R}\big(\mathrm{R}^2-2\mathcal{M}\mathrm{R}+\mathcal{S}^2\big)}.
\end{eqnarray}
The quartet ($b_1,b_2,b_3,b_4$) can now be easily found by
simultaneously solving Eqs.\eqref{g21}-\eqref{g22}. Their values are
\begin{eqnarray}\label{g23}
b_1&=&\frac{\mathrm{R}^4\big(2\mathcal{M}\mathrm{R}-\mathcal{S}^2\big)}
{4\big(\mathcal{M}\mathrm{R}-\mathcal{S}^2)^2},\\\label{g24}
b_2&=&\frac{\mathcal{M}\mathrm{R}-\mathcal{S}^2}{2\mathrm{R}^2\big(\mathrm{R}^2
-2\mathcal{M}\mathrm{R}+\mathcal{S}^2\big)},\\\label{g25}
b_3&=&\bigg(\frac{\mathrm{R}^2-2\mathcal{M}\mathrm{R}+\mathcal{S}^2}{\mathrm{R}^2}\bigg)
e^{\frac{\mathcal{M}\mathrm{R}-\mathcal{S}^2}{2\mathcal{M}\mathrm{R}-\mathrm{R}^2-\mathcal{S}^2}},\\\label{g25a}
b_4&=&\frac{2\big(2\mathcal{M}\mathrm{R}-\mathcal{S}^2\big)}{\mathcal{M}\mathrm{R}-\mathcal{S}^2}
e^{\frac{\mathcal{M}\mathrm{R}-\mathcal{S}^2}{2\mathcal{M}\mathrm{R}-\mathrm{R}^2-\mathcal{S}^2}}.
\end{eqnarray}
Determining the dimension of these constants is significant in such
analysis of the compact stars. We find that the constant $b_1$ has a
dimension of $\ell^2$ and $b_2$ having $\frac{1}{\ell^2}$. However,
the other constants, i.e., $b_3$ and $b_4$ have null dimensions. The
graphical interpretation of the solutions (which shall be obtained
later) needs some definite values of these constants. In order to
make this possible, a star LMC X-4 is considered along with its
observed data \cite{42aa}. In the following, the numerical values of
these four constants are calculated in Tables \textbf{I} and
\textbf{II} for multiple stars by choosing the exterior charge as
$0.2$ and $0.8$, respectively with $\textbf{M}_{\bigodot}$ being the
mass of the Sun.
\begin{table}[H]
\scriptsize \centering \caption{Values of embedding class-one
constants ($b_1,b_2,b_3,b_4$) for $\mathcal{S}=0.2$.} \label{Table1}
\vspace{+0.07in} \setlength{\tabcolsep}{0.95em}
\begin{tabular}{cccccc}
\hline\hline Compact Stars & LMC X-4 & 4U 1820-30 & SMC X-4 & SAX J
1808.4-3658 & Her X-I
\\\hline $\mathcal{M}~(\textbf{M}_{\bigodot})$ & 1.04 & 1.58 & 1.29 & 0.9 & 0.85
\\\hline $\mathrm{R}~(\mathrm{km})$ & 9.1 & 7.95 & 8.1 & 8.831 & 8.301
\\\hline $b_1$ & 187.321 & 162.134 & 181.623 & 190.332 & 213.199
\\\hline $b_2$ & 0.00211845 & 0.00316083 & 0.00241950 & 0.00197398 &
0.00170023
\\\hline
$b_3$ & 0.471224 & 0.289282 & 0.390542 & 0.519452 & 0.552855
\\\hline
$b_4$ & 2.99193 & 2.37202 & 2.74590 & 3.12262 & 3.20645 \\
\hline\hline
\end{tabular}
\end{table}
\begin{table}[H]
\scriptsize \centering \caption{Values of embedding class-one
constants ($b_1,b_2,b_3,b_4$) for $\mathcal{S}=0.8$.} \label{Table2}
\vspace{+0.07in} \setlength{\tabcolsep}{0.95em}
\begin{tabular}{cccccc}
\hline\hline Compact Stars & LMC X-4 & 4U 1820-30 & SMC X-4 & SAX J
1808.4-3658 & Her X-I
\\\hline $b_1$ & 201.499 & 169.301 & 191.865 & 207.971 & 233.803
\\\hline $b_2$ & 0.00199084 & 0.00302633 & 0.00230181 & 0.00183519 &
0.00157849
\\\hline
$b_3$ & 0.486203 & 0.300188 & 0.40315 & 0.536174 & 0.569192
\\\hline
$b_4$ & 3.12067 & 2.46086 & 2.84874 & 3.27424 & 3.36101 \\
\hline\hline
\end{tabular}
\end{table}
The relation between the values of $b_1,~b_3,~b_4$ and the electric
charge is evident, as an increase in the later term is directly
associated with the variations in these three constants. However,
the value of $b_2$ is decreased as the electric charge is increased.

\section{Physical Requirements admitting by Stellar Models}

In this section, we review multiple conditions that have been
discussed in the literature whose satisfaction leads to the compact
interior models to be physically relevant \cite{ab}-\cite{af}. We
highlight some interesting and necessary conditions among them that
must be discussed while studying the stars in the following.
\begin{itemize}
\item A critical aspect involves the investigation of geometric quantities such as $e^{\varrho_1}$ and
$e^{\varrho_2}$. It must be verified that both these components are
positive to maintain physical significance. Additionally, the
regularity of these functions should be confirmed within the defined
physical domain, ensuring they do not exhibit singularities.

\item Within compact stars, the behavior of energy density and pressure, along with their first two
derivatives, is critical in understanding the internal configuration
of these astrophysical objects. Typically, as one moves from the
stellar surface towards the center, both these parameters tend to
increase, reaching their maximum values at the core. The first
derivatives with respect to radial distance capture the rate of
change of these quantities, highlighting the distribution of mass
and the response of matter to gravitational forces.

\item A point of discussion among researchers is the mass function that describes the fluid content enclosed by
a body. This helps in understanding the gravitational impact of that
structure. We express this in the form of energy density as
\begin{equation}\label{g39}
m(r)=\frac{1}{2}\int_{0}^{\mathrm{R}}\bar{r}^2\rho d\bar{r}.
\end{equation}
The strength of a field surrounding a self-gravitating structure due
to its gravity in relation with its size is measured by the
compactness. It is actually a ratio between the mass of a body and
its radius. Its expression is given by
\begin{equation}\label{g40}
\lambda(r)=\frac{m(r)}{r},
\end{equation}
which must be less than $\frac{4}{9}$ to get a physically relevant
interior \cite{42a}. The redshift characterizes the extent to which
photons are stretched as they climb out of the gravitational well of
a compact star. We describe it as
\begin{equation}\label{g41}
\textbf{z}(r)=\frac{1-\sqrt{1-2\lambda(r)}}{\sqrt{1-2\lambda(r)}}.
\end{equation}
It has been found that the redshift at the surface boundary must not
be higher than $2$ \cite{42a}, i.e., $\textbf{z}_{\Sigma} \leq 2$.
On the other hand, when Ivanov dealt with anisotropic pressure
fluid, he established this limit to be $5.211$ \cite{42c}.

\item The incorporation of energy conditions holds paramount significance
in discussions about compact stars. These conditions play a pivotal
role in constraining the matter distribution within these dense
astrophysical objects. By imposing constraints on energy density and
pressure, energy conditions ensure the physical viability of
solutions, guiding the development of realistic models for
self-gravitating structures. Upholding these conditions not only
fosters mathematical consistency but also provides crucial insights
into the nature of matter supporting these stellar objects. For the
case of charged fluid, they have the form
\begin{eqnarray}\nonumber
&&\rho+P \geq 0, \quad \rho-P+\frac{s^2}{4\pi r^4} \geq 0, \quad
\rho+3P+\frac{s^2}{4\pi r^4} \geq 0.
\end{eqnarray}

\item Various approaches have been proposed to
assess the stability of celestial systems, with one method involving
the consideration of the causality condition derived from the sound
speed, expressed as $v_{s}^{2}=\frac{dP}{d\rho}$. According to Abreu
et al. \cite{42bb}, this condition ensures that information within
the stellar medium propagates at speeds less than the speed of
light, preventing causality violations, i.e., $0 < v_{s}^{2} < 1$.
At the same time, one can check the stability by studying the
thermodynamic behavior of the celestial object. This can be
discussed through the adiabatic index, indicated by $\Gamma_{ai}$,
whose formula is given as follows
\begin{equation}\nonumber
\Gamma_{ai}=\frac{\rho+P}{P}\left(\frac{dP}{d\rho}\right).
\end{equation}
To maintain the equilibrium of a compact model, the outward pressure
must be as enough as it can counterbalance the force of gravity
acting inward. This can only be achieved if the adiabatic index gain
its value greater than $\frac{4}{3}$ everywhere \cite{42ba}.
\end{itemize}

\section{Brief Discussion on Two Different $f(\mathcal{R},\mathcal{L}_{m},\mathcal{T})$ Models}

In this section, we obtain different solutions and perform a
comprehensive analysis on their physical properties corresponding to
two distinct models of the considered modified gravitational theory.
We further extend our exploration by choosing two different forms of
the matter Lagrangian density, one in terms of the energy density
and other in the form of an isotropic pressure. A large body of
literature guarantees the formation of acceptable solutions for both
these choices. Now, we discuss them one by one in the following.

\subsection{Model I}

Two different $f(\mathcal{R},\mathcal{L}_{m},\mathcal{T})$ models
have been extensively discussed along with their cosmological
implications by Haghani and Harko \cite{25ah}. The major difference
between these models is that one is based on the minimal
fluid-geometry interaction and the other model contains product
terms, representing non-minimal coupling. We, firstly, consider a
minimal interaction model as it is much easy to handle the
corresponding calculations due to the appearance of linear-order
fluid variables. This model, containing a triplet
($\beta_0,\beta_1,\beta_2$) of real-valued parameters, has the form
\begin{equation}\label{g51}
f(\mathcal{R},\mathcal{L}_{m},\mathcal{T})=
\mathcal{R}+\beta_0f_1(\mathcal{R})+2\beta_1
f_2(\mathcal{L}_{m})+\beta_2 f_3(\mathcal{T}),
\end{equation}
whose linear, and hence, simplified form is written as
\begin{equation}\label{g51a}
f(\mathcal{R},\mathcal{L}_{m},\mathcal{T})=
\mathcal{R}+2\beta_1\mathcal{L}_{m}+\beta_2\mathcal{T}.
\end{equation}

\subsubsection{Stellar Solution for $\mathcal{L}_{m}=P$}

In this case, we adopt the Lagrangian density to be
$\mathcal{L}_{m}=P$. The above model along with this choice and the
field equations \eqref{g8}-\eqref{g10} provide
\begin{align}\label{g52}
&e^{-\varrho_2}\left(\frac{\varrho_2'}{r}-\frac{1}{r^2}\right)
+\frac{1}{r^2}=\left(8\pi+\beta_1+\beta_2\right)\rho+\frac{s^2}{r^4}-\beta_2\left(P-\frac{\rho}{2}\right),\\\label{g52a}
&e^{-\varrho_2}\left(\frac{1}{r^2}+\frac{\varrho_1'}{r}\right)
-\frac{1}{r^2}=\left(8\pi+\beta_1+\beta_2\right)P-\frac{s^2}{r^4}+\beta_2\left(P-\frac{\rho}{2}\right),\\\label{g52b}
&\frac{e^{-\varrho_2}}{4}\left[\varrho_1'^2-\varrho_2'\varrho_1'+2\varrho_1''-\frac{2\varrho_2'}{r}+\frac{2\varrho_1'}{r}\right]
=\left(8\pi+\beta_1+\beta_2\right)P+\frac{s^2}{r^4}+\beta_2\left(P-\frac{\rho}{2}\right).
\end{align}
Since we have three equations in three unknowns (two fluid
parameters and the charge), it is easy enough to calculate their
explicit expressions and then merge them with Eqs.\eqref{g14i} and
\eqref{g15}. This manipulation gives
\begin{align}\nonumber
\rho&=\frac{b_2}{\big(b_2 b_4 r^2 e^{2 b_2 r^2}+1\big)^2\big\{5
\beta_2 ^2+7 \beta_2  \beta_1 +8 \pi  (7 \beta_2 +4 \beta_1 )+2
\beta_1 ^2+128 \pi ^2\big\}}\\\nonumber &\times \big[8 \beta_2 +b_2
r^2 \big\{b_4^2 (\beta_2 +\beta_1 +8 \pi ) e^{4 b_2 r^2}-4 (\beta_2
+\beta_1 +8 \pi )+4 b_4 e^{2 b_2 r^2}\\\label{g53} &\times (7
\beta_2 +3 \beta_1 +24 \pi )\big\}+2 b_4 (5 \beta_2 +3 \beta_1 +24
\pi ) e^{2 b_2 r^2}\big],\\\nonumber P&=\frac{-b_2}{\big(b_2 b_4 r^2
e^{2 b_2 r^2}+1\big)^2\big\{5 \beta_2 ^2+7 \beta_2  \beta_1 +8 \pi
(7 \beta_2 +4 \beta_1 )+2 \beta_1 ^2+128 \pi ^2\big\}}\\\nonumber
&\times \big[b_2 r^2 \big\{b_4^2 (\beta_2 +\beta_1 +8 \pi ) e^{4 b_2
r^2}-4 (\beta_2 +\beta_1 +8 \pi )-4 b_4 e^{2 b_2 r^2}\\\label{g53a}
&\times (3 \beta_2 +\beta_1 +8 \pi )\big\}+2 \big\{(\beta_1 +8 \pi )
b_4 e^{2 b_2 r^2}-2 (3 \beta_2 +2 \beta_1 +16 \pi
)\big\}\big],\\\label{g53b} s&=\frac{b_2 r^3 \big(b_4 e^{2 b_2
r^2}-2\big)}{\sqrt{2} \big(b_2 b_4 r^2 e^{2 b_2 r^2}+1\big)} .
\end{align}

\subsubsection{Stellar Solution for $\mathcal{L}_{m}=-\rho$}

The field equations are now calculated for the other choice as
$\mathcal{L}_{m}=-\rho$. When we join this with
Eqs.\eqref{g8}-\eqref{g10} and \eqref{g51a}, this results in
\begin{align}\label{g54}
&e^{-\varrho_2}\left(\frac{\varrho_2'}{r}-\frac{1}{r^2}\right)
+\frac{1}{r^2}=\left(8\pi+\beta_1+\beta_2\right)\rho+\frac{s^2}{r^4}-\frac{\beta_2}{2}\left(3P+\rho\right),\\\label{g54a}
&e^{-\varrho_2}\left(\frac{1}{r^2}+\frac{\varrho_1'}{r}\right)
-\frac{1}{r^2}=\left(8\pi+\beta_1+\beta_2\right)P-\frac{s^2}{r^4}+\frac{\beta_2}{2}\left(3P+\rho\right),\\\label{g54b}
&\frac{e^{-\varrho_2}}{4}\left[\varrho_1'^2-\varrho_2'\varrho_1'+2\varrho_1''-\frac{2\varrho_2'}{r}+\frac{2\varrho_1'}{r}\right]
=\left(8\pi+\beta_1+\beta_2\right)P+\frac{s^2}{r^4}+\frac{\beta_2}{2}\left(3P+\rho\right).
\end{align}
The isotropic fluid parameters can explicitly be obtained by using
only Eqs.\eqref{g54} and \eqref{g54a}. Using them with components
\eqref{g14i} and \eqref{g15} leads to
\begin{align}\nonumber
\rho&=\frac{b_2}{2\big(b_2 b_4 r^2 e^{2 b_2 r^2}+1\big)^2\big\{2
\beta_2 ^2+3 \beta_2  \beta_1 +8 \pi  \big(3 \beta_2 +2 \beta_1
\big)+\beta_1 ^2+64 \pi ^2\big\}}\\\nonumber &\times \big[b_2 r^2
\big\{b_4^2 (\beta_2 +\beta_1 +8 \pi ) e^{4 b_2 r^2}-4 (\beta_2
+\beta_1 +8 \pi )+12 b_4 e^{2 b_2 r^2}\\\label{g55} &\times (3
\beta_2 +\beta_1 +8 \pi ) \big\}+6 \big\{2 \beta_2 +b_4 (2 \beta_2
+\beta_1 +8 \pi ) e^{2 b_2 r^2}\big\}\big],\\\nonumber
P&=\frac{-b_2}{2 \big(\beta_2 +\beta_1 +8 \pi \big) \big(2 \beta_2
+\beta_1 +8 \pi \big)\big(b_2 b_4 r^2 e^{2 b_2
r^2}+1\big)^2}\\\nonumber &\times \big[b_2 r^2 \big\{b_4^2 (\beta_2
+\beta_1 +8 \pi ) e^{4 b_2 r^2}-4 (\beta_2 +\beta_1 +8 \pi )-4
b_4e^{2 b_2 r^2} \\\label{g55a} &\times (\beta_1-\beta_2 +8 \pi )
\big\}+2 \big\{b_4 (2 \beta_2 +\beta_1 +8 \pi ) e^{2 b_2 r^2}-2
(\beta_2 +2 \beta_1 +16 \pi )\big\}\big].
\end{align}
When we solve Eqs.\eqref{g54a} and \eqref{g54b}, the value of the
charge is found to be the same that is already provided in
\eqref{g53b}.

We now perform a graphical check to explore the physical relevancy
of the obtained minimally coupled solutions. For this, we plot
several physical properties (that have been discussed earlier) which
are basically the requirements to be fulfilled. Since there are two
parameters involved in the considered modified model along with
charge, we adopt their numerical values or ranges to analyze the
impact on the stellar models as $\beta_2=0.1,~0.8$,
$\mathcal{S}=0.3$ and $\beta_1 \in [0.1,2]$. The question arises
here is why we choose these particular values of the model
parameters? Haghani and Harko \cite{25ah} performed a comprehensive
analysis in the context of model I and built some cosmological
solutions, i.e., radiation dominated and dust universe. They used
different combinations of parametric values such as both positive,
both negative or alternative choices, etc. From this, they deduced
that the non-negative values of both these parameters provide a best
fit with the observational data. So, we initially choose both values
and observe that only positive values of $\beta_2$ yield promising
results. For instance, its negative choices produce negative radial
pressure near the spherical junction, which is in contrast with the
requirement of physically existing compact stellar structures.

We confirm the behavior of potentials \eqref{g14i} and \eqref{g15},
and found them in agreement with the needed criterion. However, we
do not add their plots here. Further, the exploration of the fluid
sector (such as isotropic pressure and energy density in this case)
is also performed through plotting the corresponding variables in
Figures \textbf{1} and \textbf{2}. We notice their required behavior
everywhere from the center of a compact star to its boundary
surface. From these plotting, we also observe that when the
parameters $\beta_1$ and $\beta_2$ increase, both the fluid
parameters gain less values. This implies that the higher, the
values of these parameters, the less dense, the interiors are. The
isotropic pressure needs to be null at the spherical interface which
is also ensured for each case.

There exist two approaches to calculate the interior mass of any
self-gravitating fluid distribution, one in terms of the geometry
and other in the form of matter. The former approach is failed to
analyze how the modified theory affects the interior mass,
therefore, we are left with the later choice \eqref{g39}. We plot
this in Figure \textbf{3} and find it to be a rising function of
$r$. When the parameters $\beta_1$ and $\beta_2$ take smaller
values, we get the structures with higher mass. Two other factors
are also shown in the same Figure, indicating themselves consistent
with the required behavior. Figures \textbf{4} and \textbf{5} admit
the positive behavior of energy bounds, naming the developed models
as physically viable structures. Finally, both the causality and
thermodynamic variations are observed in Figures \textbf{6} and
\textbf{7}, indicating the stability of the obtained modified
stellar solutions.
\begin{figure}\center
\epsfig{file=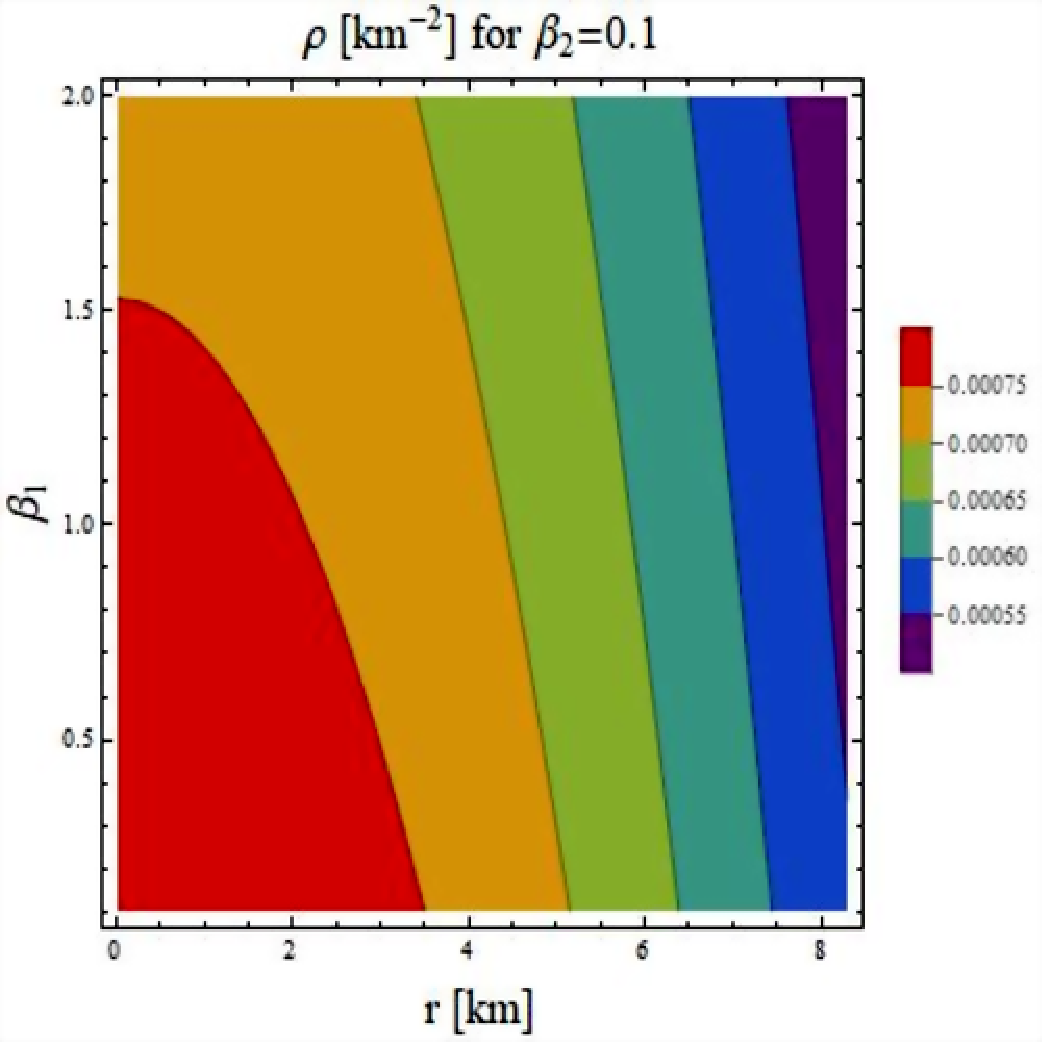,width=0.4\linewidth}\epsfig{file=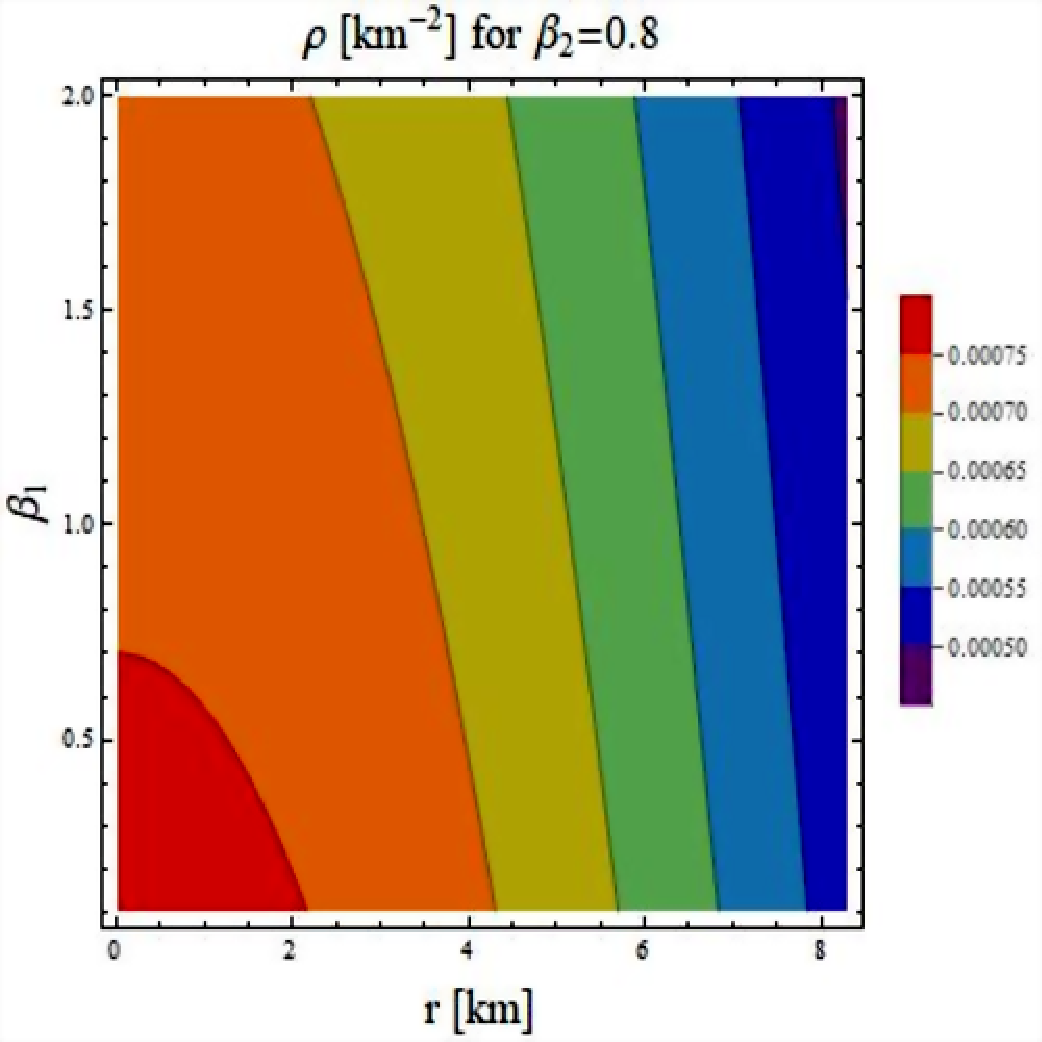,width=0.4\linewidth}
\epsfig{file=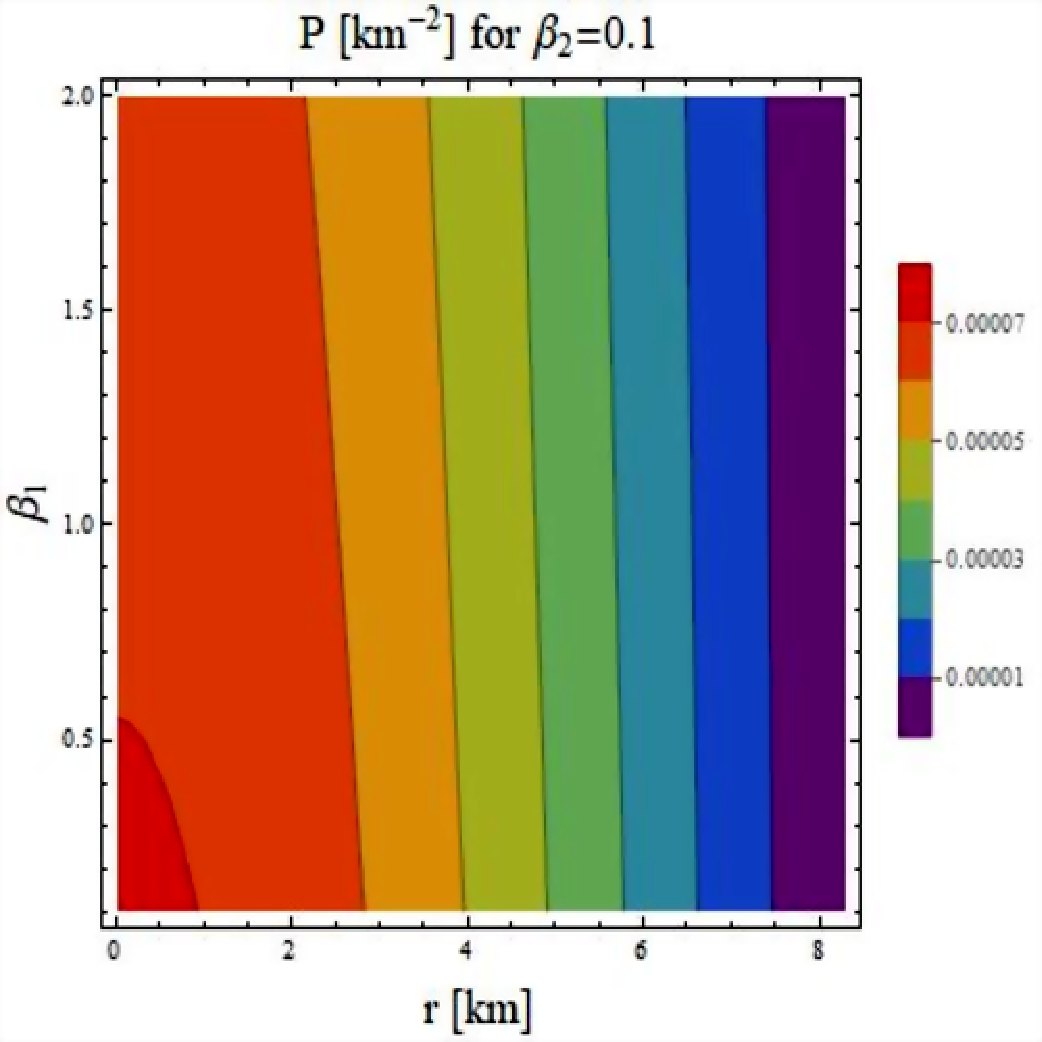,width=0.4\linewidth}\epsfig{file=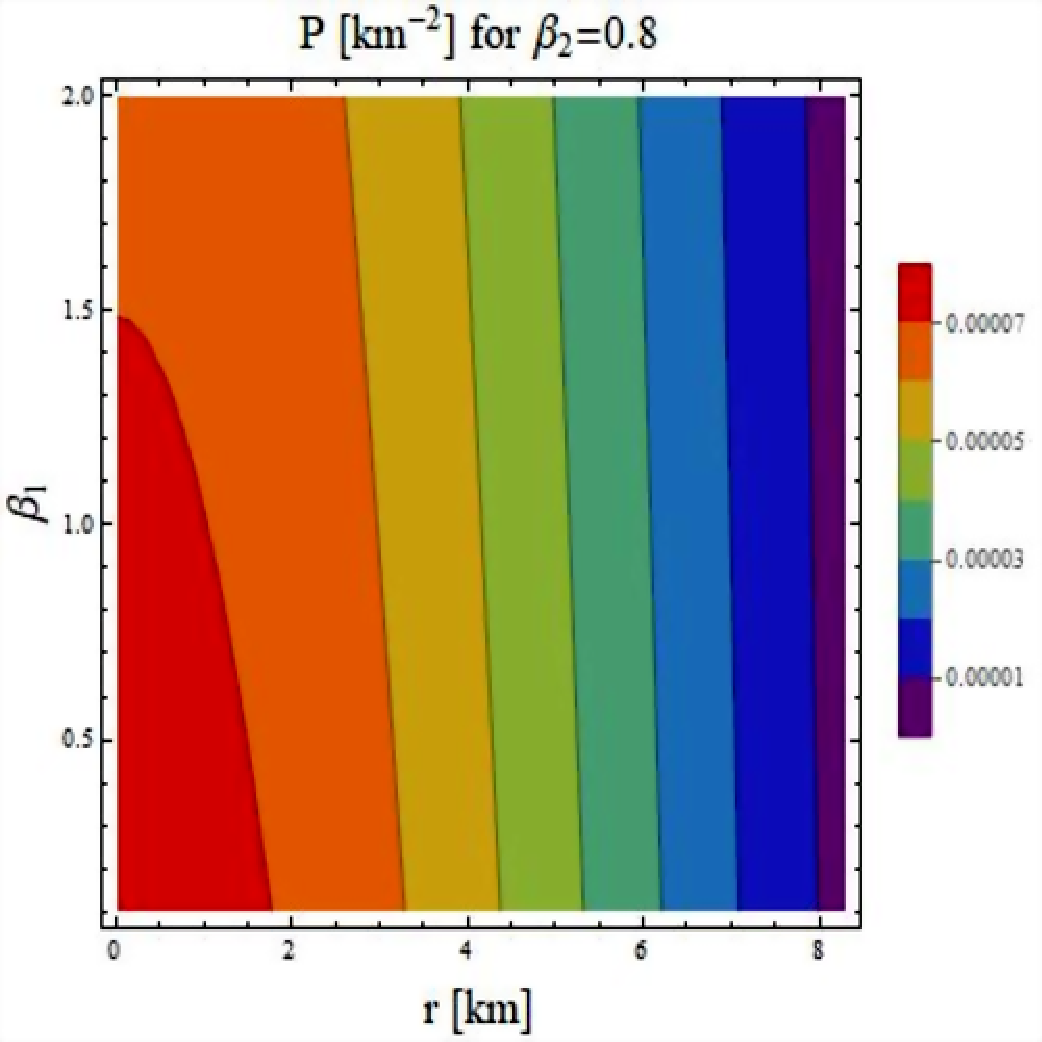,width=0.4\linewidth}
\caption{Energy density and pressure for model I with
$\mathcal{L}_{m}=P$.}
\end{figure}
\begin{figure}\center
\epsfig{file=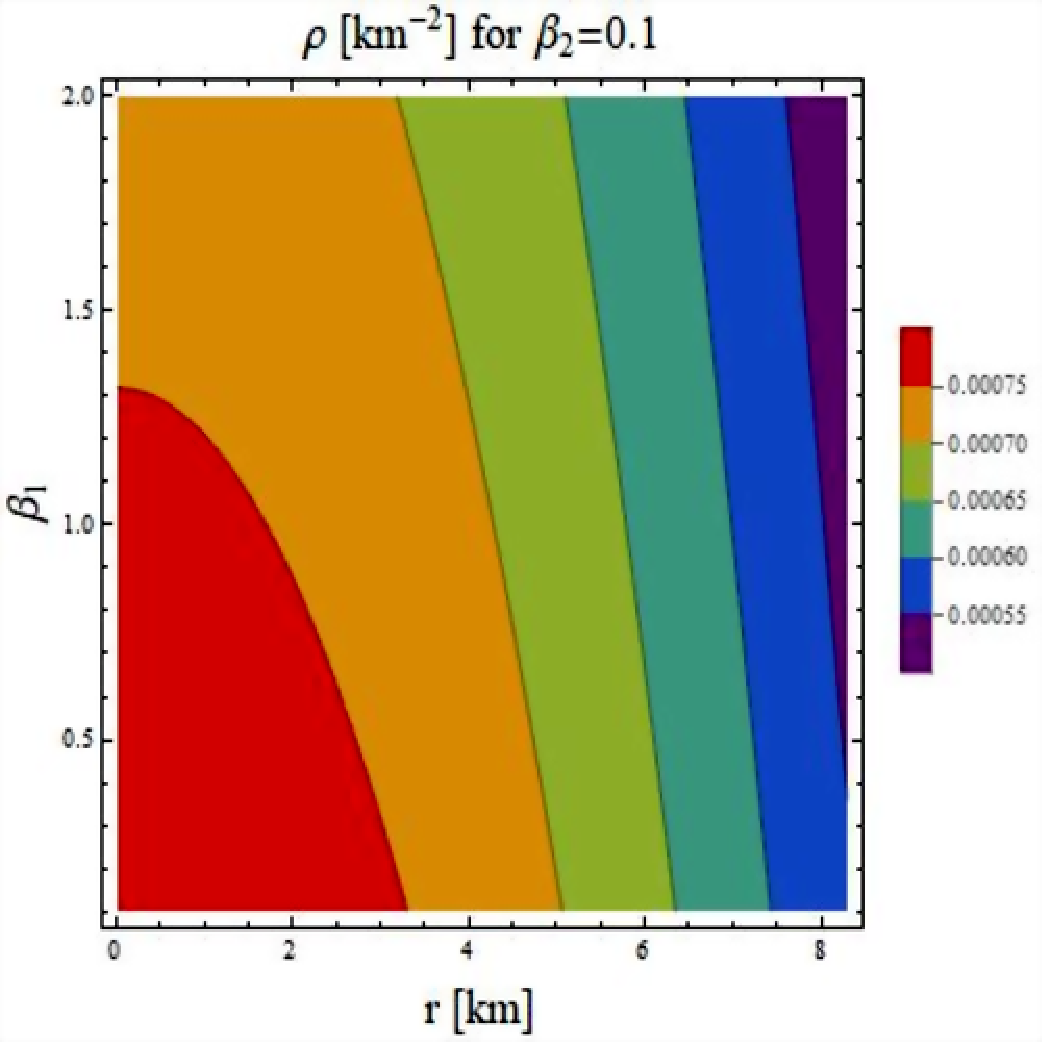,width=0.4\linewidth}\epsfig{file=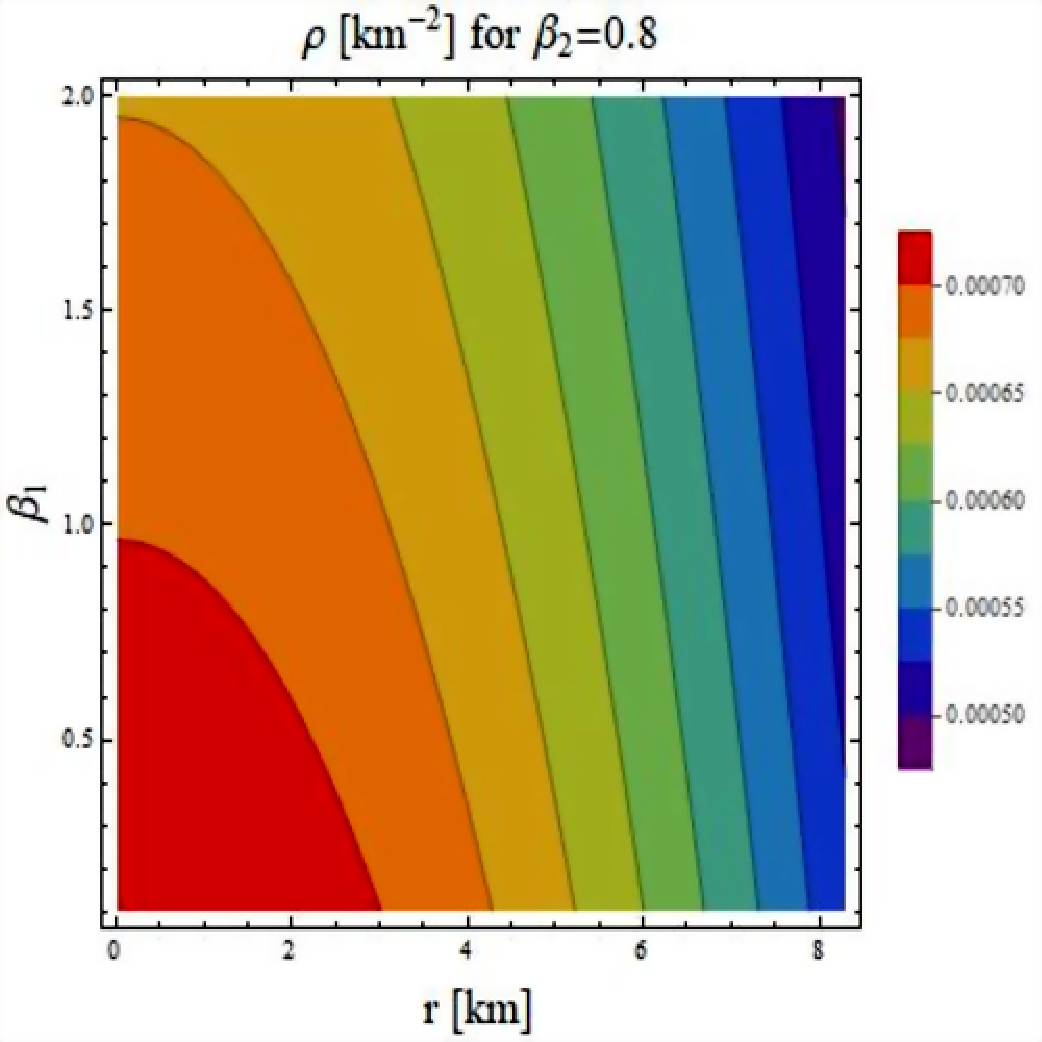,width=0.4\linewidth}
\epsfig{file=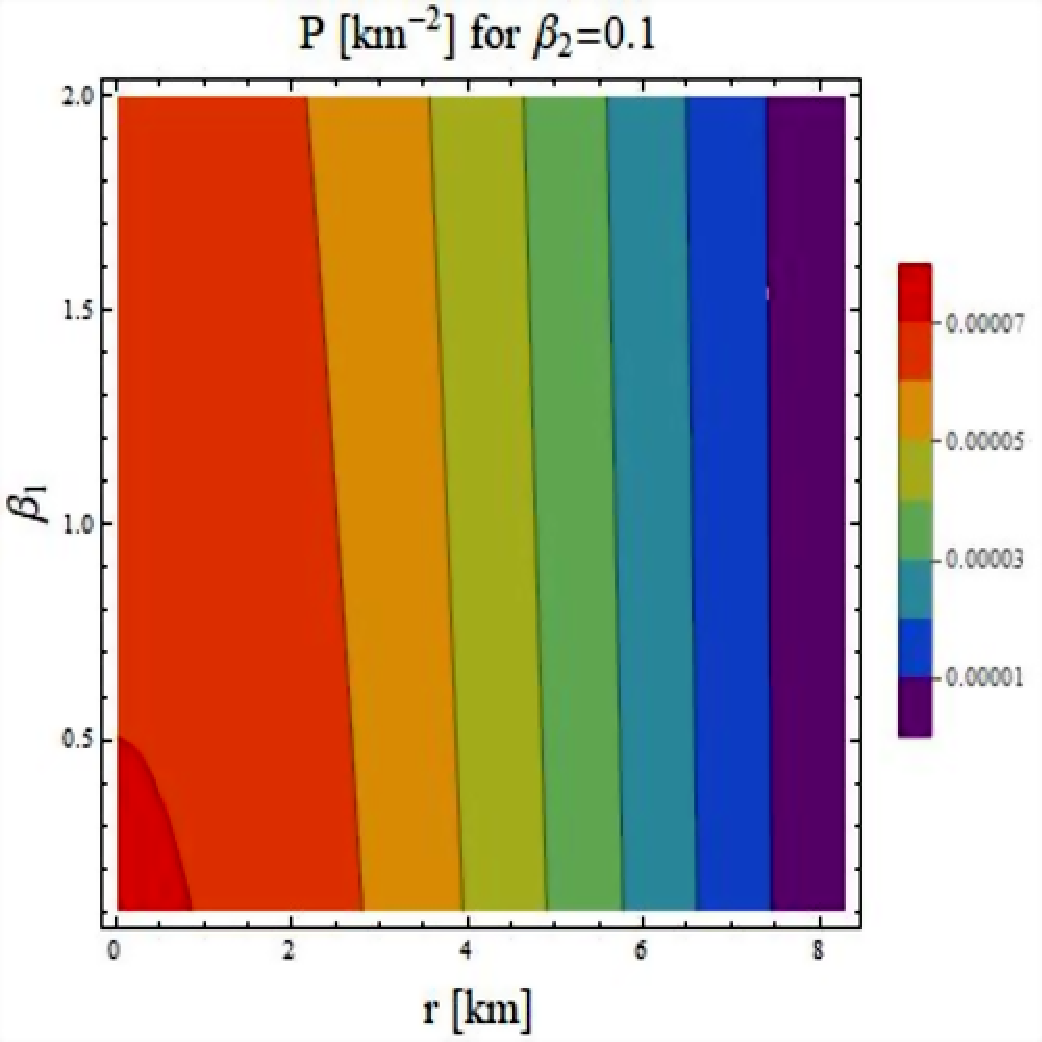,width=0.4\linewidth}\epsfig{file=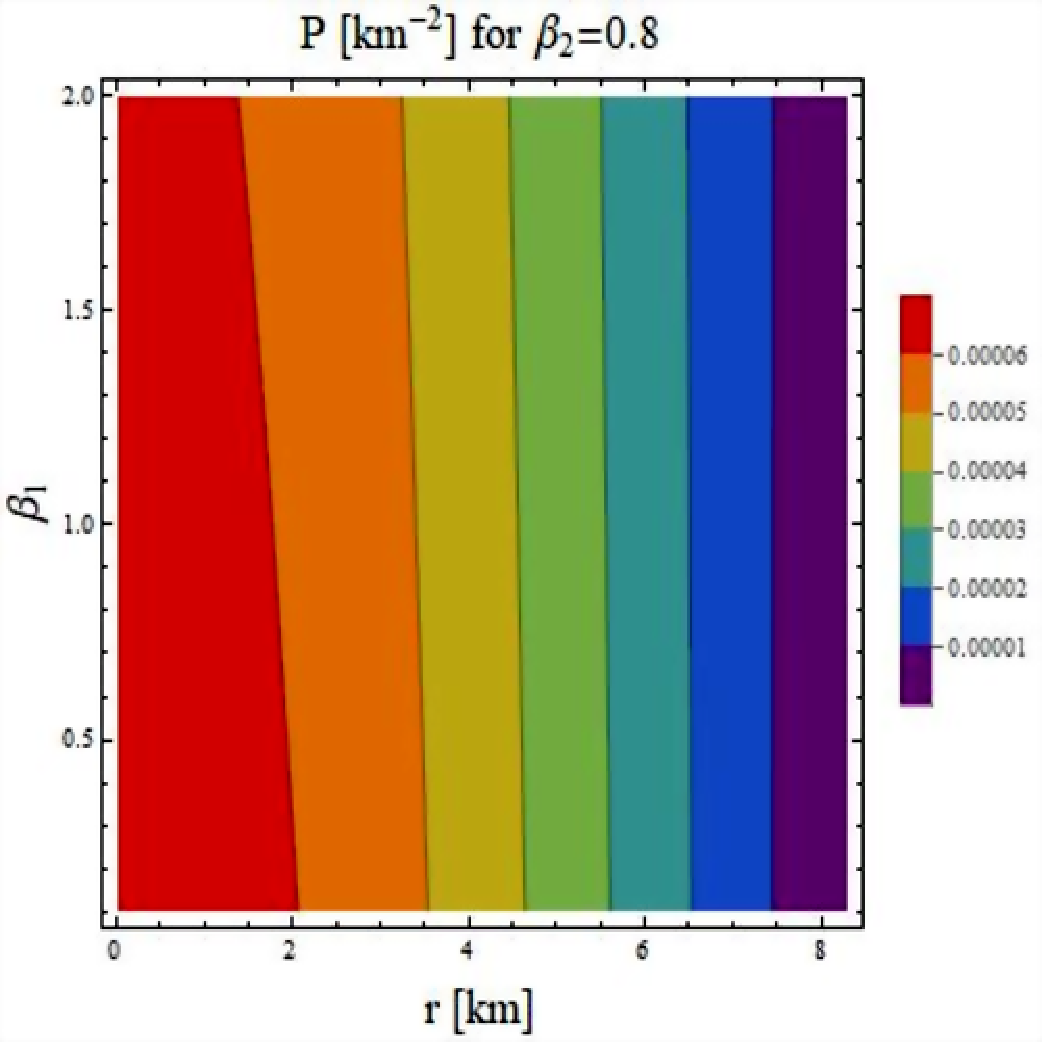,width=0.4\linewidth}
\caption{Energy density and pressure for model I with
$\mathcal{L}_{m}=-\rho$.}
\end{figure}
\begin{figure}\center
\epsfig{file=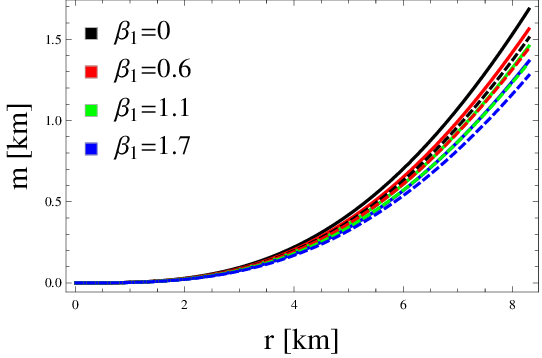,width=0.4\linewidth}\epsfig{file=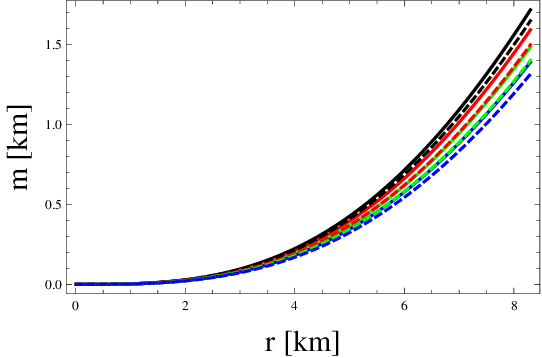,width=0.4\linewidth}
\epsfig{file=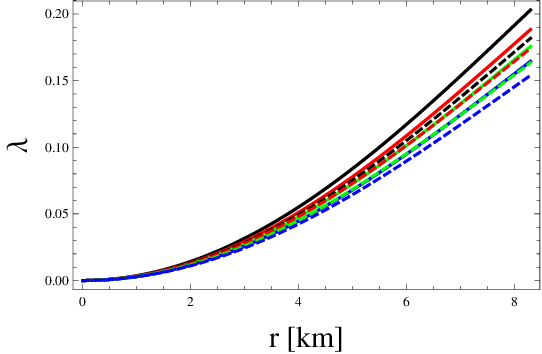,width=0.4\linewidth}\epsfig{file=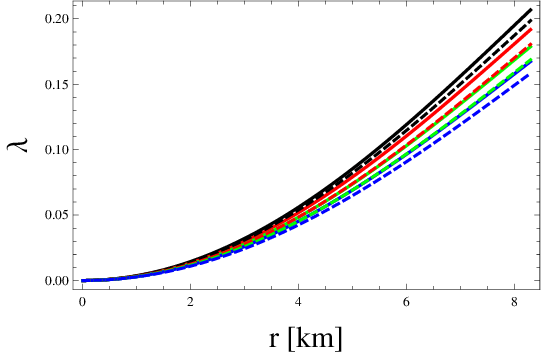,width=0.4\linewidth}
\epsfig{file=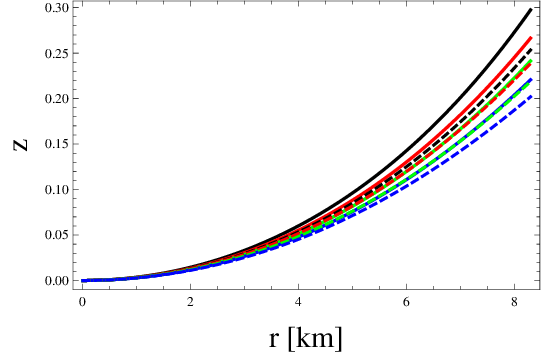,width=0.4\linewidth}\epsfig{file=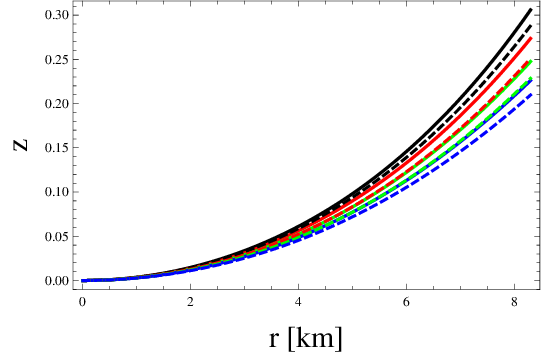,width=0.4\linewidth}
\caption{Physical factors for model I with $\mathcal{L}_{m}=P$
(left) and $-\rho$ (right).}
\end{figure}
\begin{figure}\center
\epsfig{file=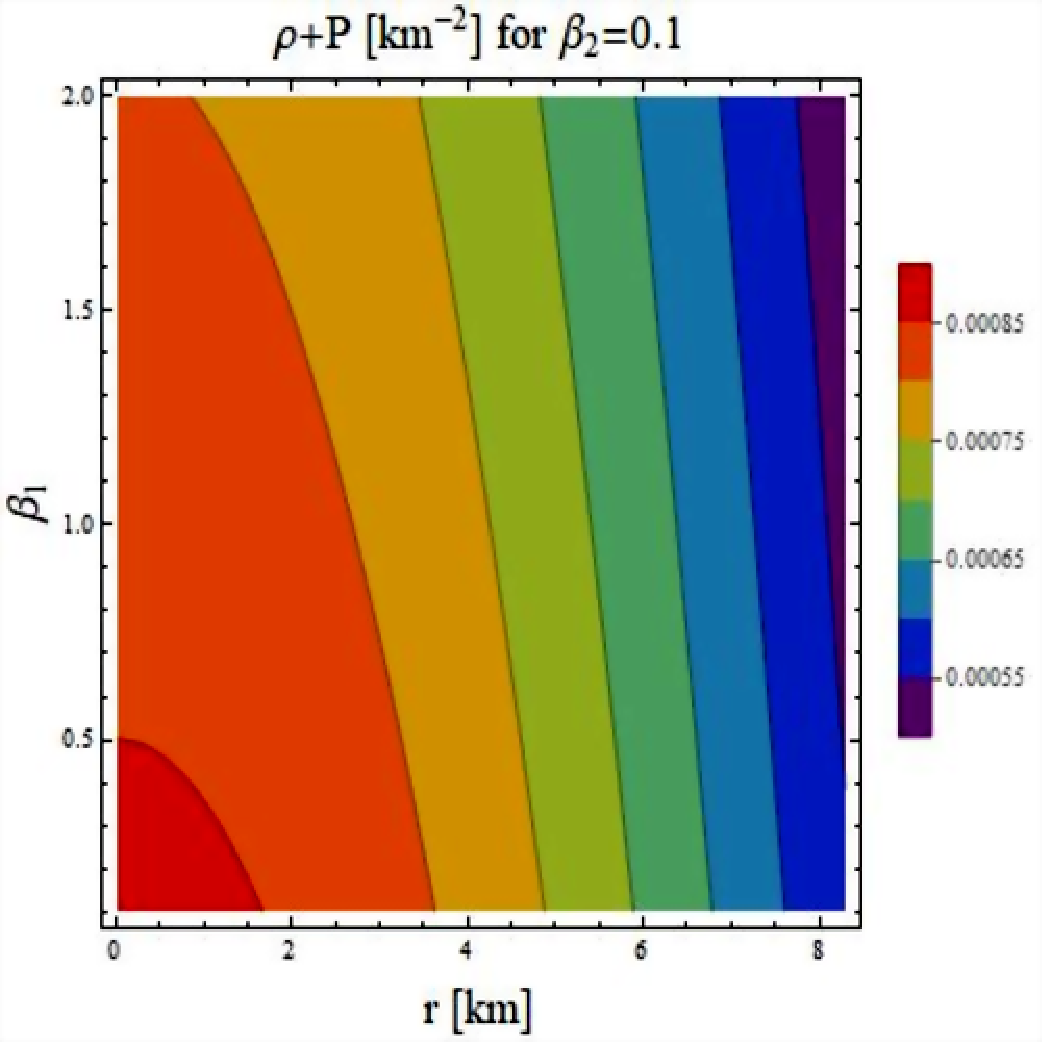,width=0.4\linewidth}\epsfig{file=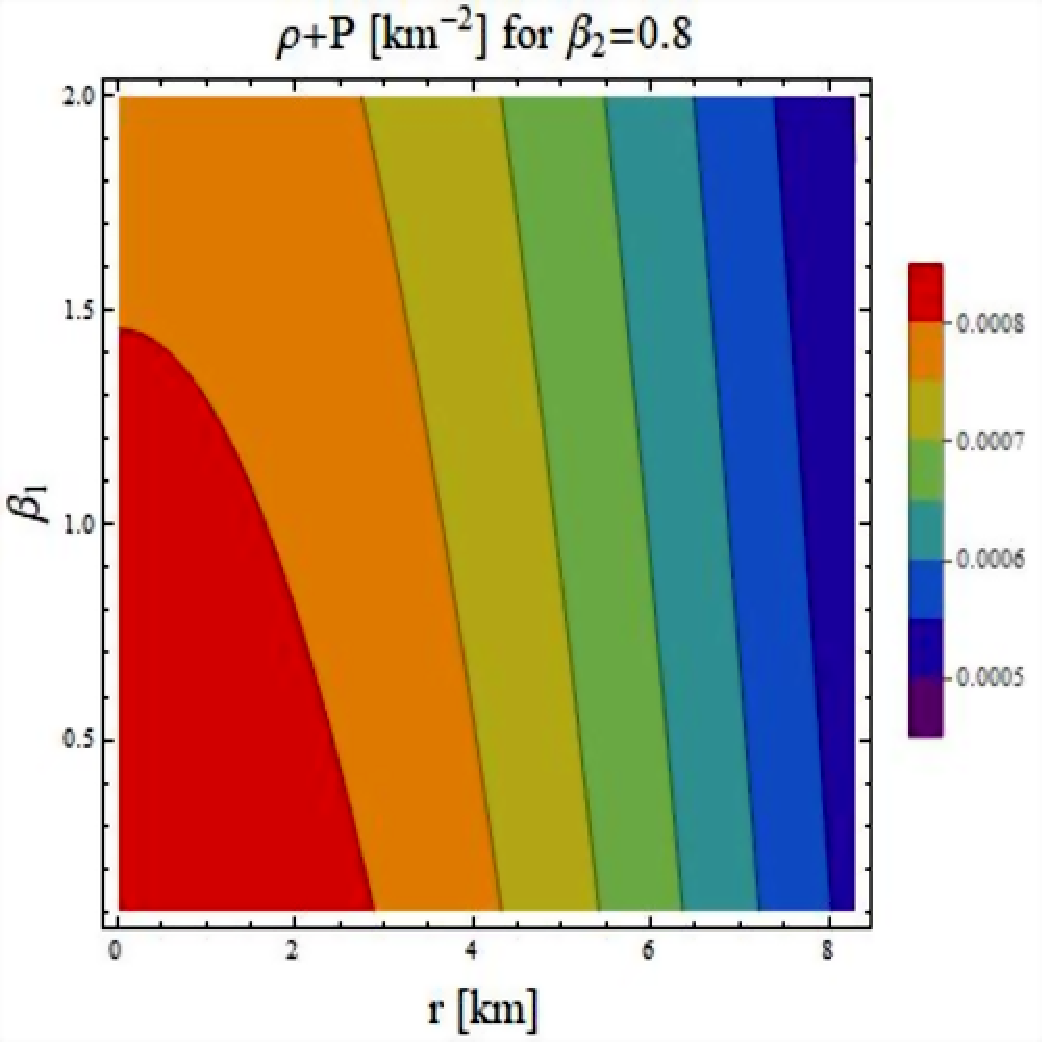,width=0.4\linewidth}
\epsfig{file=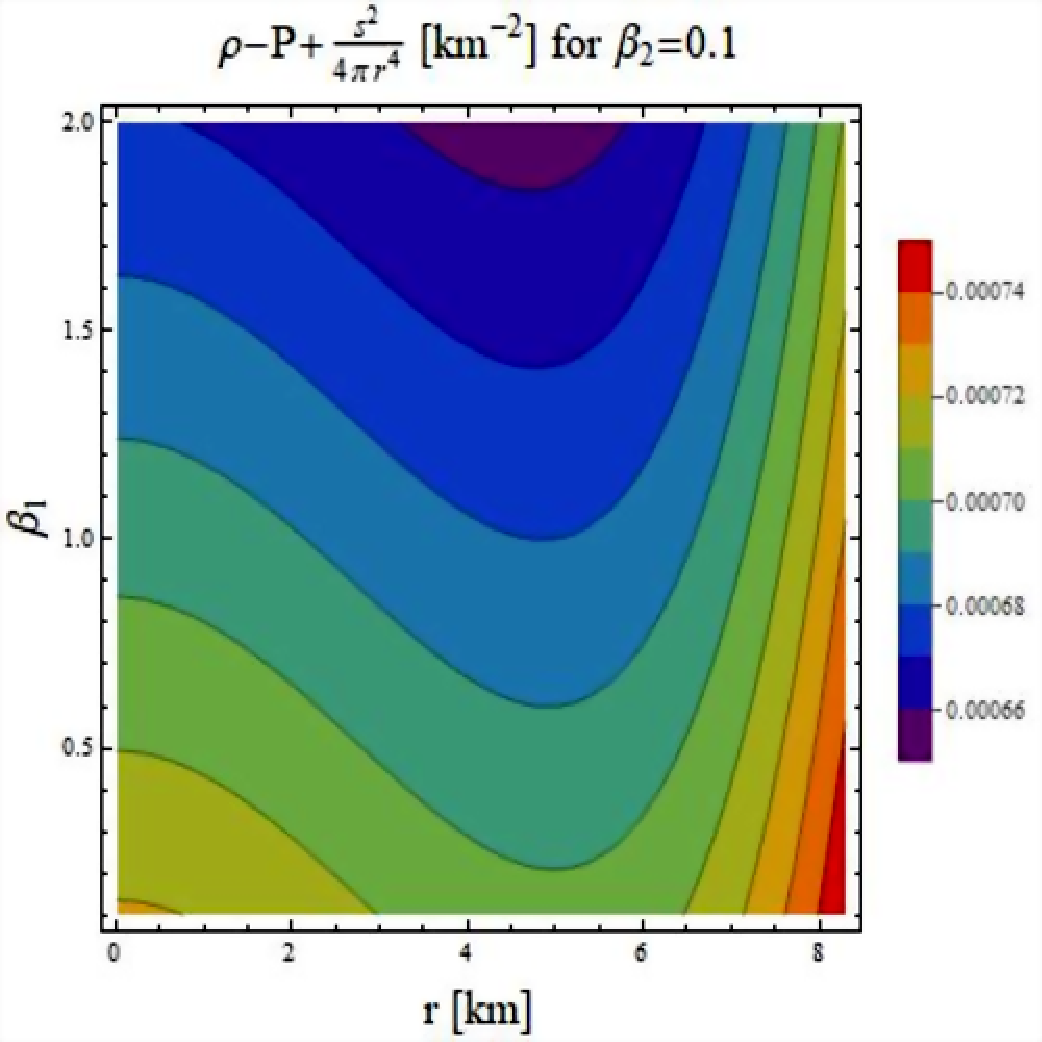,width=0.4\linewidth}\epsfig{file=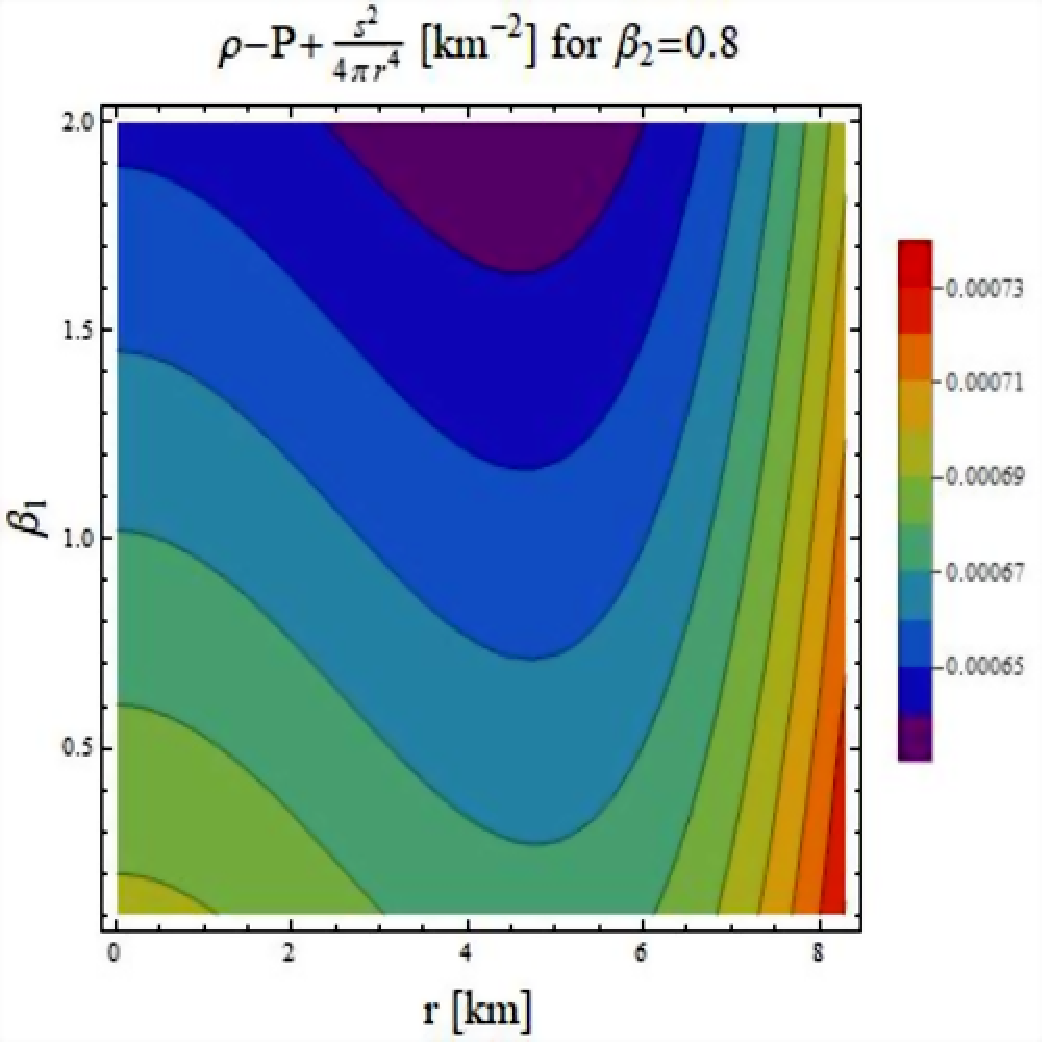,width=0.4\linewidth}
\epsfig{file=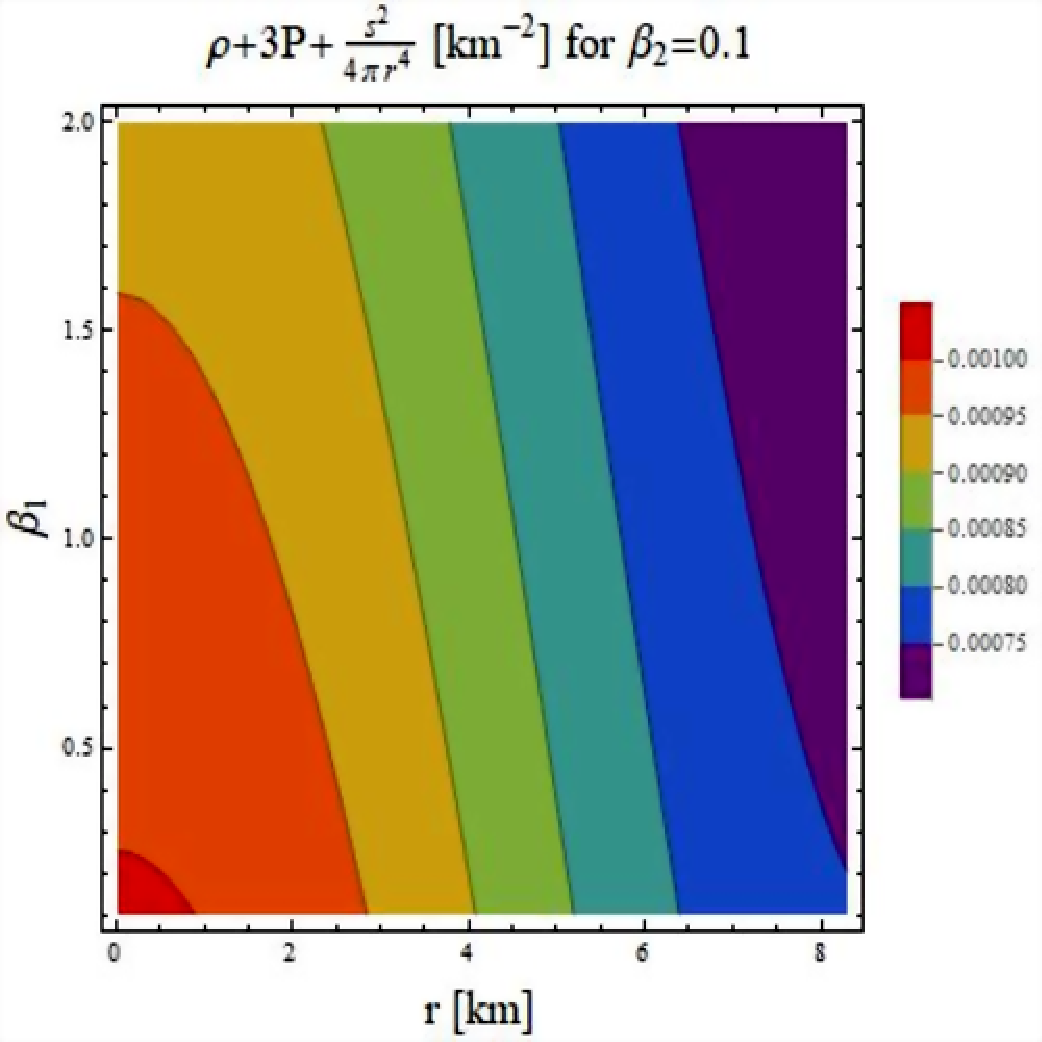,width=0.4\linewidth}\epsfig{file=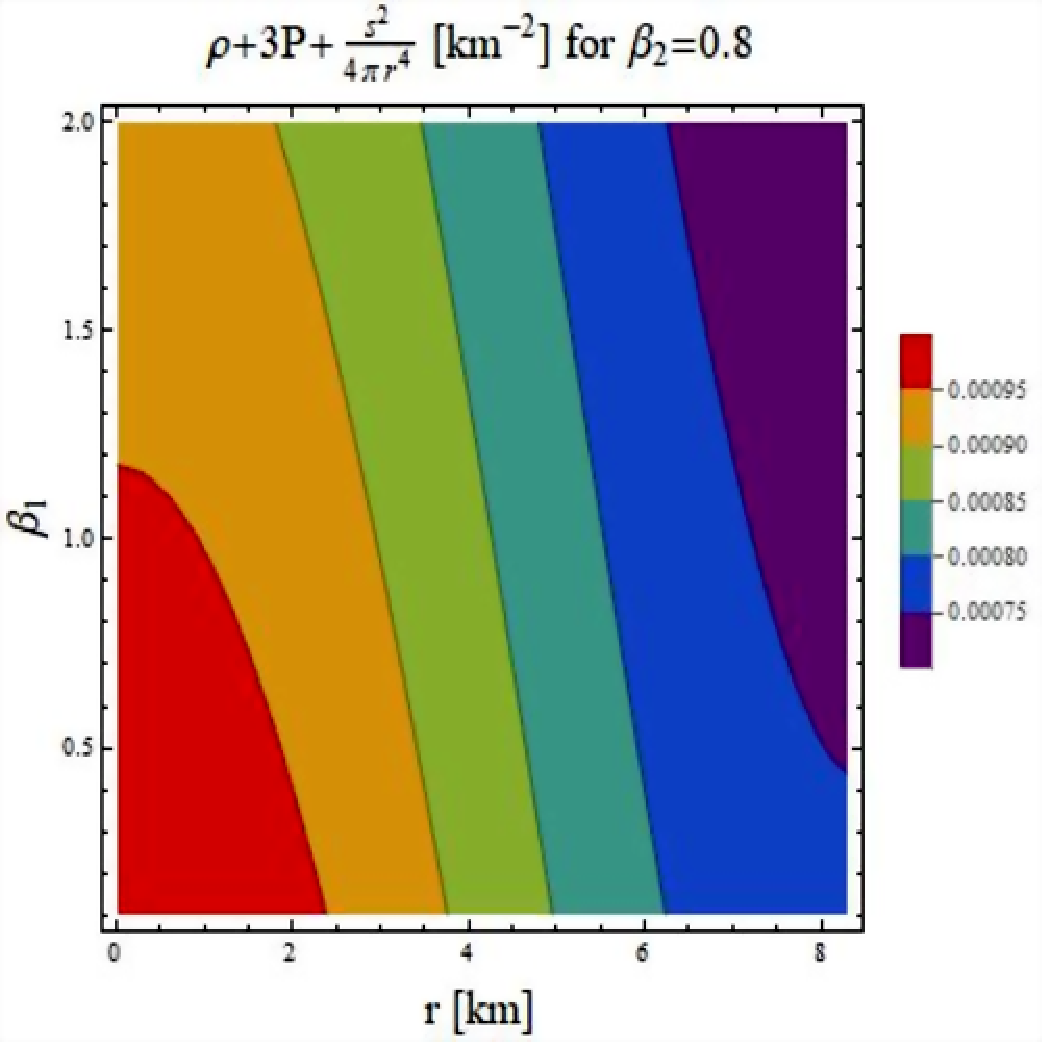,width=0.4\linewidth}
\caption{Energy conditions for model I with $\mathcal{L}_{m}=P$.}
\end{figure}
\begin{figure}\center
\epsfig{file=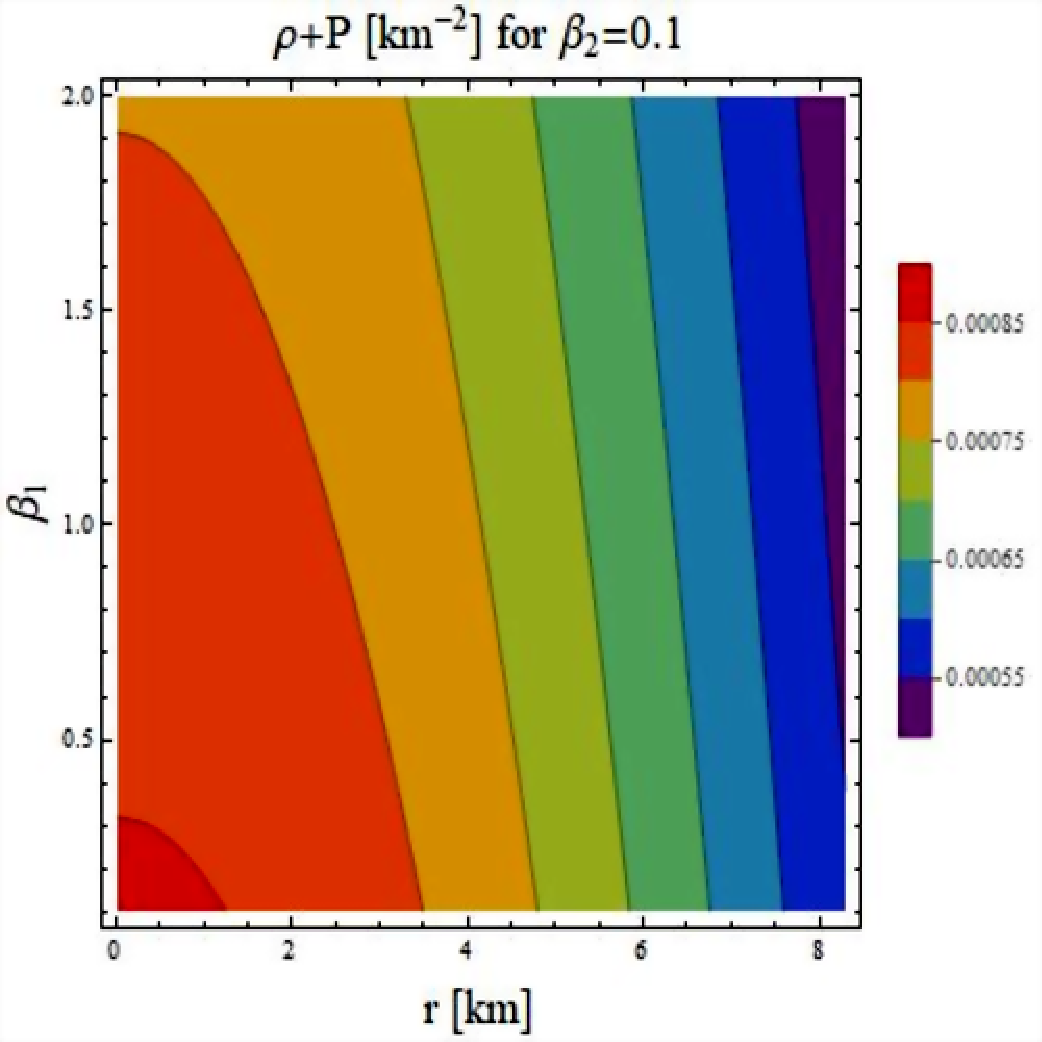,width=0.4\linewidth}\epsfig{file=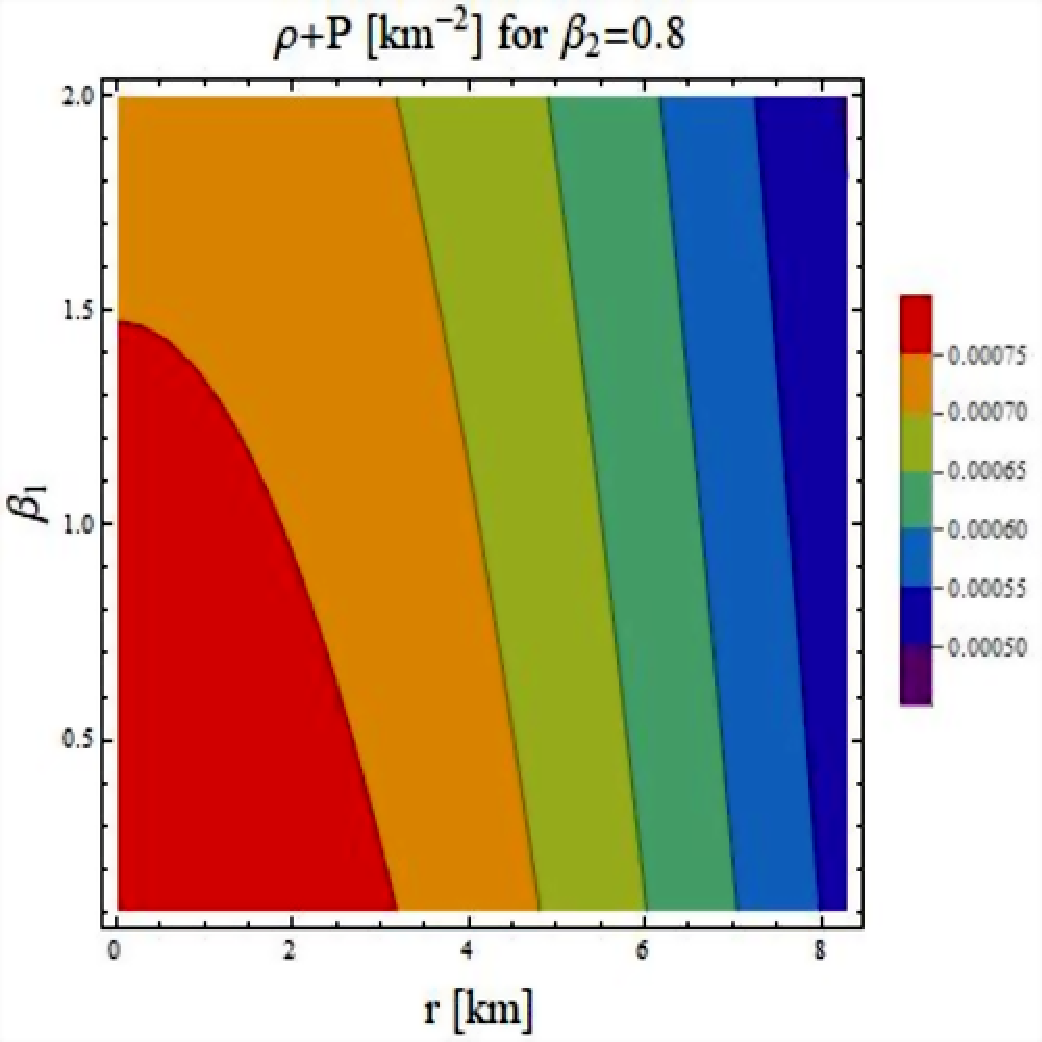,width=0.4\linewidth}
\epsfig{file=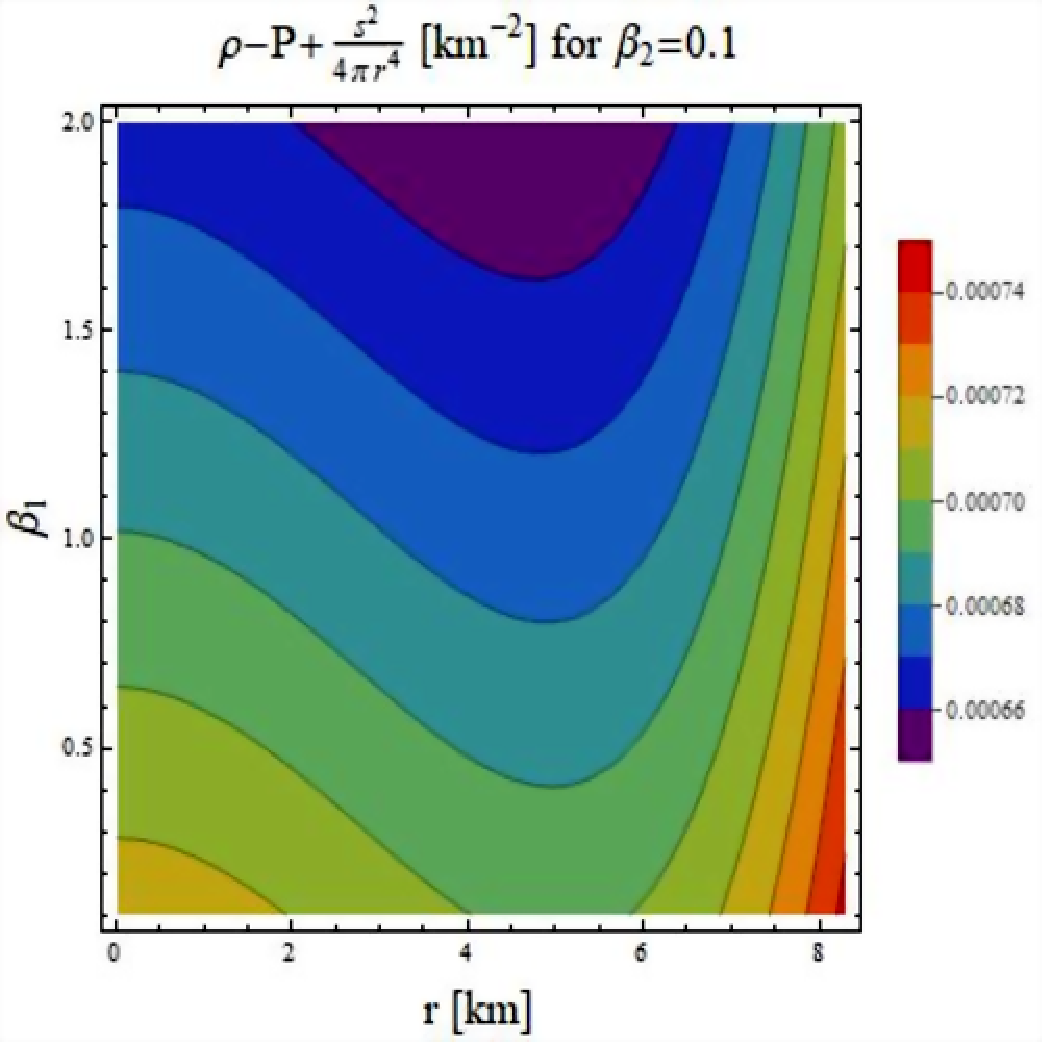,width=0.4\linewidth}\epsfig{file=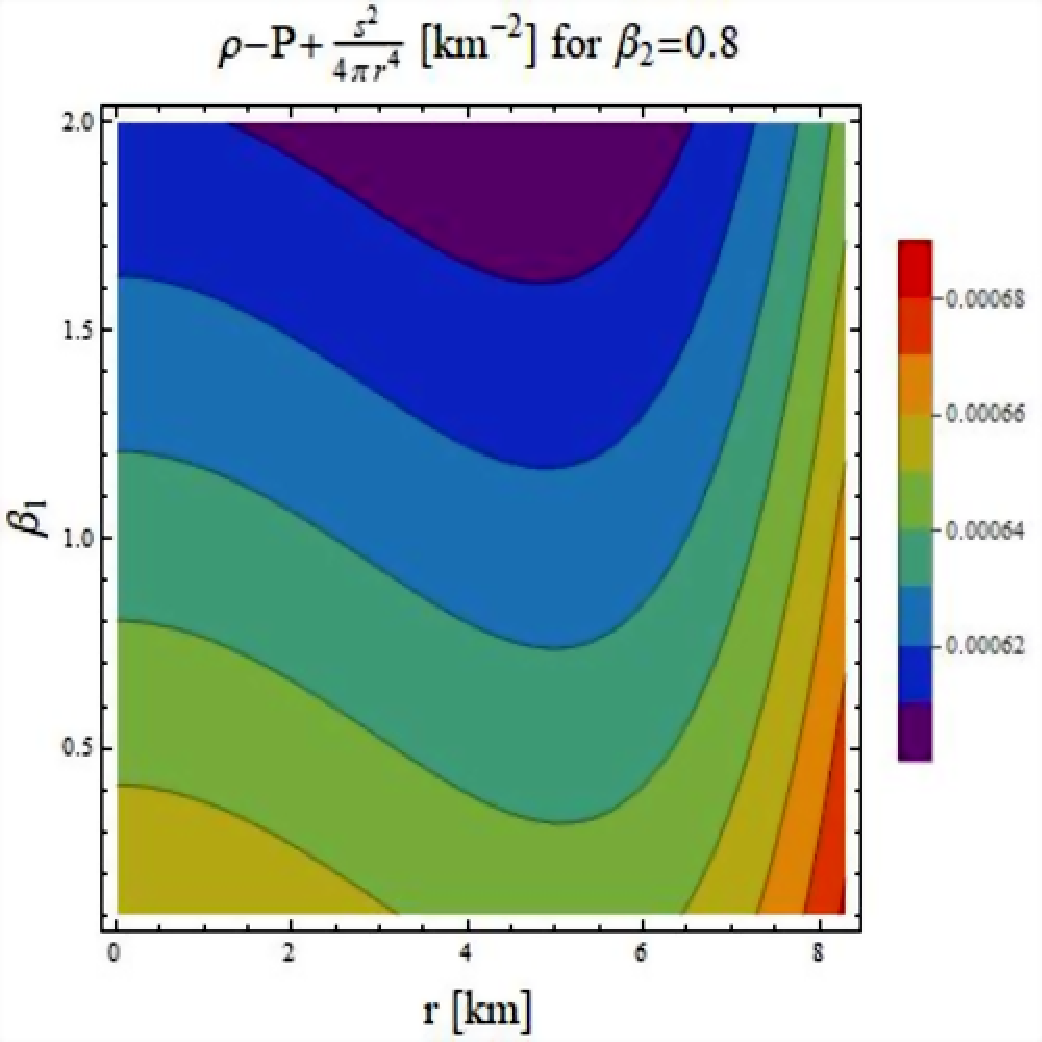,width=0.4\linewidth}
\epsfig{file=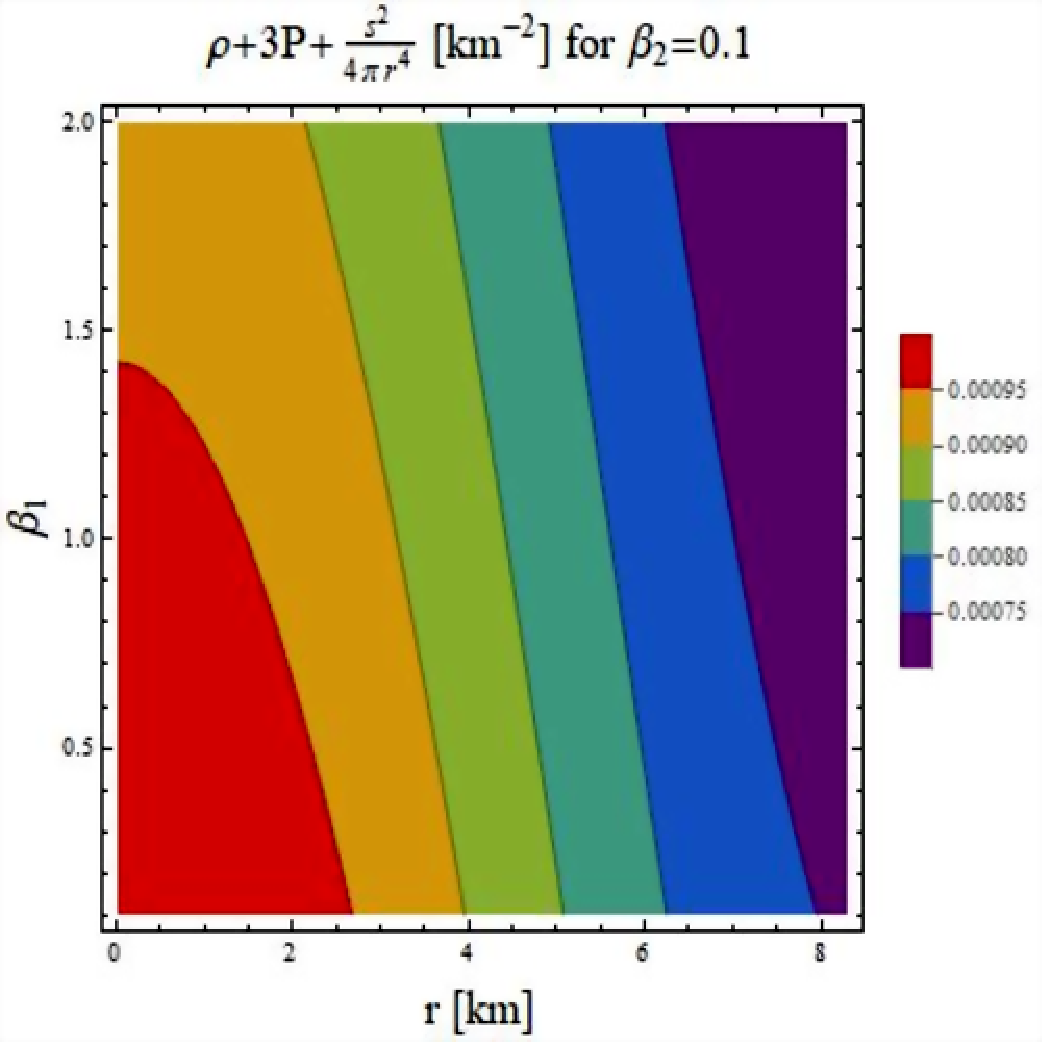,width=0.4\linewidth}\epsfig{file=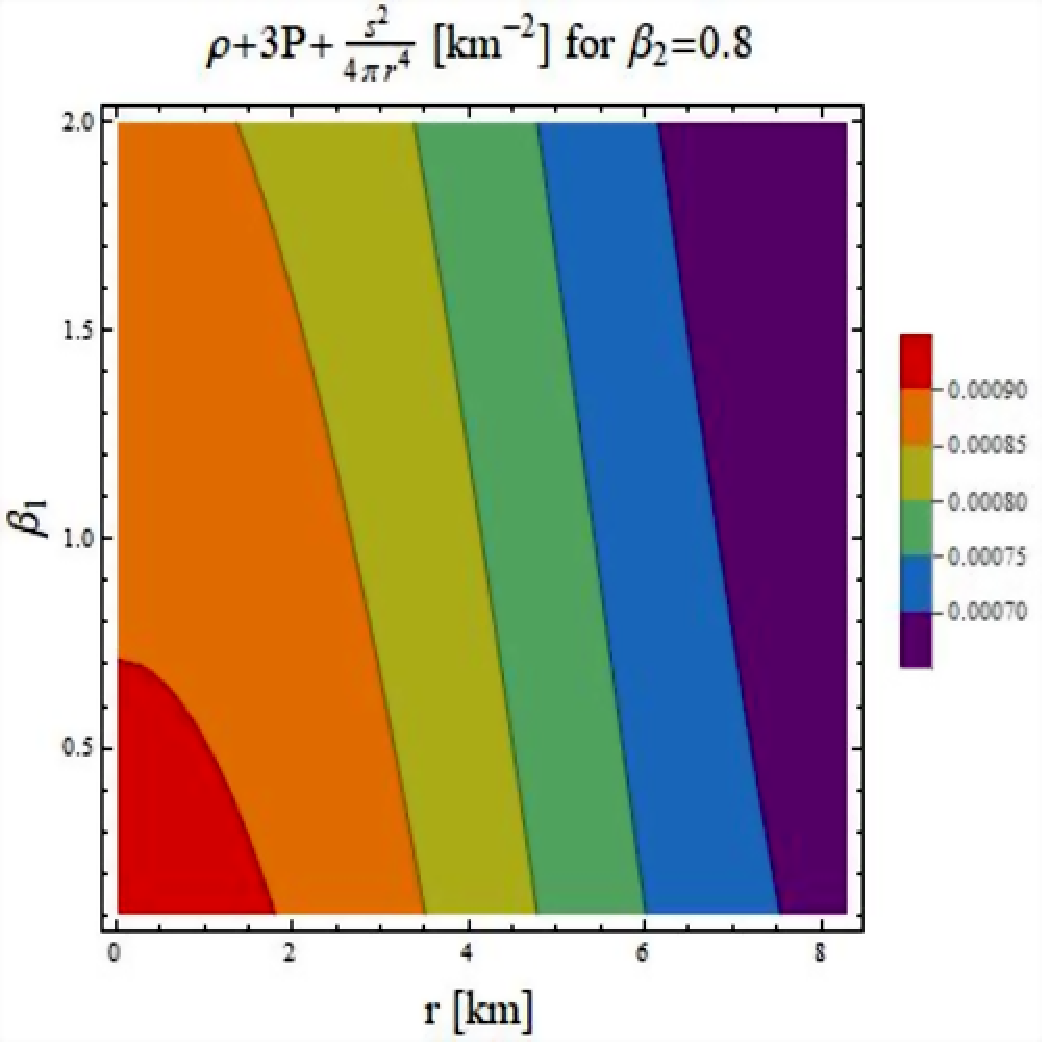,width=0.4\linewidth}
\caption{Energy conditions for model I with
$\mathcal{L}_{m}=-\rho$.}
\end{figure}
\begin{figure}\centering
\epsfig{file=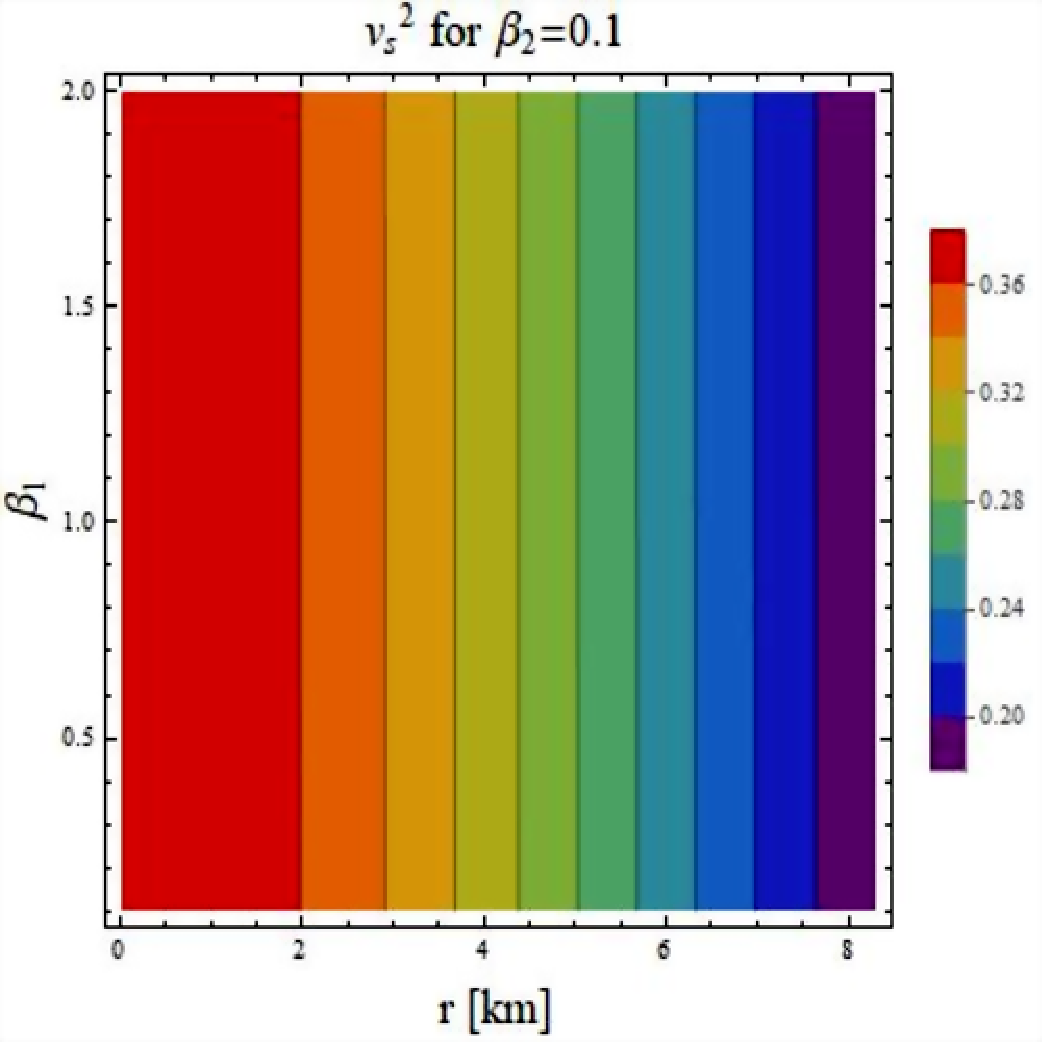,width=0.4\linewidth}\epsfig{file=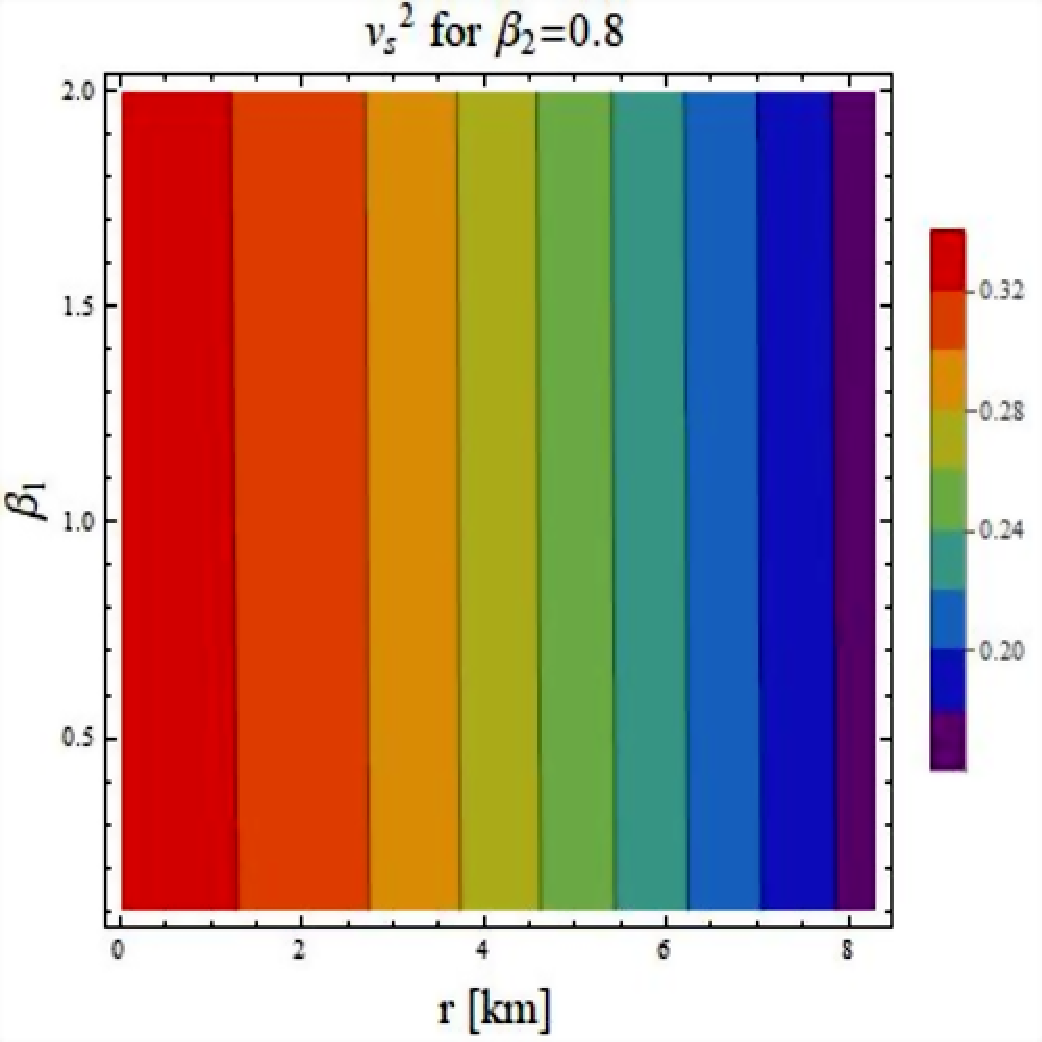,width=0.4\linewidth}
\epsfig{file=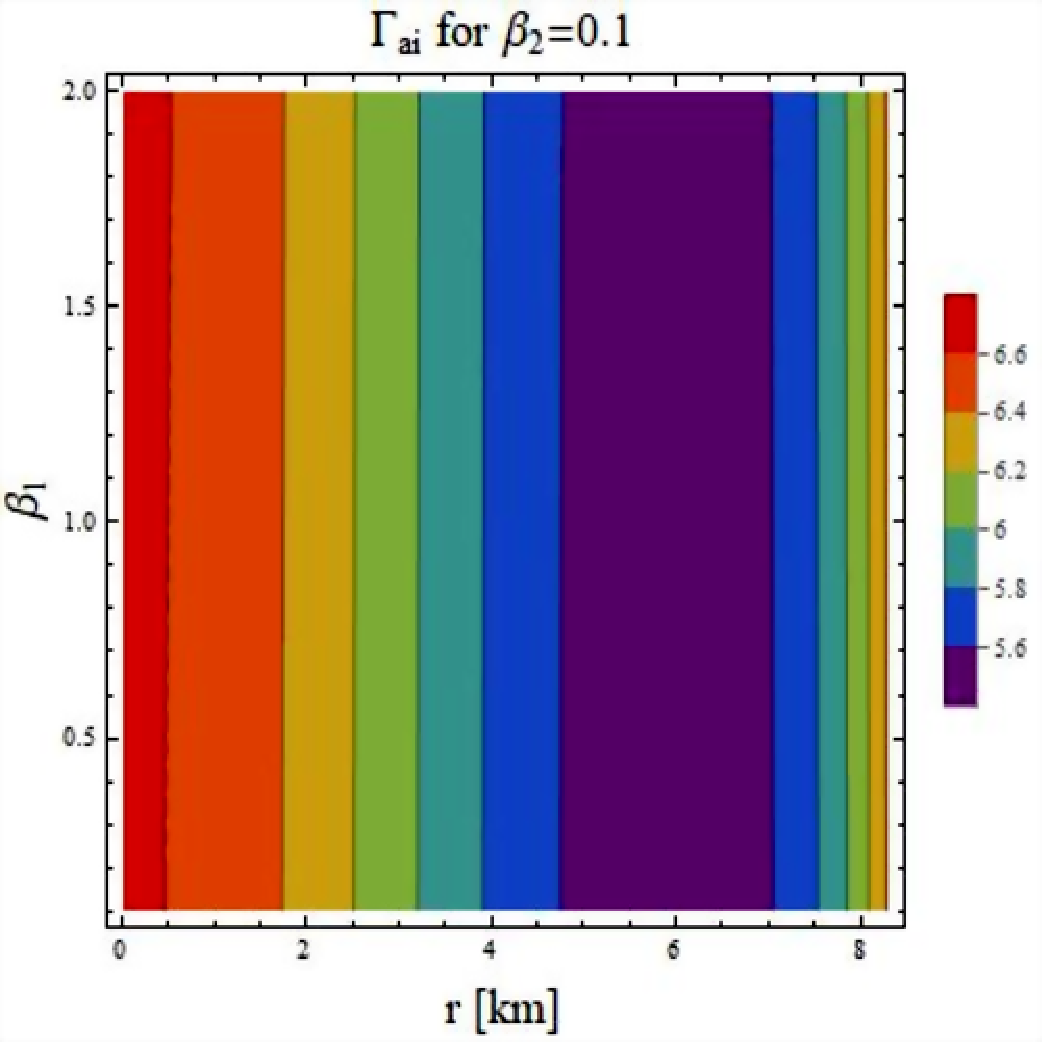,width=0.4\linewidth}\epsfig{file=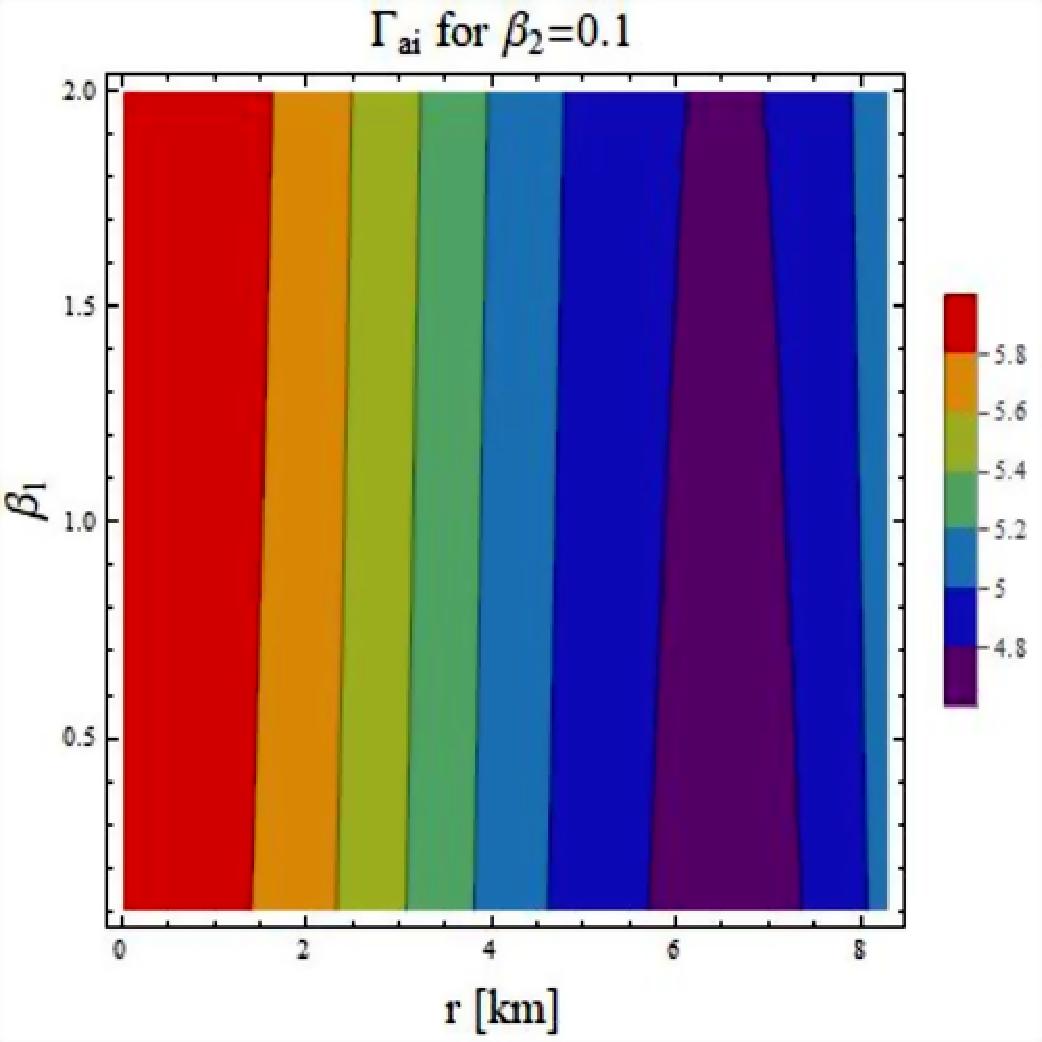,width=0.4\linewidth}
\caption{Stability analysis for model I with $\mathcal{L}_{m}=P$.}
\end{figure}
\begin{figure}\center
\epsfig{file=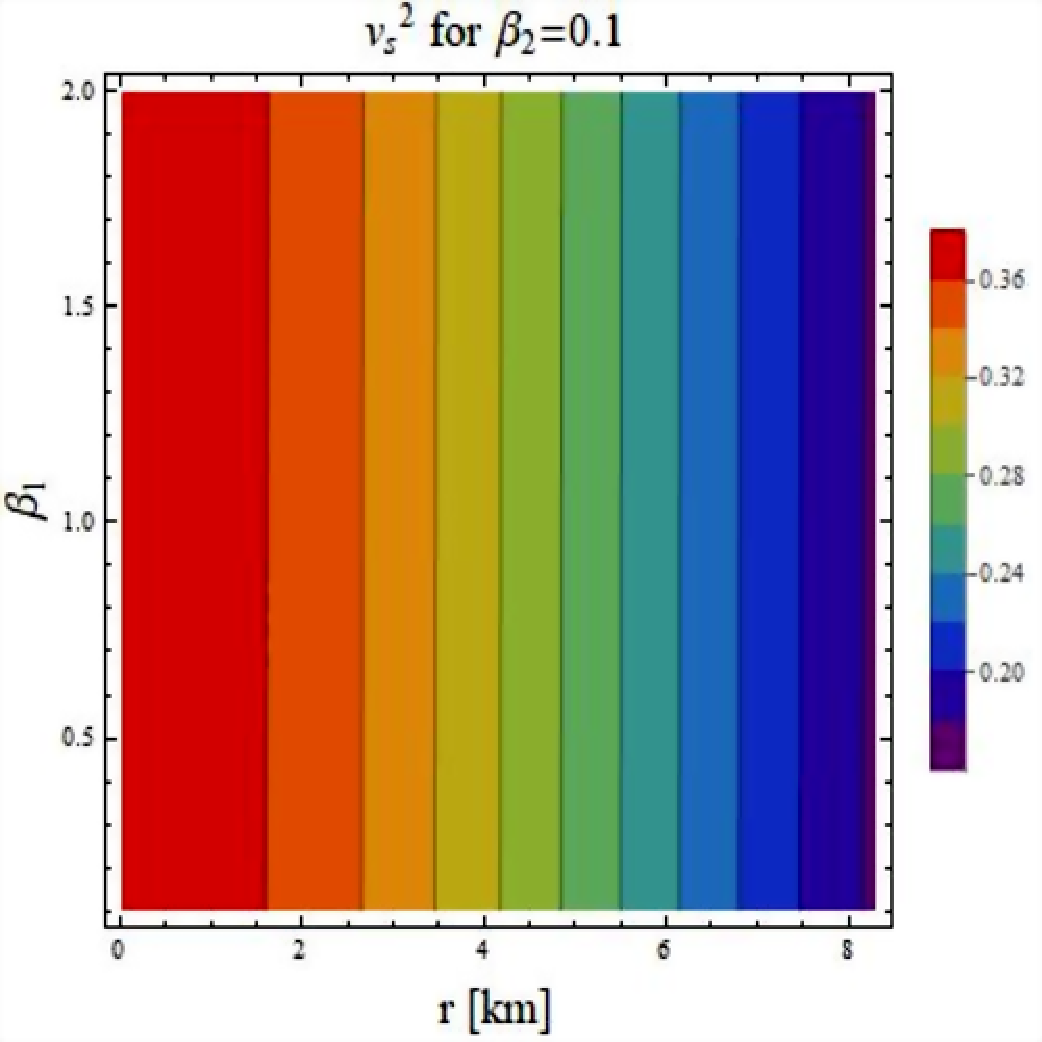,width=0.4\linewidth}\epsfig{file=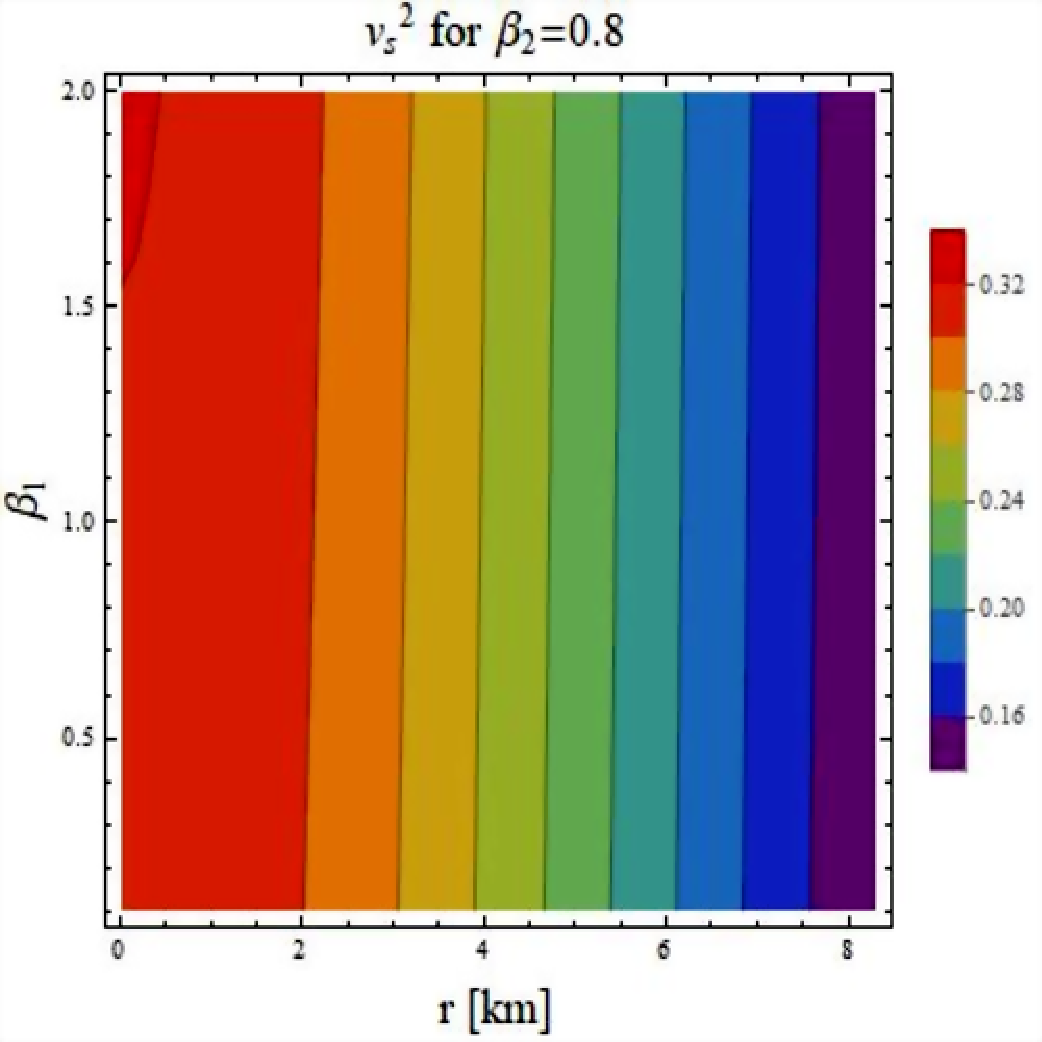,width=0.4\linewidth}
\epsfig{file=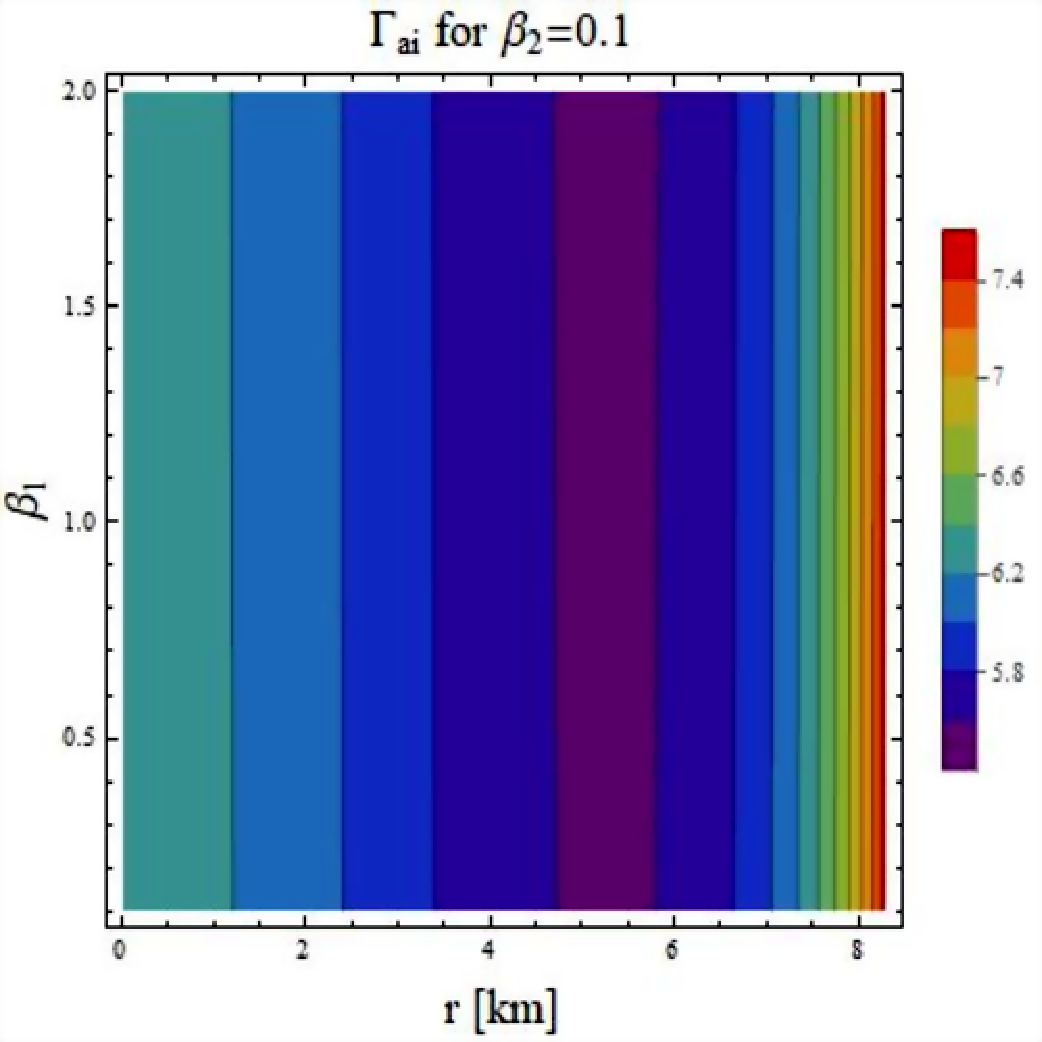,width=0.4\linewidth}\epsfig{file=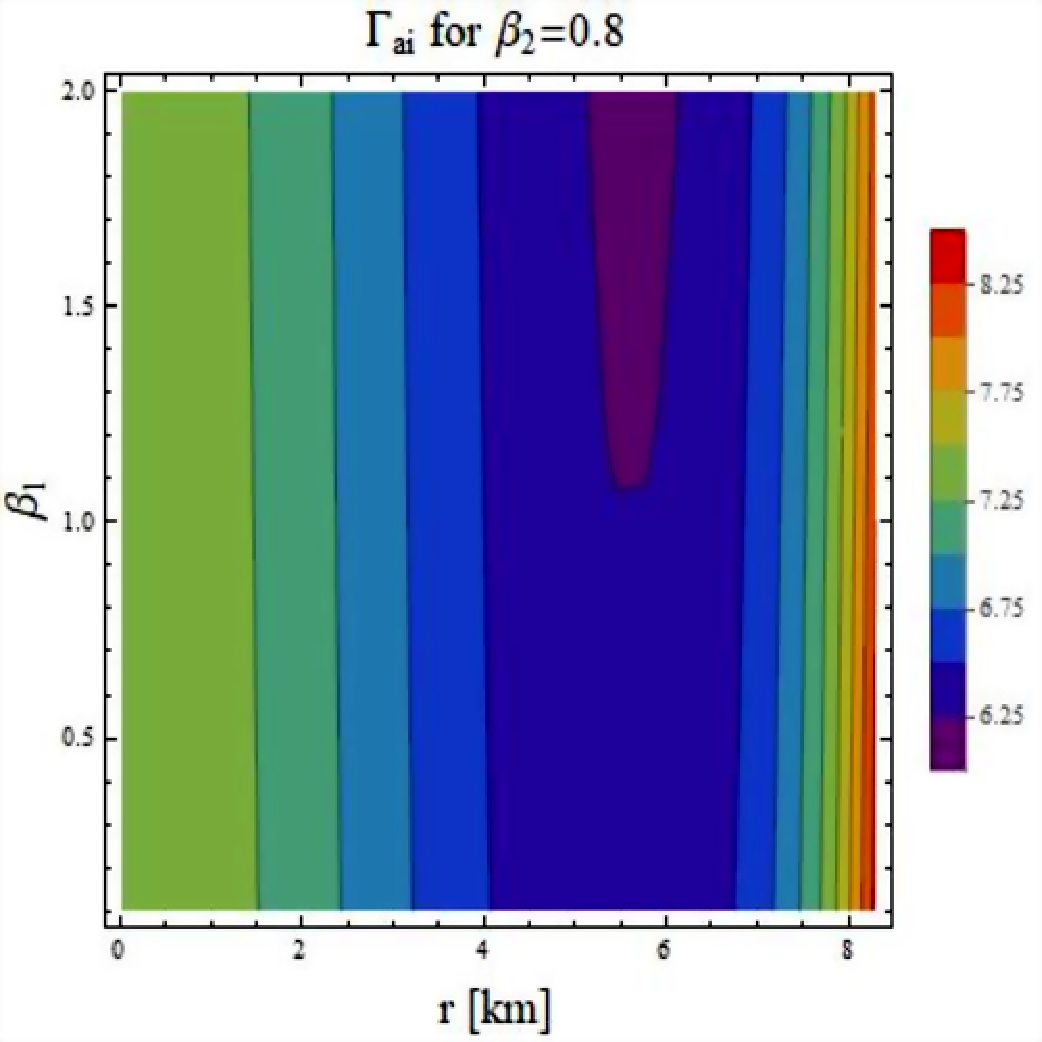,width=0.4\linewidth}
\caption{Stability analysis for model I with
$\mathcal{L}_{m}=-\rho$.}
\end{figure}

\subsection{Model II}

This subsection discusses the dynamics of a spherically symmetric
interior by adopting a strong non-minimal model of
$f(\mathcal{R},\mathcal{L}_{m},\mathcal{T})$ gravity that is given
in the following
\begin{equation}\label{g56}
f(\mathcal{R},\mathcal{L}_{m},\mathcal{T})=
\mathcal{R}+\delta_0f_1(\mathcal{R})+\delta_1
f_2(\mathcal{L}_{m},\mathcal{T}),
\end{equation}
where the two terms $\delta_0$ and $\delta_1$ symbolize arbitrary
constants. We adopt a particular form of the functionals $f_1$ and
$f_2$ that make the above model as
\begin{equation}\label{g56a}
f(\mathcal{R},\mathcal{L}_{m},\mathcal{T})=
\mathcal{R}+\delta_1\mathcal{L}_{m}\mathcal{T}.
\end{equation}

\subsubsection{Stellar Solution for $\mathcal{L}_{m}=P$}

Here, we follow the same pattern again as we already discuss in the
previous subsection. The equations of motion for the considered
geometry are explored for the matter Lagrangian as
$\mathcal{L}_{m}=P$. Putting this choice with the model \eqref{g56a}
in the field equations \eqref{g8}-\eqref{g10} and performing some
manipulation leads to
\begin{align}\label{g57}
&e^{-\varrho_2}\left(\frac{\varrho_2'}{r}-\frac{1}{r^2}\right)
+\frac{1}{r^2}=\left\{8\pi+\frac{\delta_1}{2}\big(5P-\rho\big)\right\}\rho+\delta_1
P^2+\frac{s^2}{r^4},\\\label{g57a}
&e^{-\varrho_2}\left(\frac{1}{r^2}+\frac{\varrho_1'}{r}\right)
-\frac{1}{r^2}=\left\{8\pi+\frac{\delta_1}{2}\big(5P-\rho\big)\right\}P-\delta_1
P^2-\frac{s^2}{r^4},\\\label{g57b}
&\frac{e^{-\varrho_2}}{4}\left[\varrho_1'^2-\varrho_2'\varrho_1'+2\varrho_1''-\frac{2\varrho_2'}{r}+\frac{2\varrho_1'}{r}\right]
=\left\{8\pi+\frac{\delta_1}{2}\big(5P-\rho\big)\right\}P-\delta_1
P^2+\frac{s^2}{r^4}.
\end{align}
Solving last two equations provides the same value of charge as
defined in Eq.\eqref{g53b}. Further, the isotropic system can be
completely characterized by the first two equations, however, it is
not possible to find $\rho$ and $P$ explicitly due to the appearance
of second-order fluid terms. To resolve this issue, we consider a
barotropic equation of state represented by
\begin{align}\label{g57c}
P=\delta_2\rho,
\end{align}
where $\delta_2 \in (0,1)$. After using this equation of state in
\eqref{g57} and \eqref{g57a}, we express $\rho$ and $P$ as
\begin{align}\nonumber
\rho&=\frac{r^2}{\delta_1 r^2\big(2 \delta_2 ^2+5 \delta_2
-1\big)\big(b_2 b_4 r^2 e^{2 b_2 r^2}+1\big)}\big[\big\{b_2^2 r^2
\big(b_4^2 e^{4 b_2 r^2} \big(\delta_1  \big(2 \delta_2 ^2+5
\delta_2 -1\big)\\\nonumber &+64 \pi ^2 r^2\big)-4 \delta_1  \big(2
\delta_2 ^2+5 \delta_2 -1\big)+12 \delta_1  b_4 \big(2 \delta_2 ^2+5
\delta_2 -1\big) e^{2 b_2 r^2}\big)+2 b_4 b_2 e^{2 b_2
r^2}\\\label{g58} &\times \big(3 \delta_1 \big(2 \delta_2 ^2+5
\delta_2 -1\big)+64 \pi ^2 r^2\big)+64 \pi ^2\big\}^{\frac{1}{2}}-8
\pi b_2 b_4 r^2 e^{2 b_2 r^2}-8 \pi\big],\\\nonumber
P&=\frac{\sqrt{\delta_2} r^2}{\delta_1
r^2\big(3\delta_2-1\big)\big(b_2 b_4 r^2 e^{2 b_2
r^2}+1\big)}\big[\big\{b_2^2 r^2 \big(4 \delta_1 (3 \delta_2 -1)+4
\delta_1  b_4 (3 \delta_2 -1) e^{2 b_2 r^2}\\\nonumber &+b_4^2 e^{4
b_2 r^2} \big(\delta_1-3 \delta_1  \delta_2 +64 \pi ^2 r^2 \delta_2
\big)\big)+b_2 \big(2 b_4 e^{2 b_2 r^2} \big(\delta_1-3 \delta_1
\delta_2 +64 \pi ^2 r^2 \delta_2 \big)\\\label{g58a} &+8 \delta_1 (3
\delta_2-1)\big)+64\pi^2\delta_2\big\}^{\frac{1}{2}}-\sqrt{\delta_2}\big(8
\pi b_2 b_4 r^2 e^{2 b_2 r^2}+8 \pi\big)\big],
\end{align}
where Eqs.\eqref{g14i} and \eqref{g15} are also used.

\subsubsection{Stellar Solution for $\mathcal{L}_{m}=-\rho$}

Another choice of the matter Lagrangian as $\mathcal{L}_{m}=-\rho$
is considered that makes Eqs.\eqref{g8}-\eqref{g10} when combined
with the model \eqref{g56a} as
\begin{align}\label{g59}
&e^{-\varrho_2}\left(\frac{\varrho_2'}{r}-\frac{1}{r^2}\right)
+\frac{1}{r^2}=\left\{8\pi+\frac{3\delta_1}{2}\big(P-\rho\big)\right\}\rho+\delta_1
\rho^2+\frac{s^2}{r^4},\\\label{g59a}
&e^{-\varrho_2}\left(\frac{1}{r^2}+\frac{\varrho_1'}{r}\right)
-\frac{1}{r^2}=\left\{8\pi+\frac{3\delta_1}{2}\big(P-\rho\big)\right\}\rho-\delta_1
\rho^2-\frac{s^2}{r^4},\\\label{g59b}
&\frac{e^{-\varrho_2}}{4}\left[\varrho_1'^2-\varrho_2'\varrho_1'+2\varrho_1''-\frac{2\varrho_2'}{r}+\frac{2\varrho_1'}{r}\right]
=\left\{8\pi+\frac{3\delta_1}{2}\big(P-\rho\big)\right\}\rho-\delta_1
\rho^2+\frac{s^2}{r^4}.
\end{align}

Simultaneous use of Eqs.\eqref{g14i}, \eqref{g15}, \eqref{g57c},
\eqref{g59} and \eqref{g59a} results in the following expressions of
$\rho$ and $P$ as
\begin{align}\nonumber
\rho&=\frac{r^2}{\delta_1 r^2\big(3\delta_2-1\big)\big(b_2 b_4 r^2
e^{2 b_2 r^2}+1\big)}\big[\big\{b_2^2 r^2 \big(4 \delta_1  (1-3
\delta_2 )+12 \delta_1  b_4 (3 \delta_2 -1) e^{2 b_2 r^2}\\\nonumber
& +b_4^2 e^{4 b_2 r^2}\big(3 \delta_1 \delta_2 -\delta_1 +64 \pi ^2
r^2\big)\big)+2 b_4 b_2 e^{2 b_2 r^2} \big(9 \delta_1 \delta_2 -3
\delta_1 +64 \pi ^2 r^2\big)+64 \pi
^2\big\}^{\frac{1}{2}}\\\label{g60} &-8 \pi b_2 b_4 r^2 e^{2 b_2
r^2}-8 \pi\big],\\\nonumber P&=\frac{\delta_2 r^2}{\delta_1
r^2\big(3 \delta_2 ^2-3 \delta_2 -2\big)\big(b_2 b_4 r^2 e^{2 b_2
r^2}+1\big)}\big[\big\{b_2 \big(8 \delta_1 \big(3 \delta_2 ^2-3
\delta_2 -2\big)+2 b_4 e^{2 b_2 r^2}\\\nonumber &\times
\big(\delta_1 \big(3 \delta_2-3 \delta_2 ^2 +2\big)+64 \pi ^2 r^2
\delta_2 ^2\big)\big)+b_2^2 r^2 \big(4 \delta_1 \big(3 \delta_2 ^2-3
\delta_2 -2\big)+b_4^2 e^{4 b_2 r^2}\\\nonumber &\times
\big(\delta_1 \big(3 \delta_2-3 \delta_2 ^2 +2\big)+64 \pi ^2 r^2
\delta_2 ^2\big)+4 \delta_1 b_4 \big(3 \delta_2 ^2-3 \delta_2
-2\big) e^{2 b_2 r^2}\big)+64 \pi ^2 \delta_2
^2\big\}^{\frac{1}{2}}\\\label{g60a} &-\delta_2\big(8 \pi b_2 b_4
r^2 e^{2 b_2 r^2}+8 \pi\big)\big].
\end{align}

As the graphical exploration for the solution corresponding to model
II is concerned, we choose the parametric values as $\delta_1 \in
[0.1,2]$ and $\delta_2=0.01,0.95$. It has been observed that only
positive values of $\delta_1$ produce the accelerating solution, but
the parameter $\delta_2$ could either be positive or negative
\cite{25ah}. However, when we plot physical properties corresponding
to our developed solution for its negative choices, the results are
not so well behaved. Hence, we are left with positive values of both
parameters.

The value of the exterior charge remains same as considered for
model I. The same properties (already plotted for the first model)
are again explored for this model to check its physical significance
in the framework of astronomical structures. Figures \textbf{8} and
\textbf{9} exhibit the profiles of the fluid sector for the above
described parametric values, and we observe their acceptable nature.
We find that this model produces less dense systems in comparison
with the first model for all parametric choices. The factors which
are plotted in Figure \textbf{3} for model I are again checked for
the current scenario and we obtain almost the same results.
Therefore, we exclude their graphs from this paper. Further, our
model II is also physically viable and this is ensured by the
observations which we make in Figures \textbf{10} and \textbf{11}.
Lastly, Figures \textbf{12} and \textbf{13} present the variations
in the sound speed and adiabatic index w.r.t. $r,~\delta_1$ and
$\delta_2$. It is found that the developed solution only for
$\mathcal{L}_{m}=P$ is stable, however, the model corresponding to
the other choice of the Lagrangian density does not fulfill the
required criteria.
\begin{figure}[H]\center
\epsfig{file=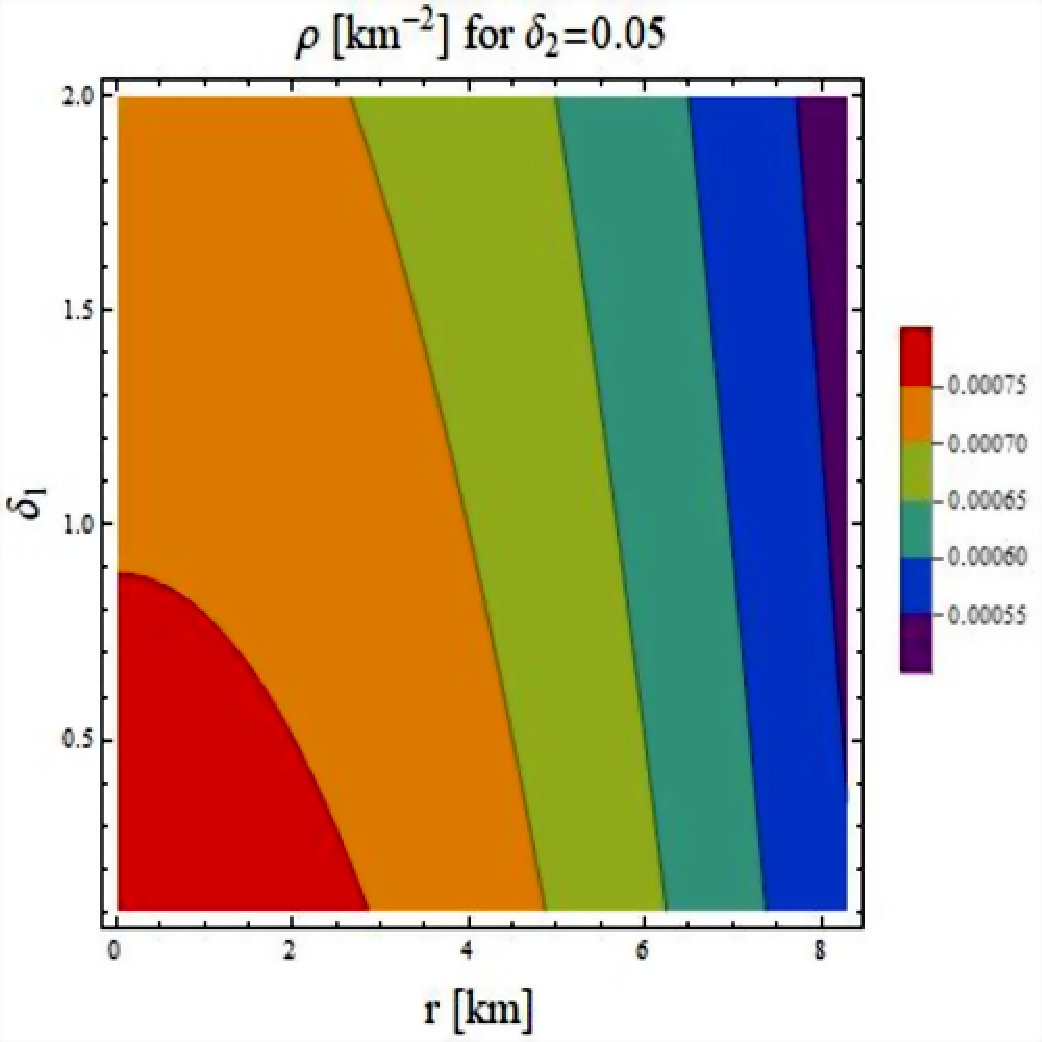,width=0.4\linewidth}\epsfig{file=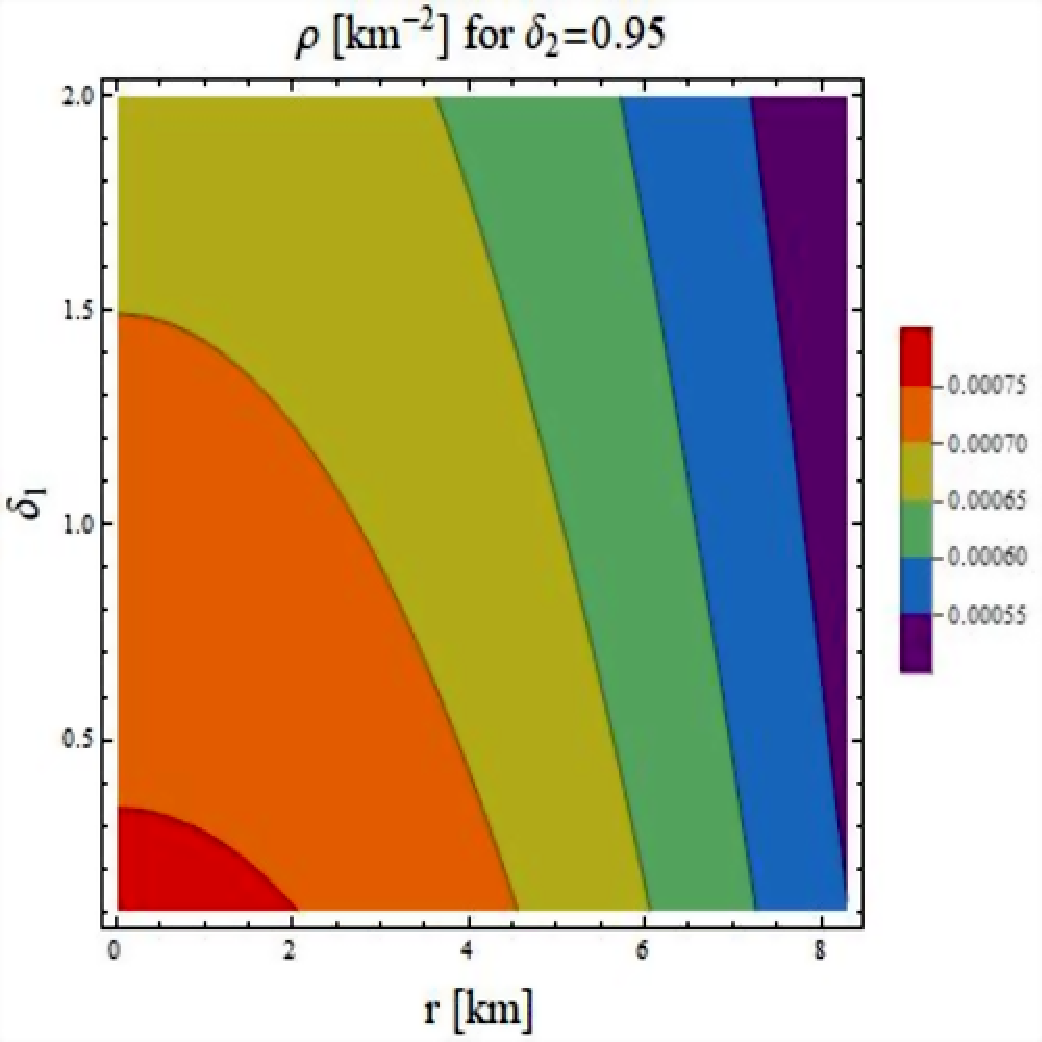,width=0.4\linewidth}
\epsfig{file=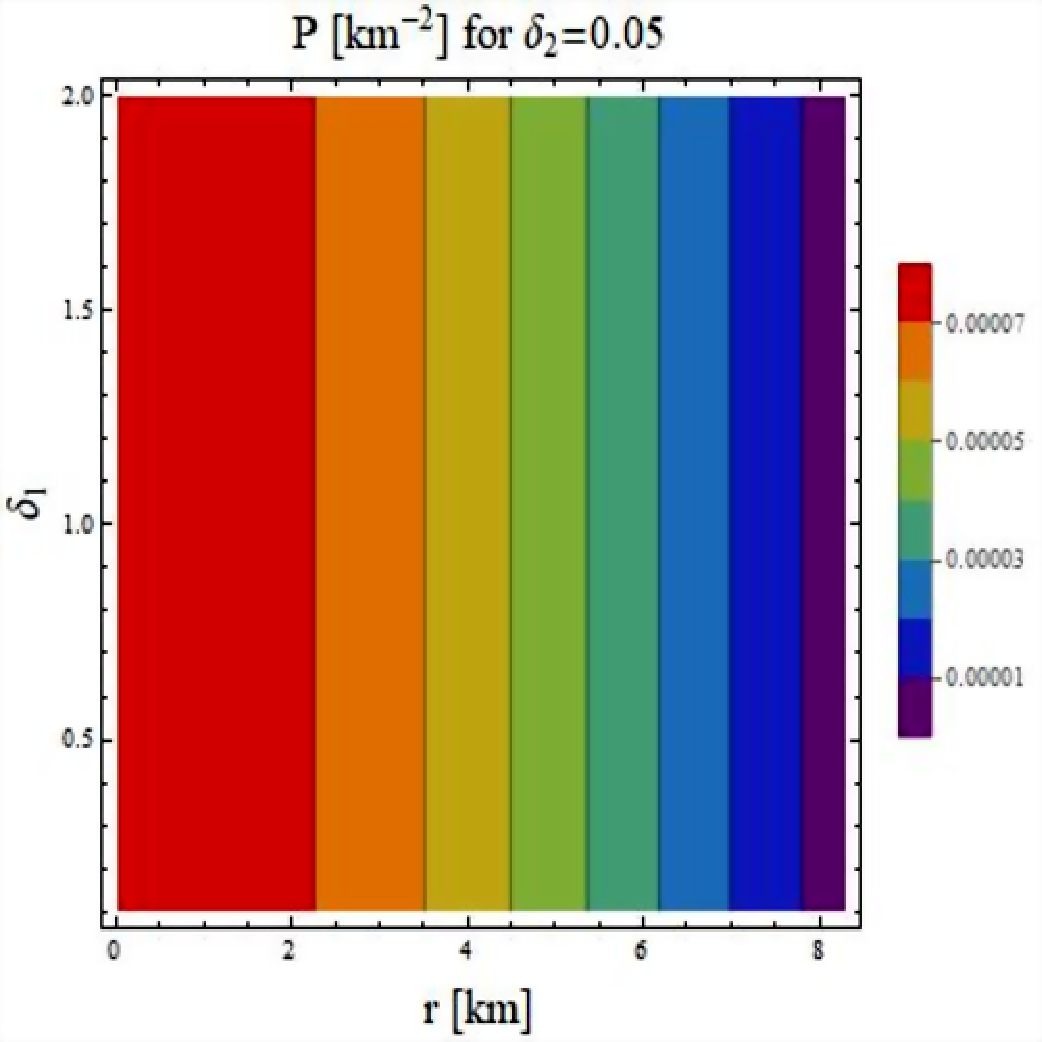,width=0.4\linewidth}\epsfig{file=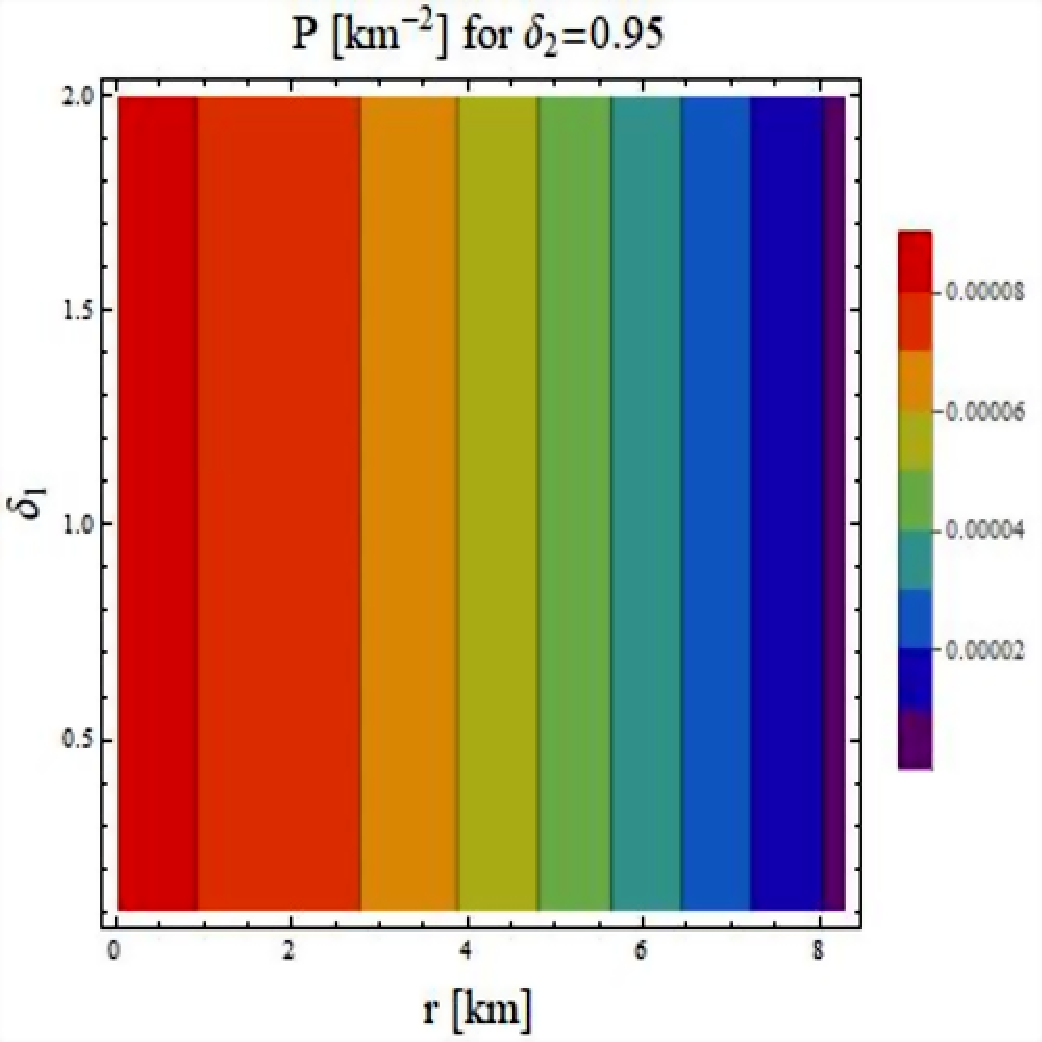,width=0.4\linewidth}
\caption{Energy density and pressure for model II with
$\mathcal{L}_{m}=P$.}
\end{figure}
\begin{figure}[H]\center
\epsfig{file=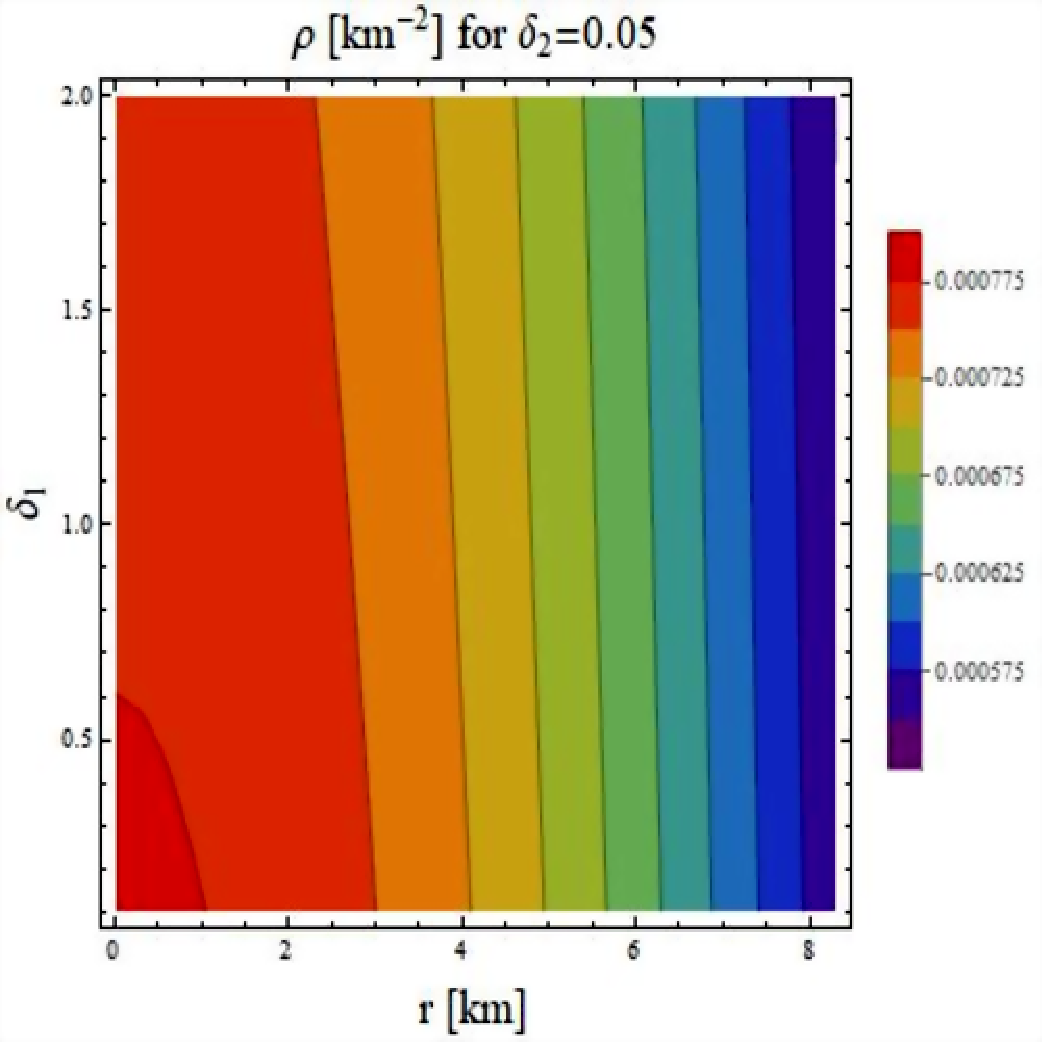,width=0.4\linewidth}\epsfig{file=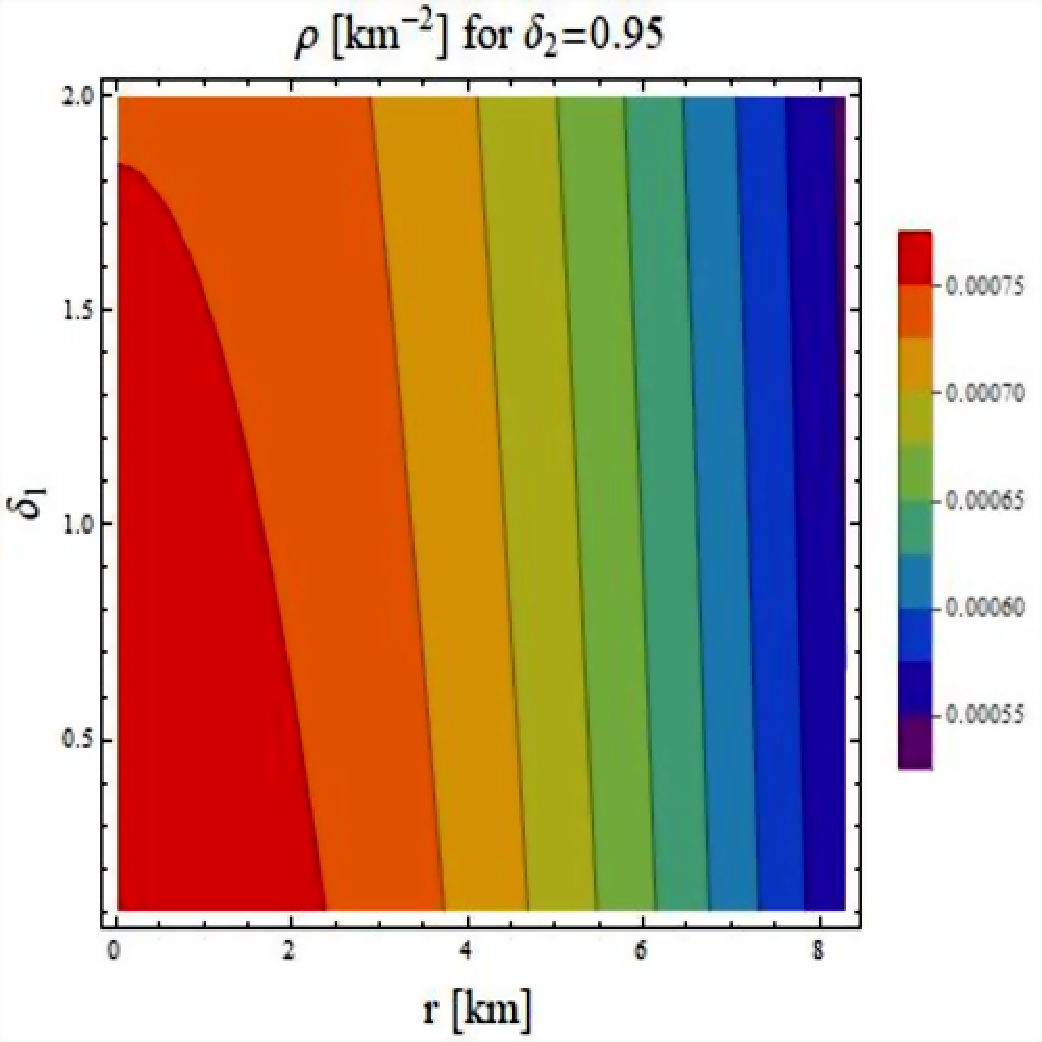,width=0.4\linewidth}
\epsfig{file=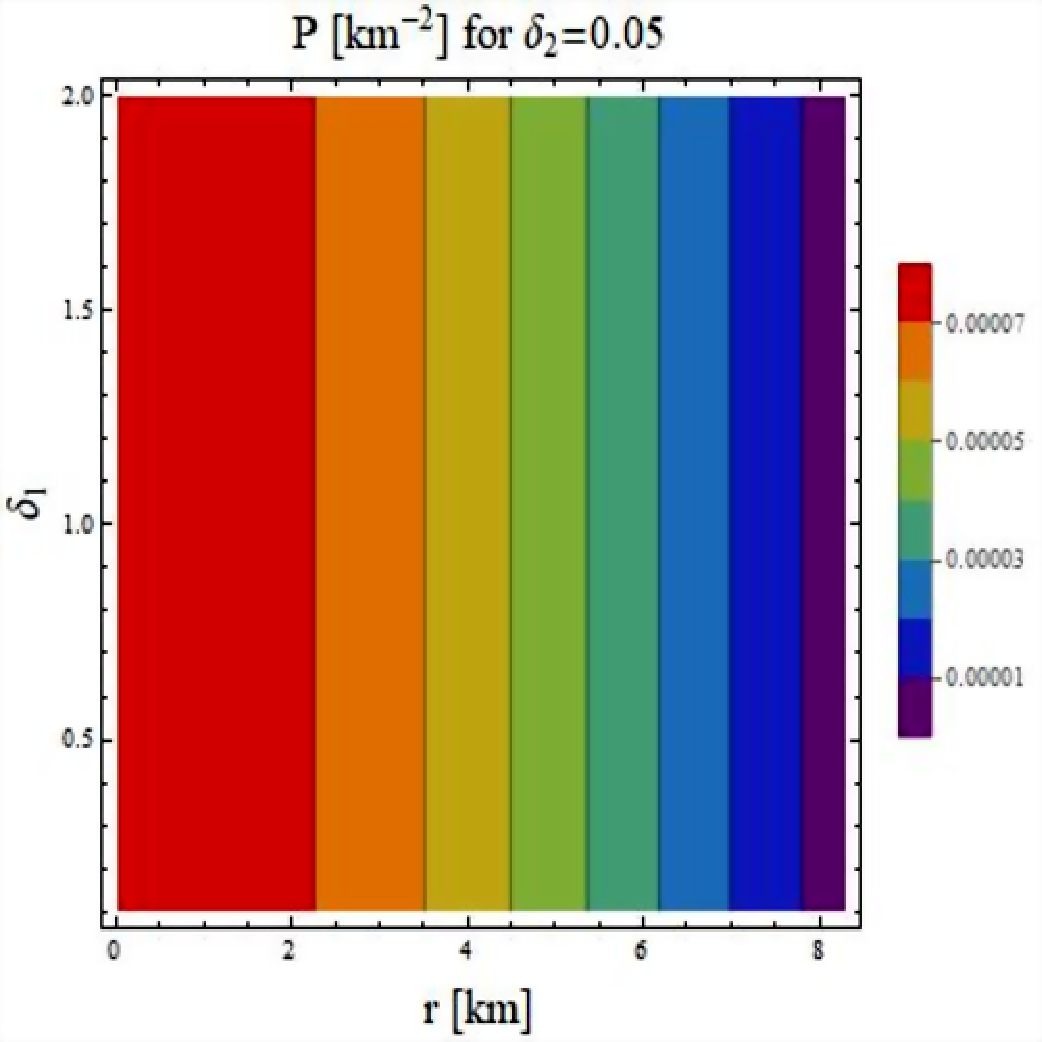,width=0.4\linewidth}\epsfig{file=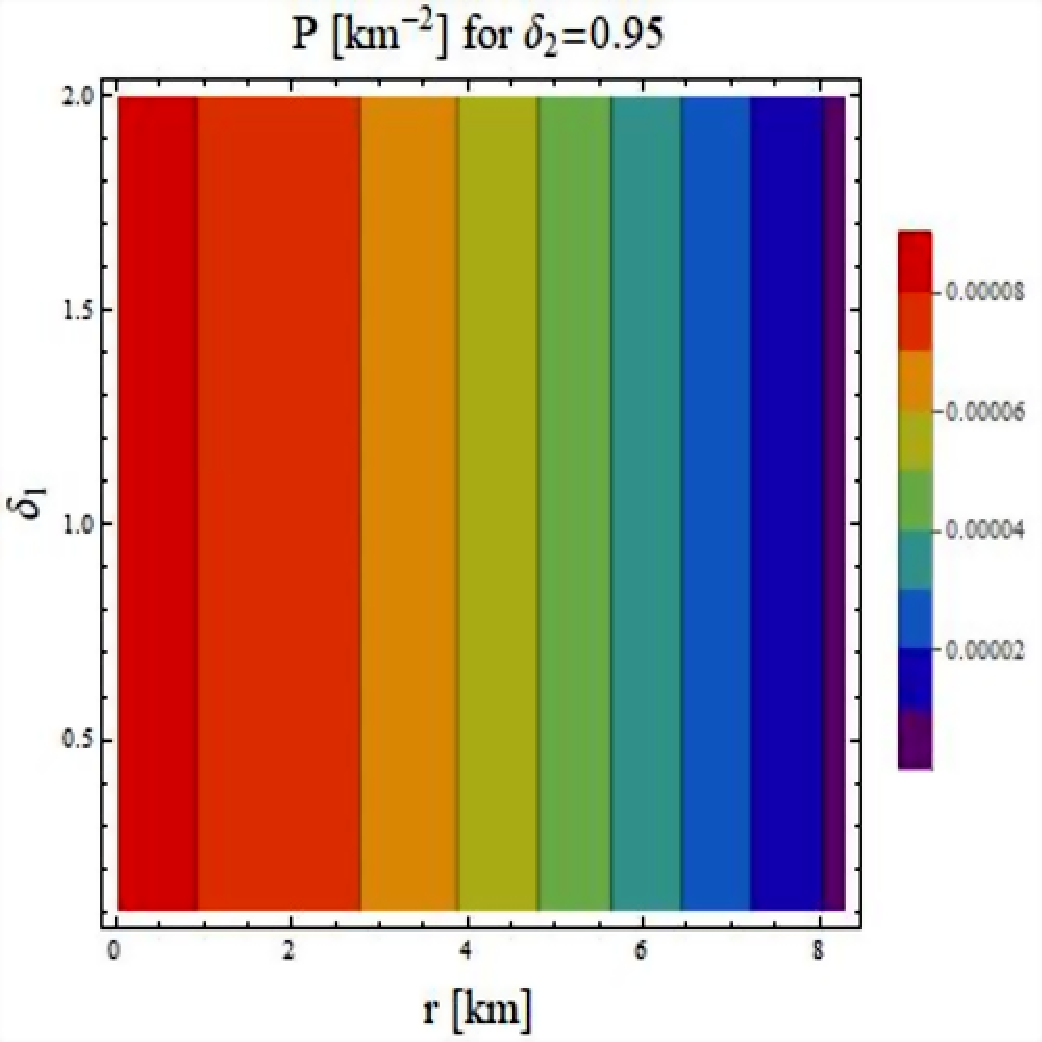,width=0.4\linewidth}
\caption{Energy density and pressure for model II with
$\mathcal{L}_{m}=-\rho$.}
\end{figure}
\begin{figure}[H]\center
\epsfig{file=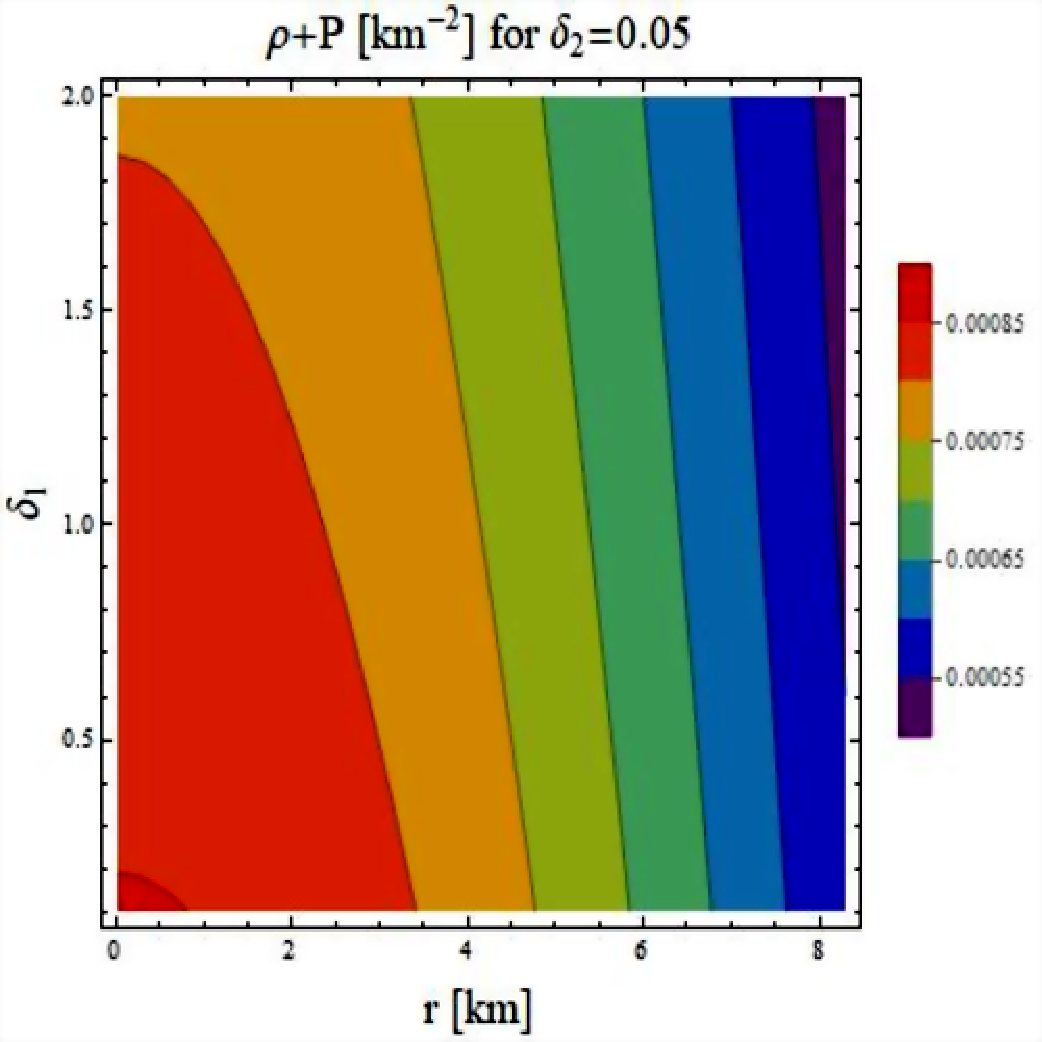,width=0.4\linewidth}\epsfig{file=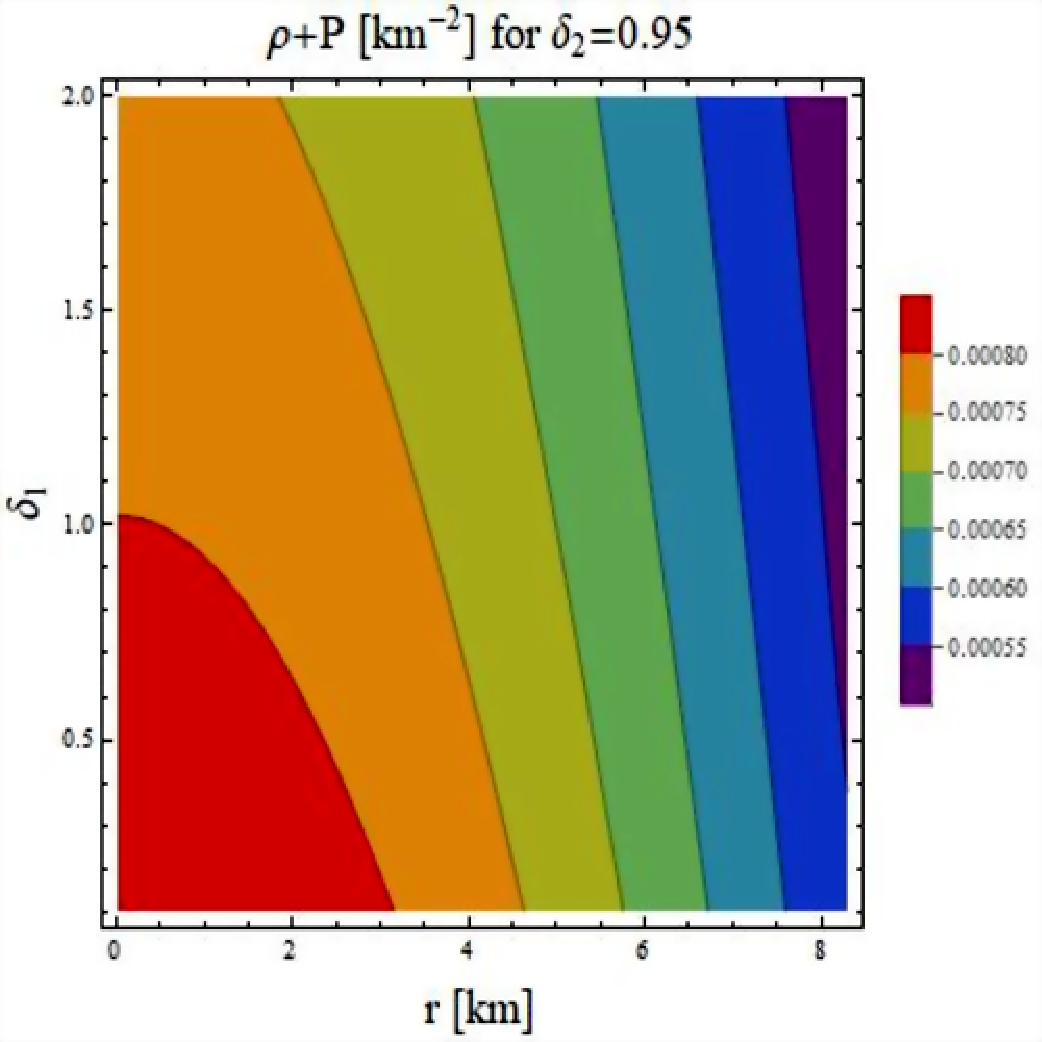,width=0.4\linewidth}
\epsfig{file=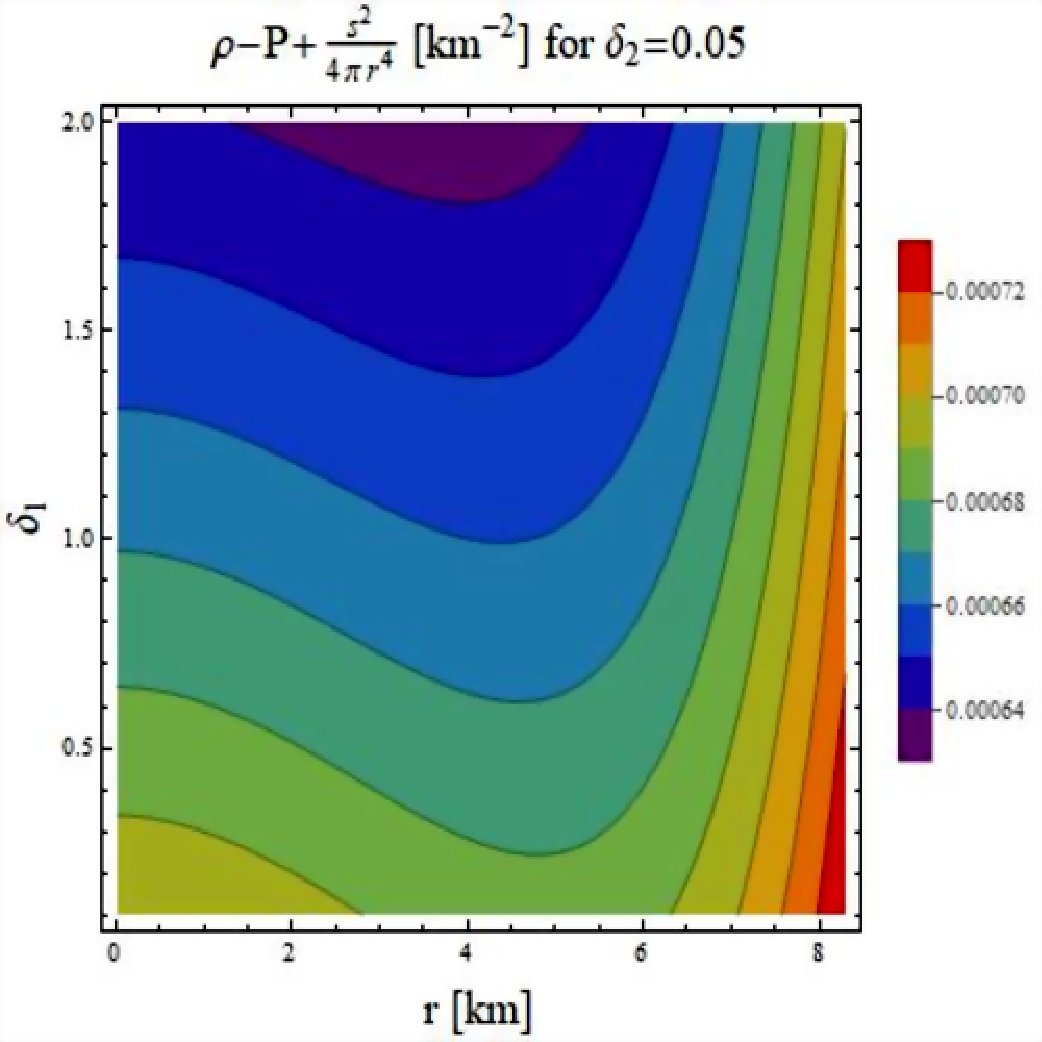,width=0.4\linewidth}\epsfig{file=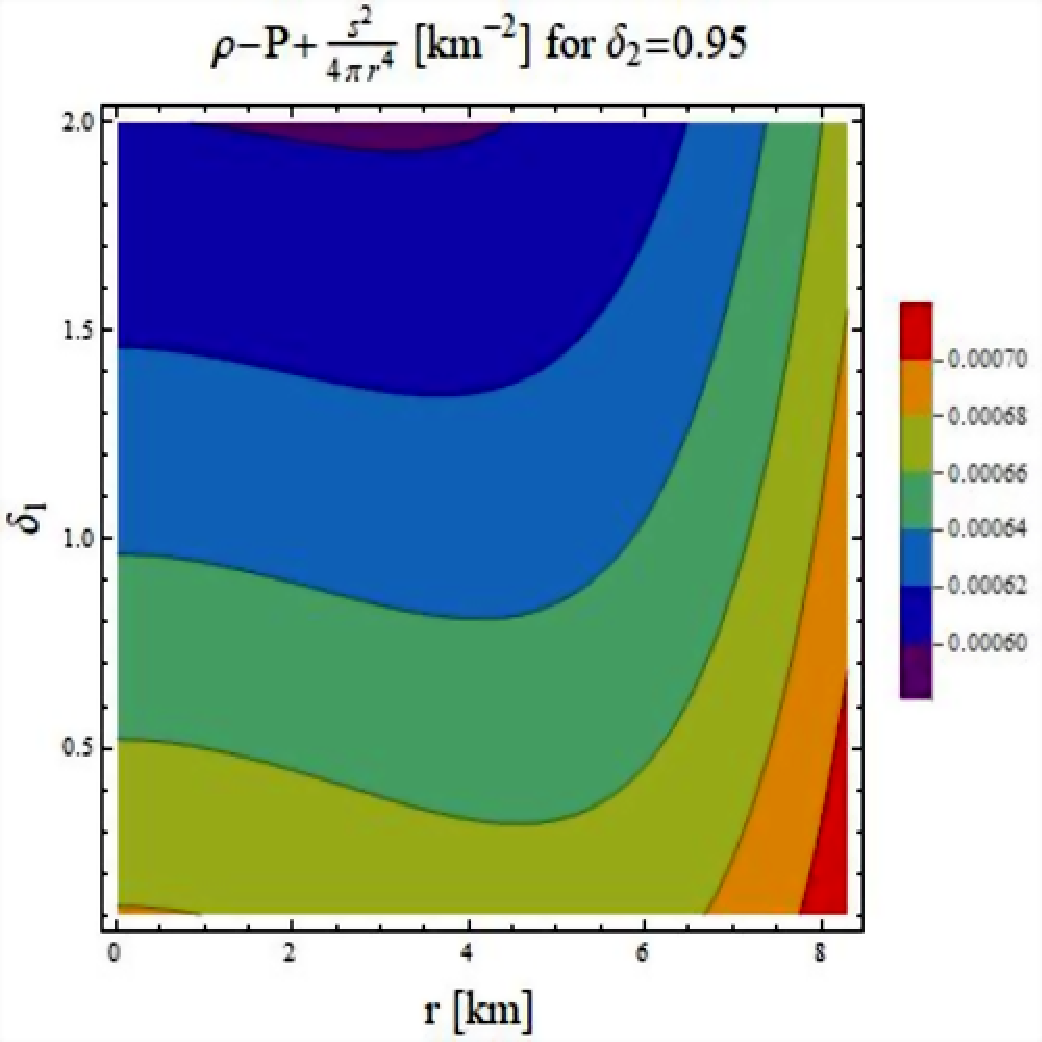,width=0.4\linewidth}
\epsfig{file=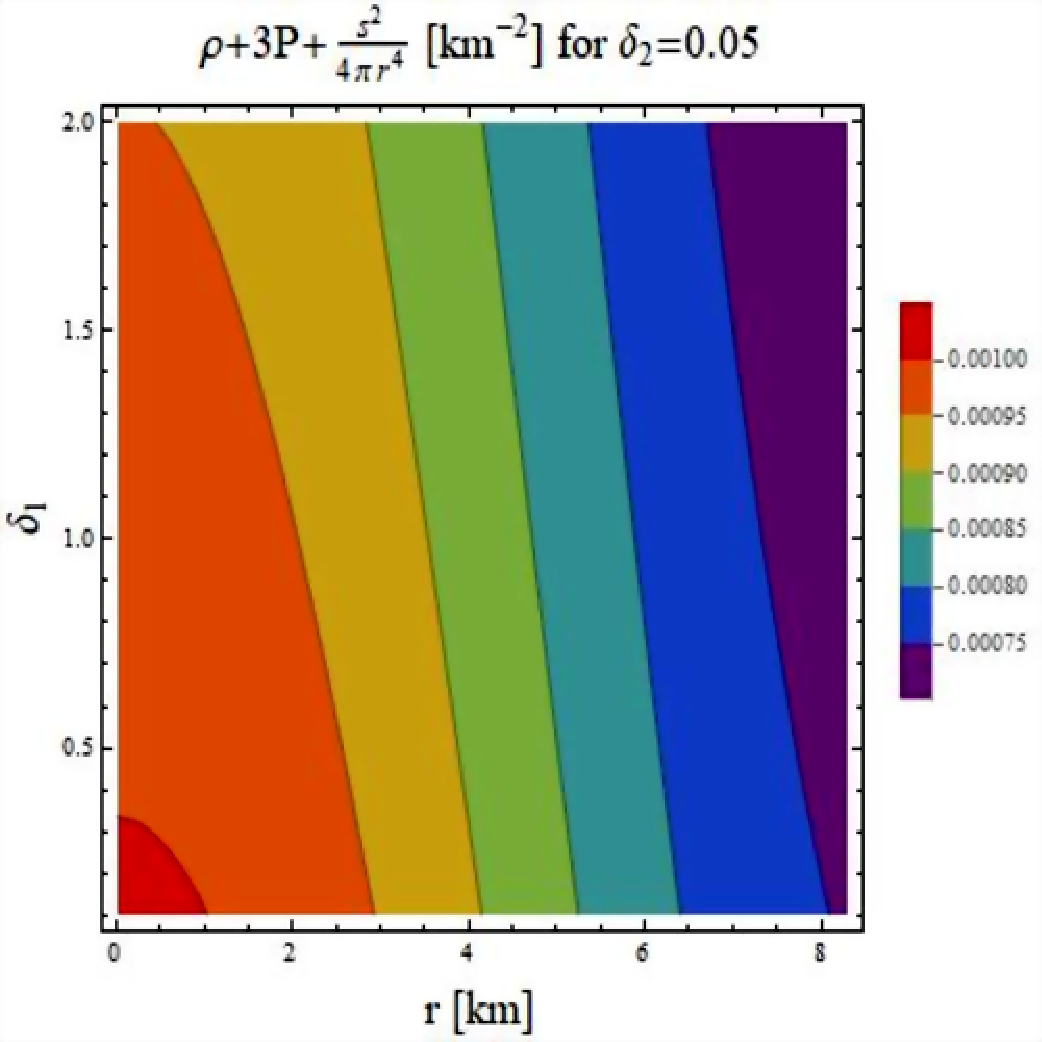,width=0.4\linewidth}\epsfig{file=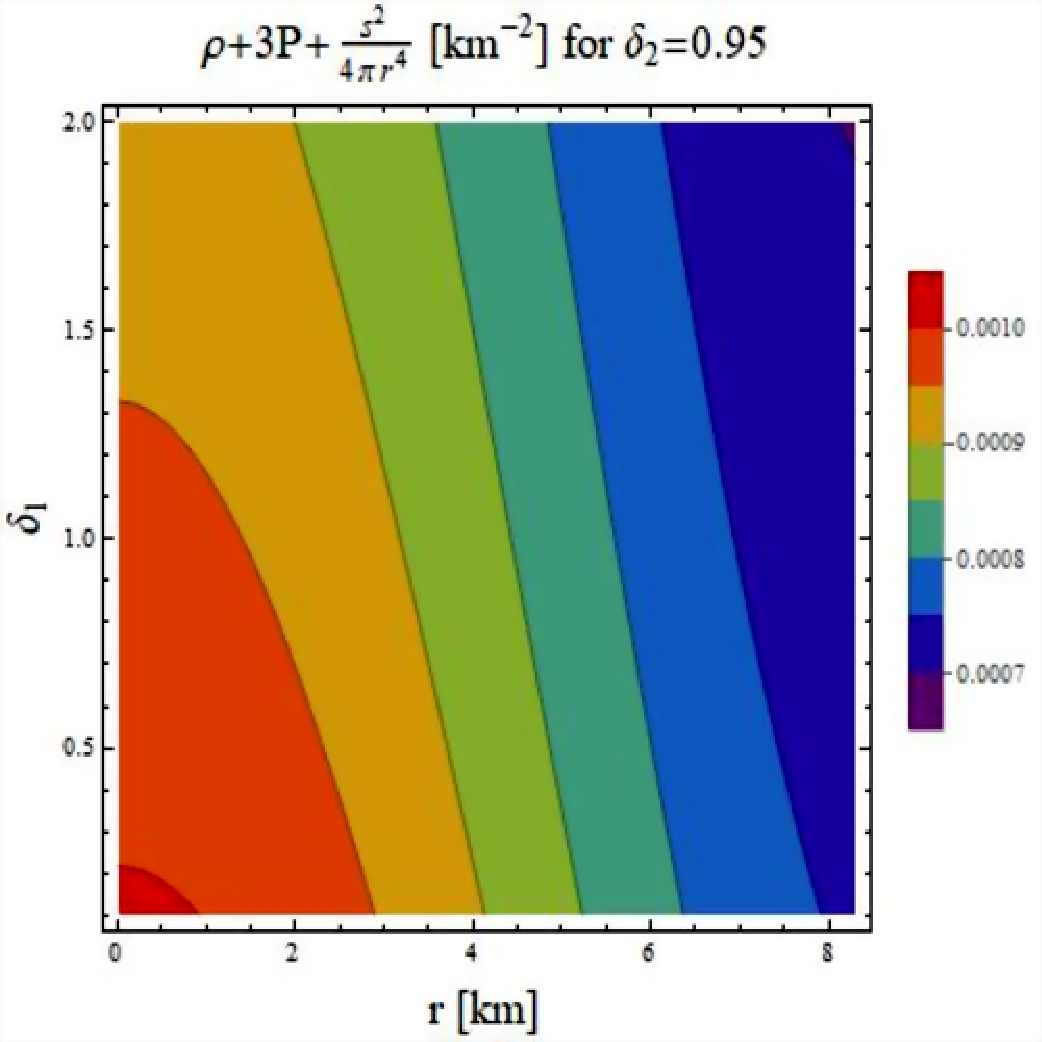,width=0.4\linewidth}
\caption{Energy conditions for model II with $\mathcal{L}_{m}=P$.}
\end{figure}
\begin{figure}[H]\center
\epsfig{file=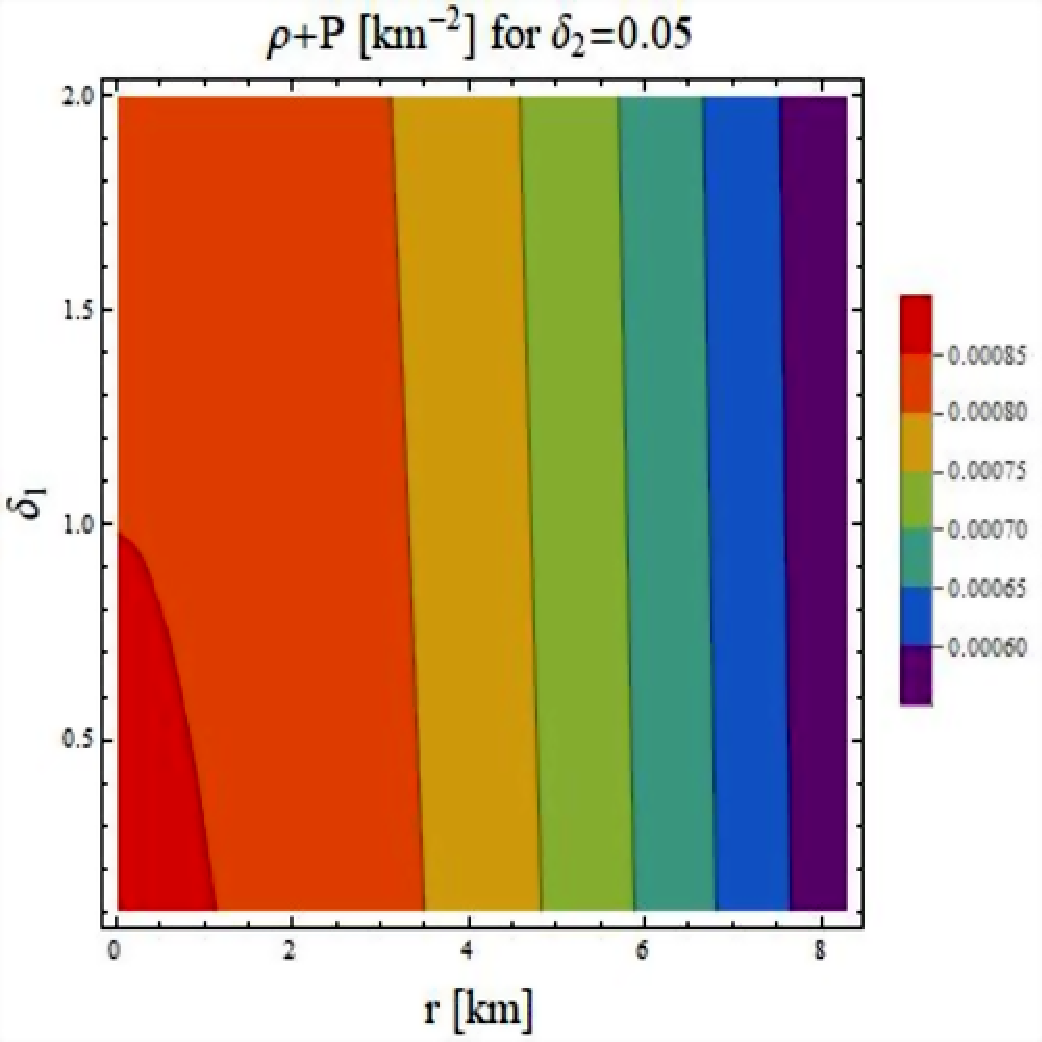,width=0.4\linewidth}\epsfig{file=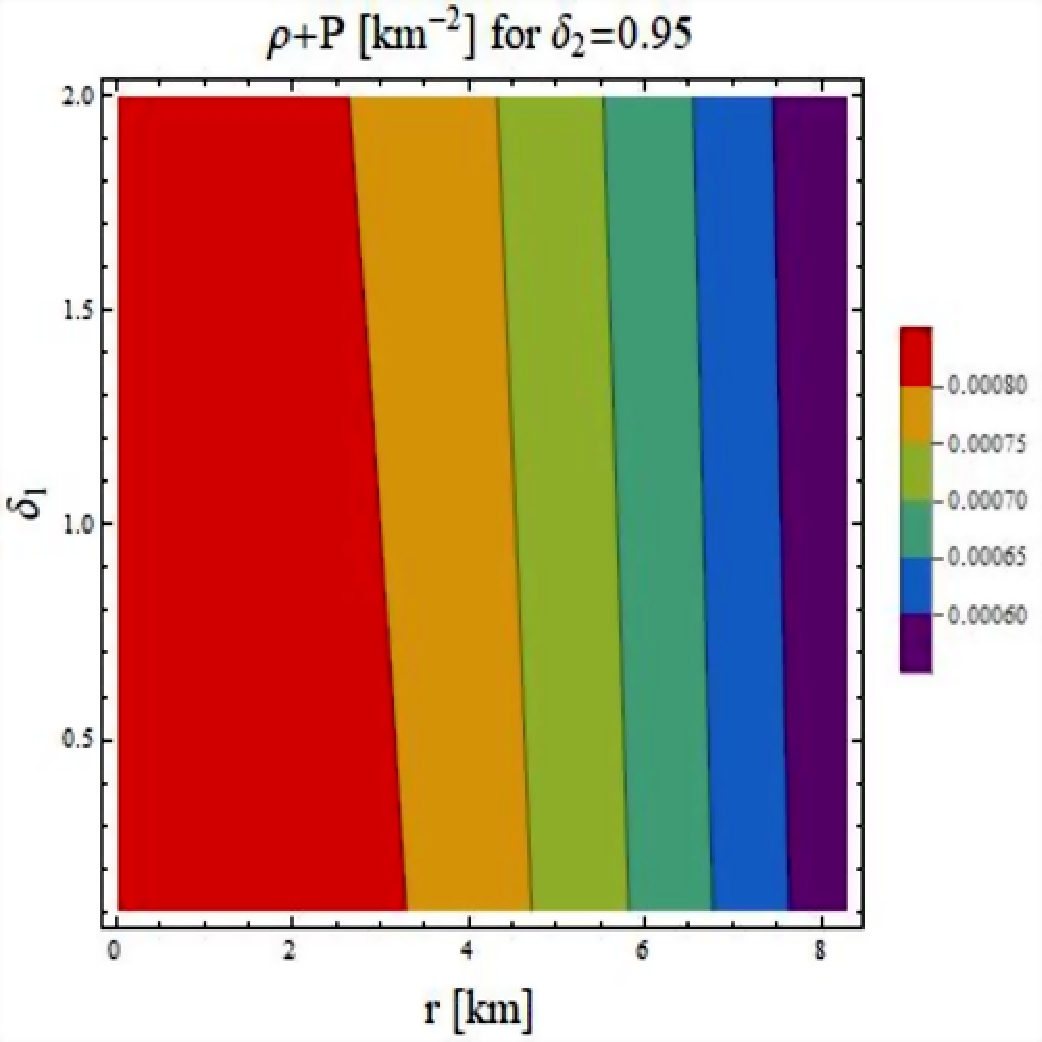,width=0.4\linewidth}
\epsfig{file=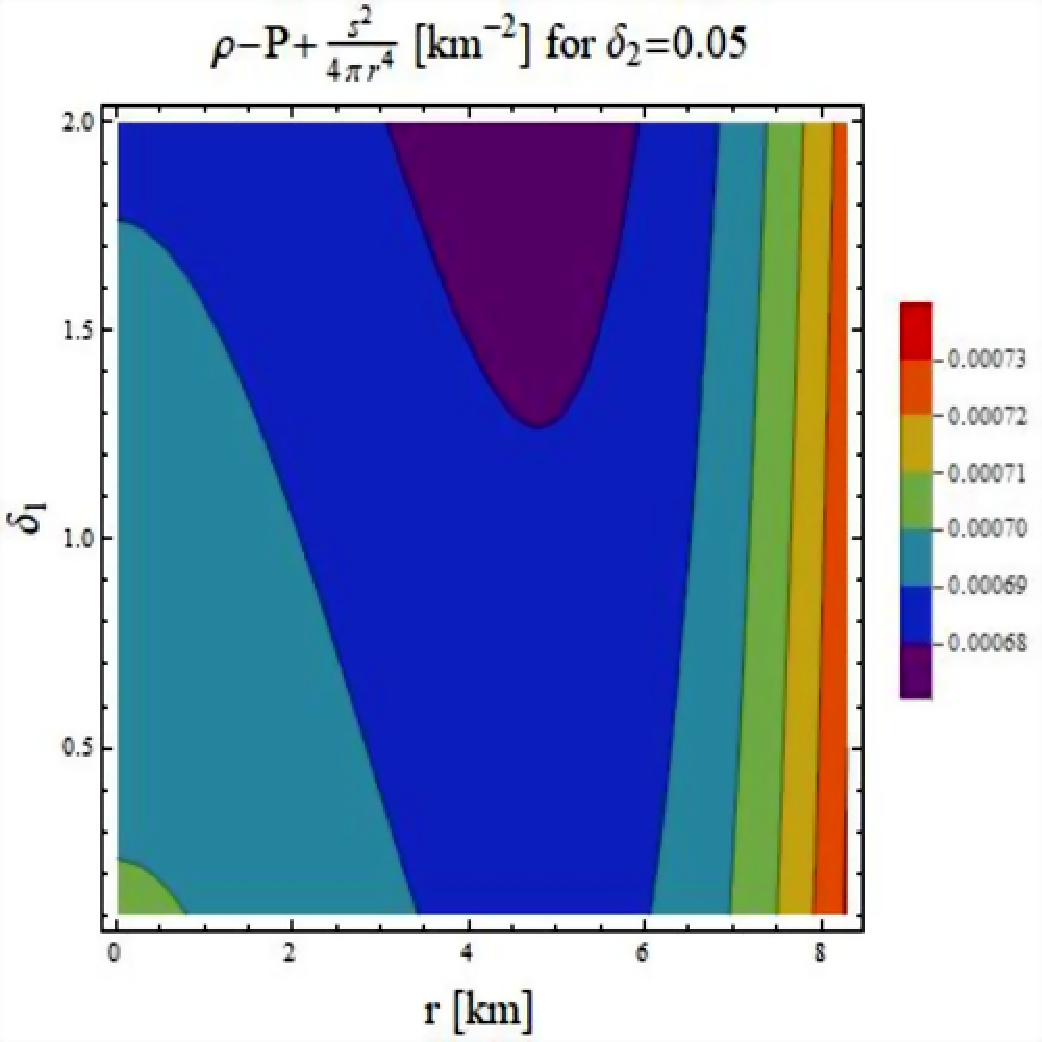,width=0.4\linewidth}\epsfig{file=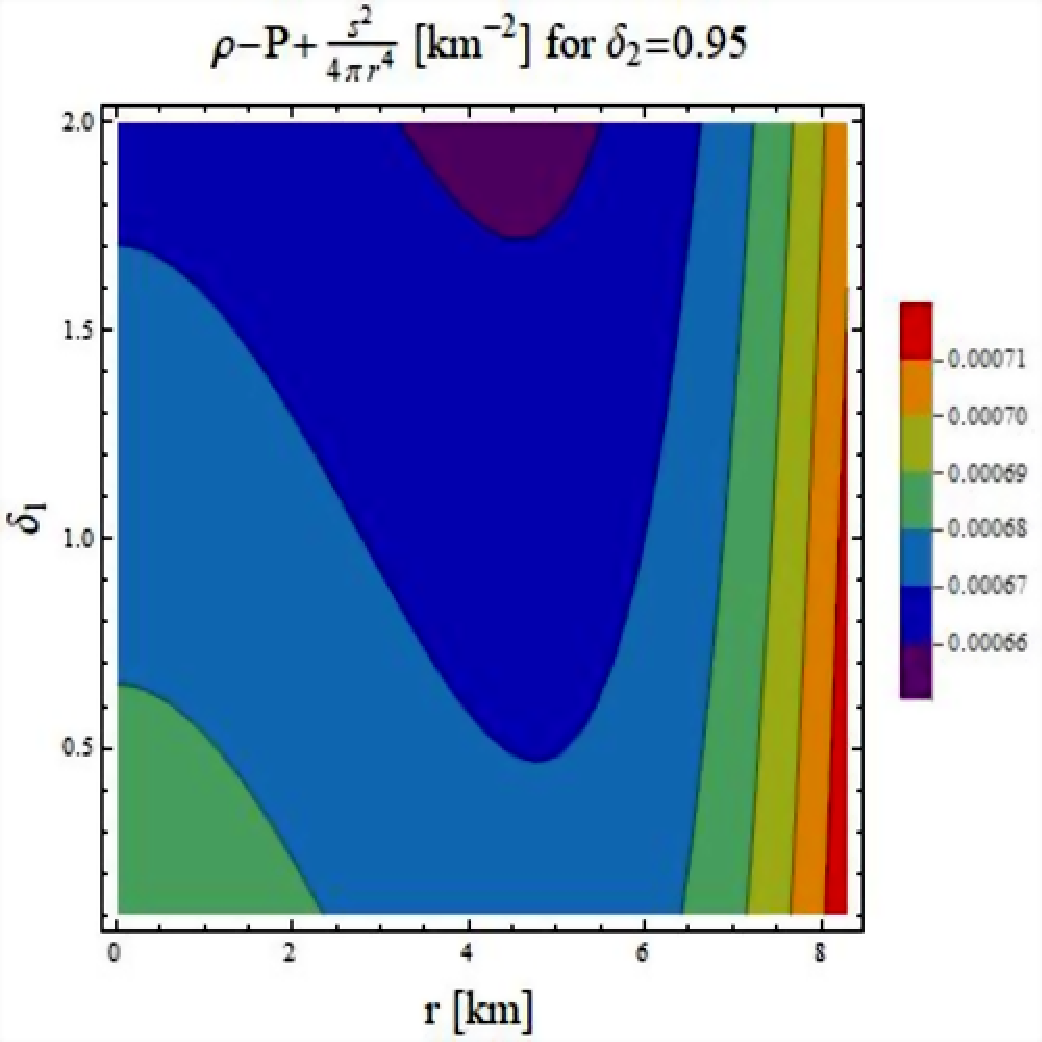,width=0.4\linewidth}
\epsfig{file=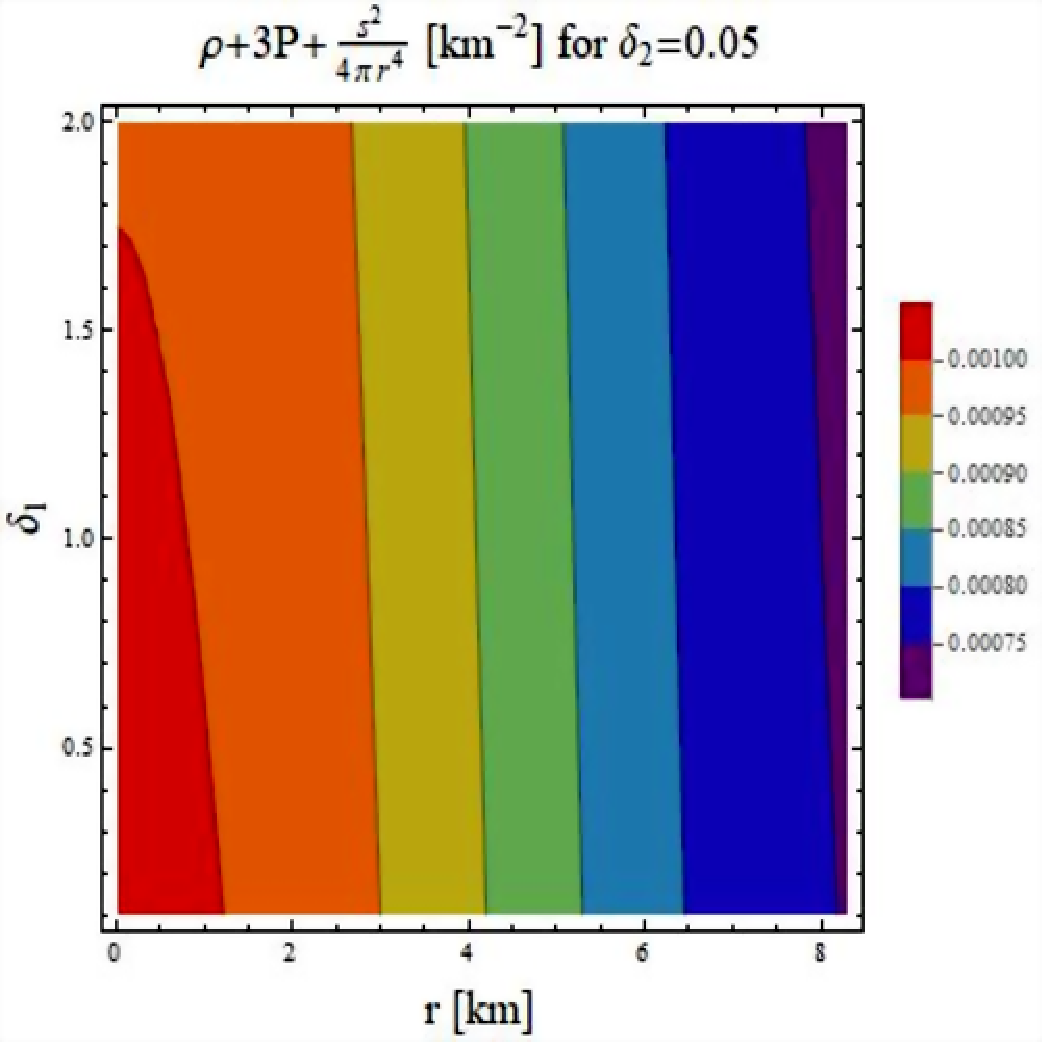,width=0.4\linewidth}\epsfig{file=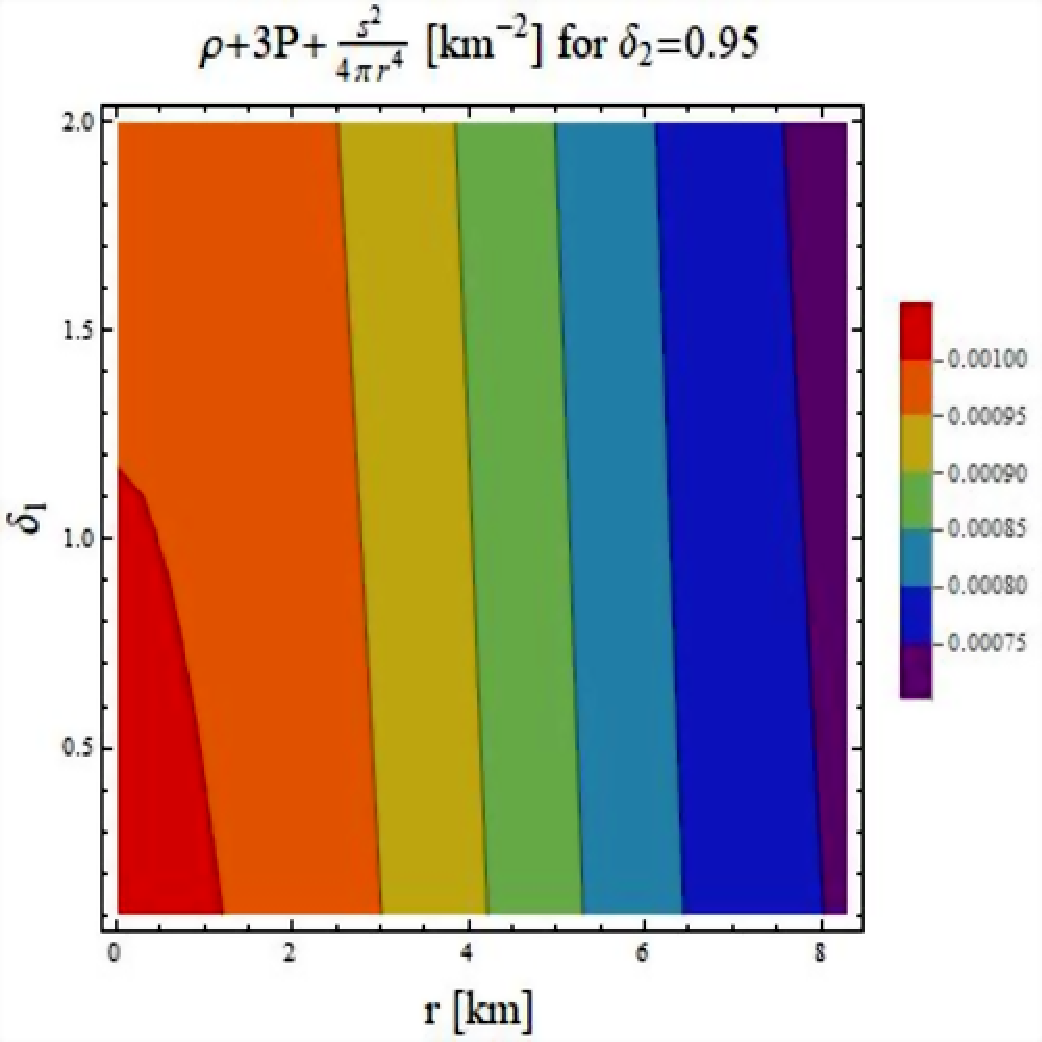,width=0.4\linewidth}
\caption{Energy conditions for model II with
$\mathcal{L}_{m}=-\rho$.}
\end{figure}
\begin{figure}[H]\center
\epsfig{file=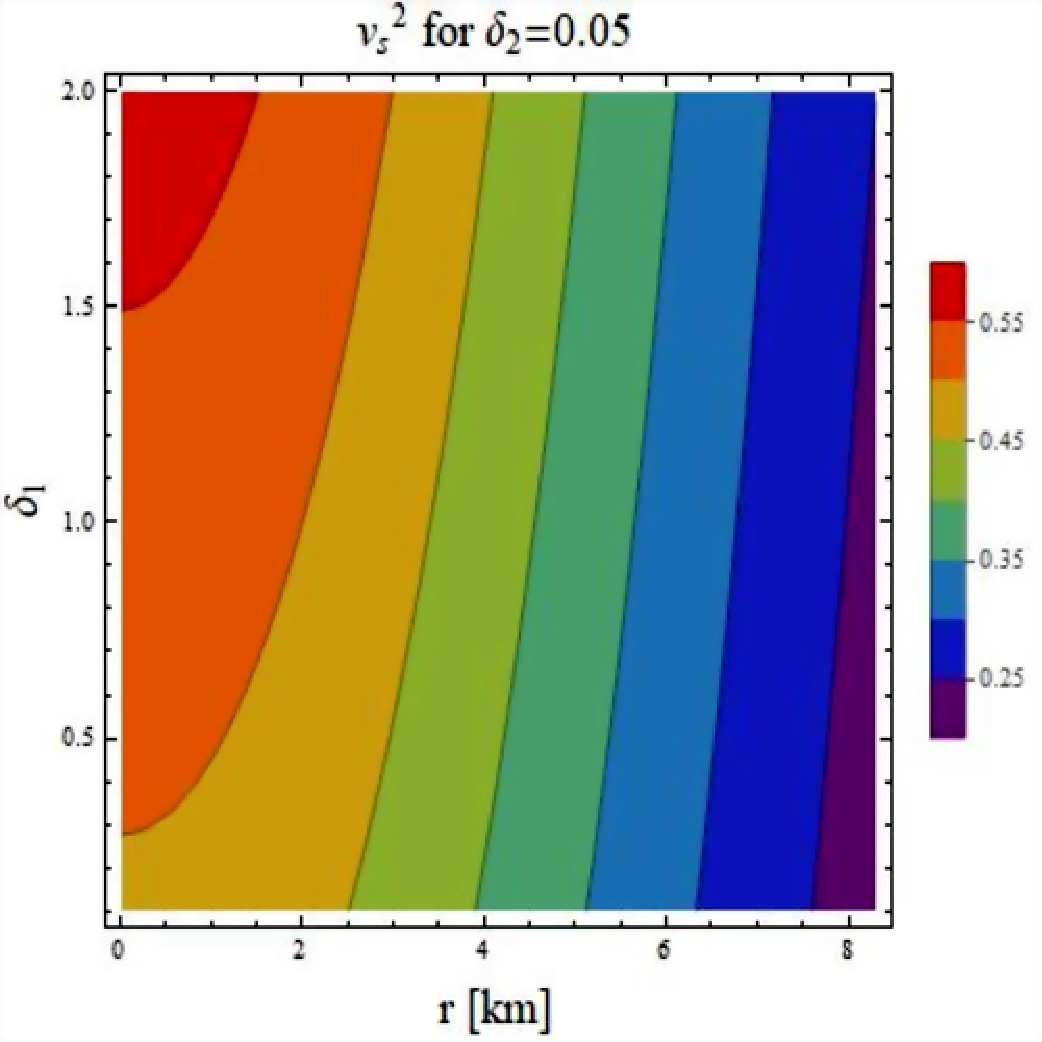,width=0.4\linewidth}\epsfig{file=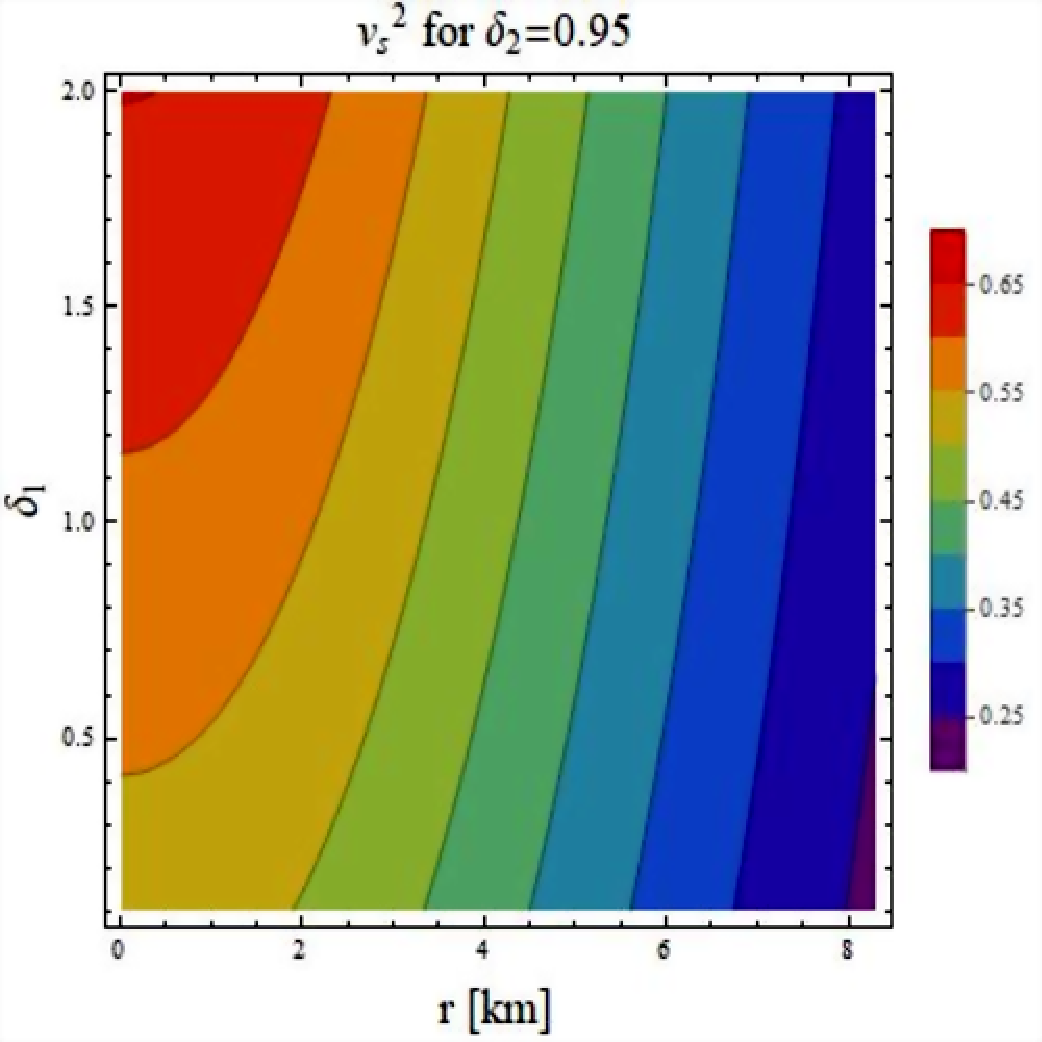,width=0.4\linewidth}
\epsfig{file=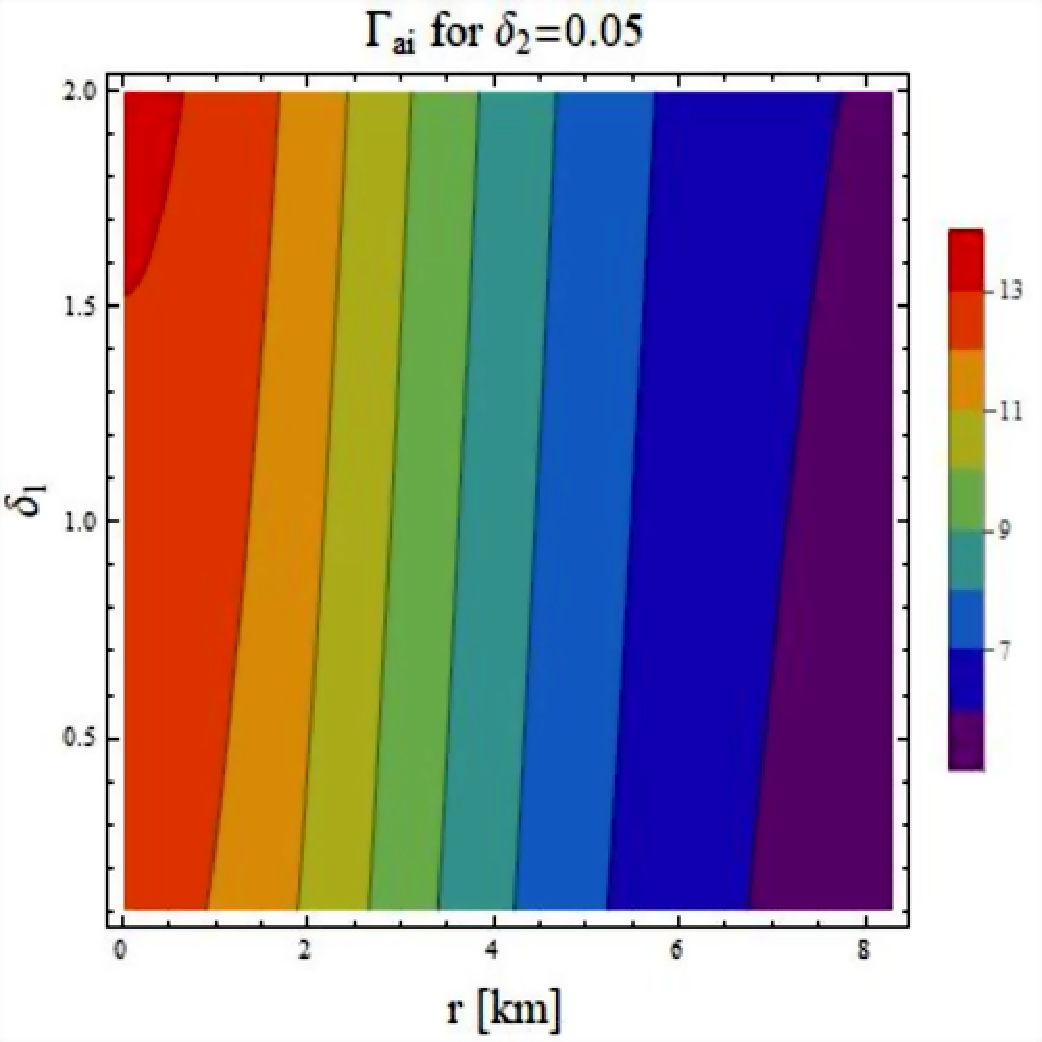,width=0.4\linewidth}\epsfig{file=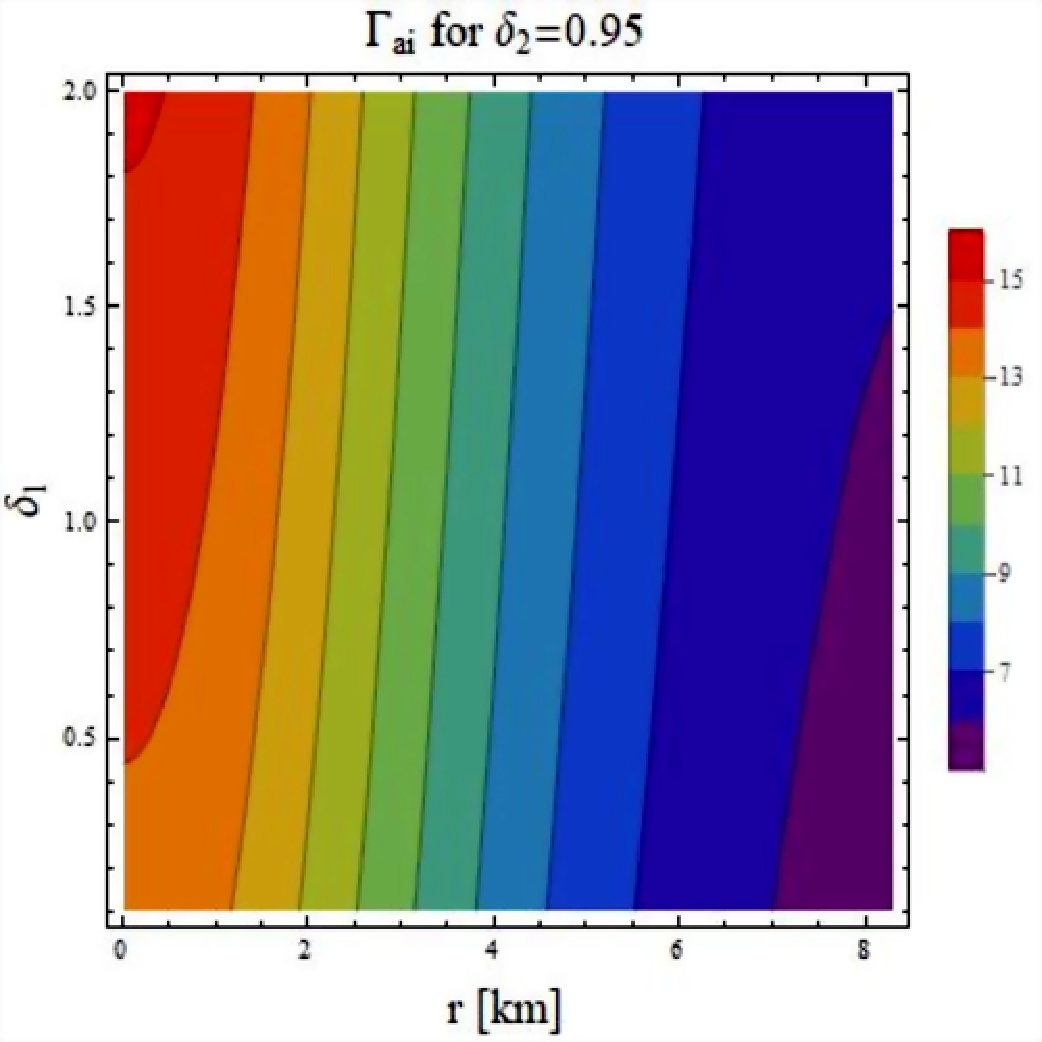,width=0.4\linewidth}
\caption{Stability analysis for model II with $\mathcal{L}_{m}=P$.}
\end{figure}
\begin{figure}[H]\center
\epsfig{file=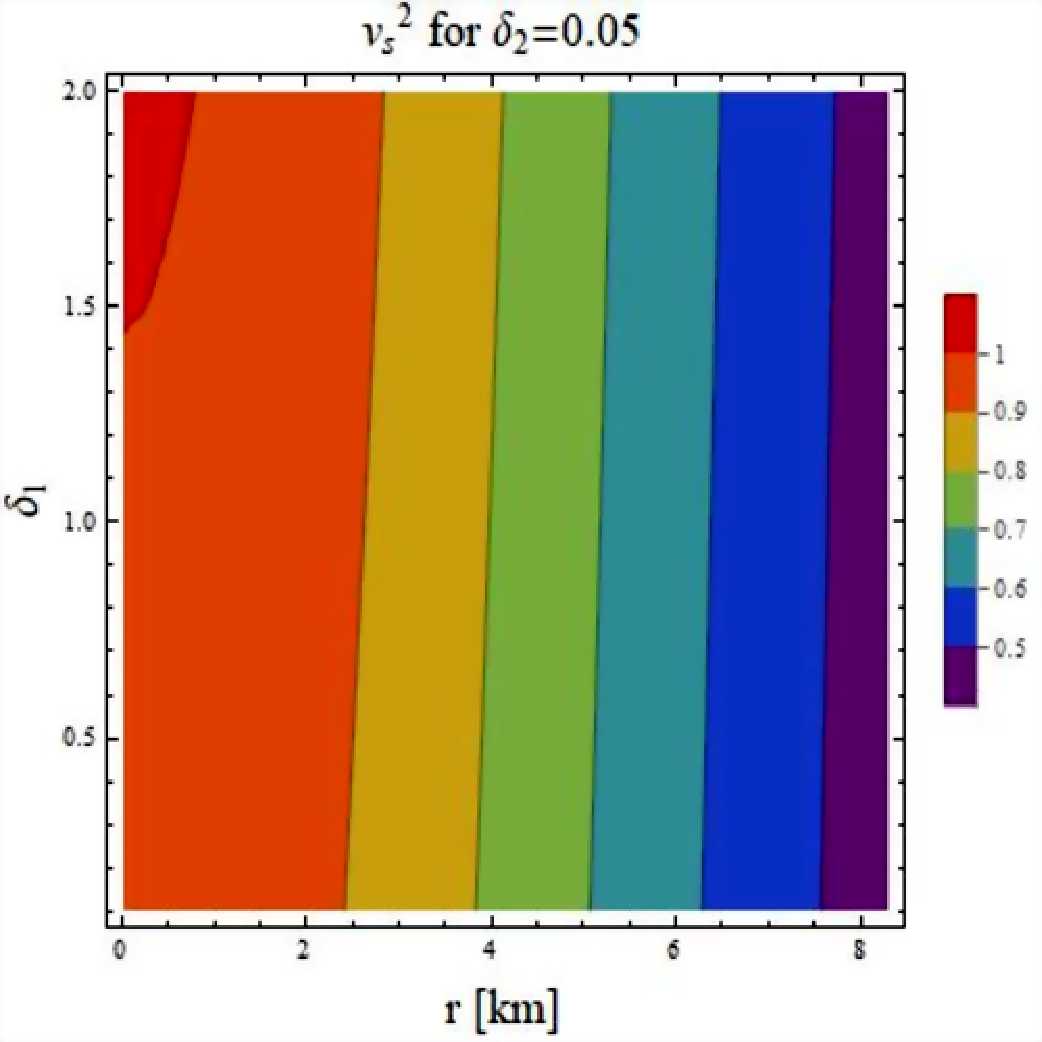,width=0.4\linewidth}\epsfig{file=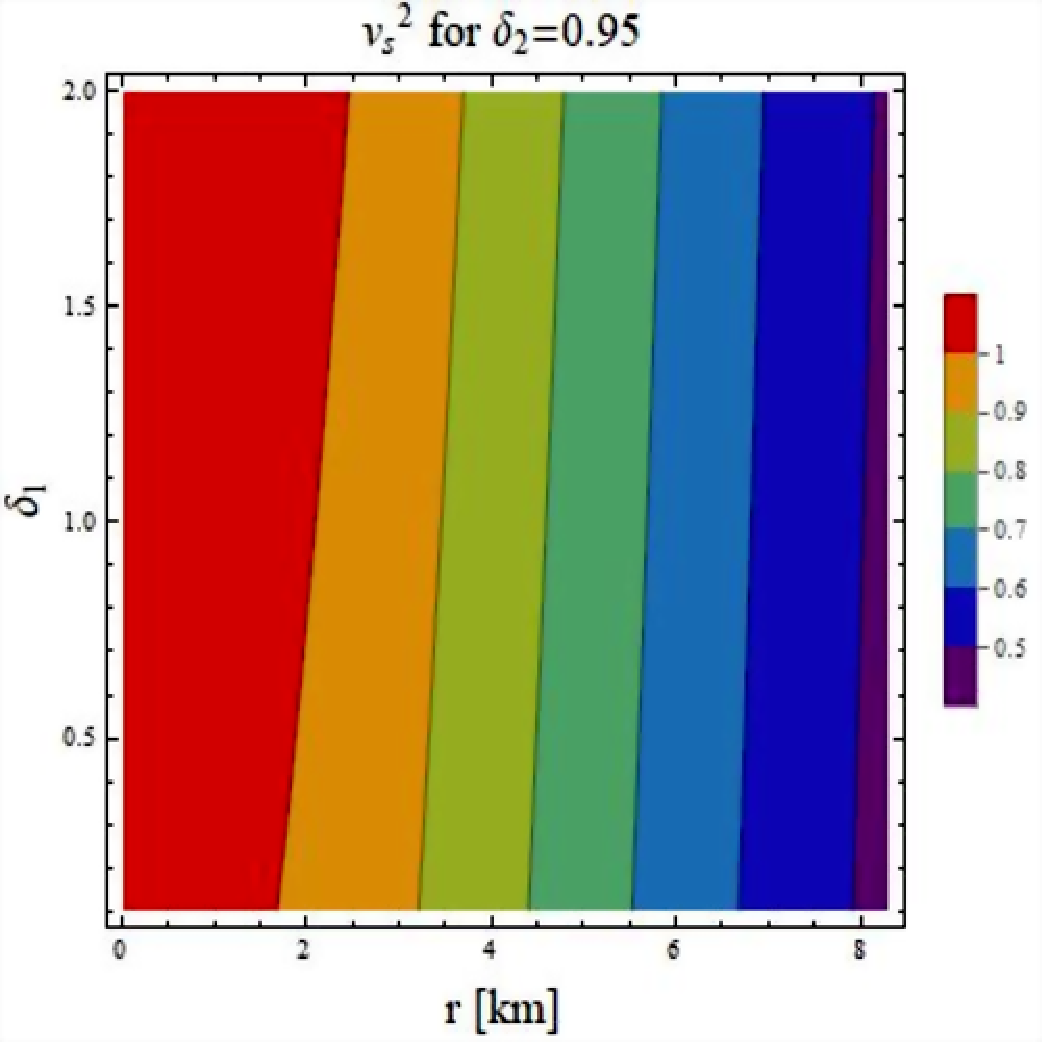,width=0.4\linewidth}
\epsfig{file=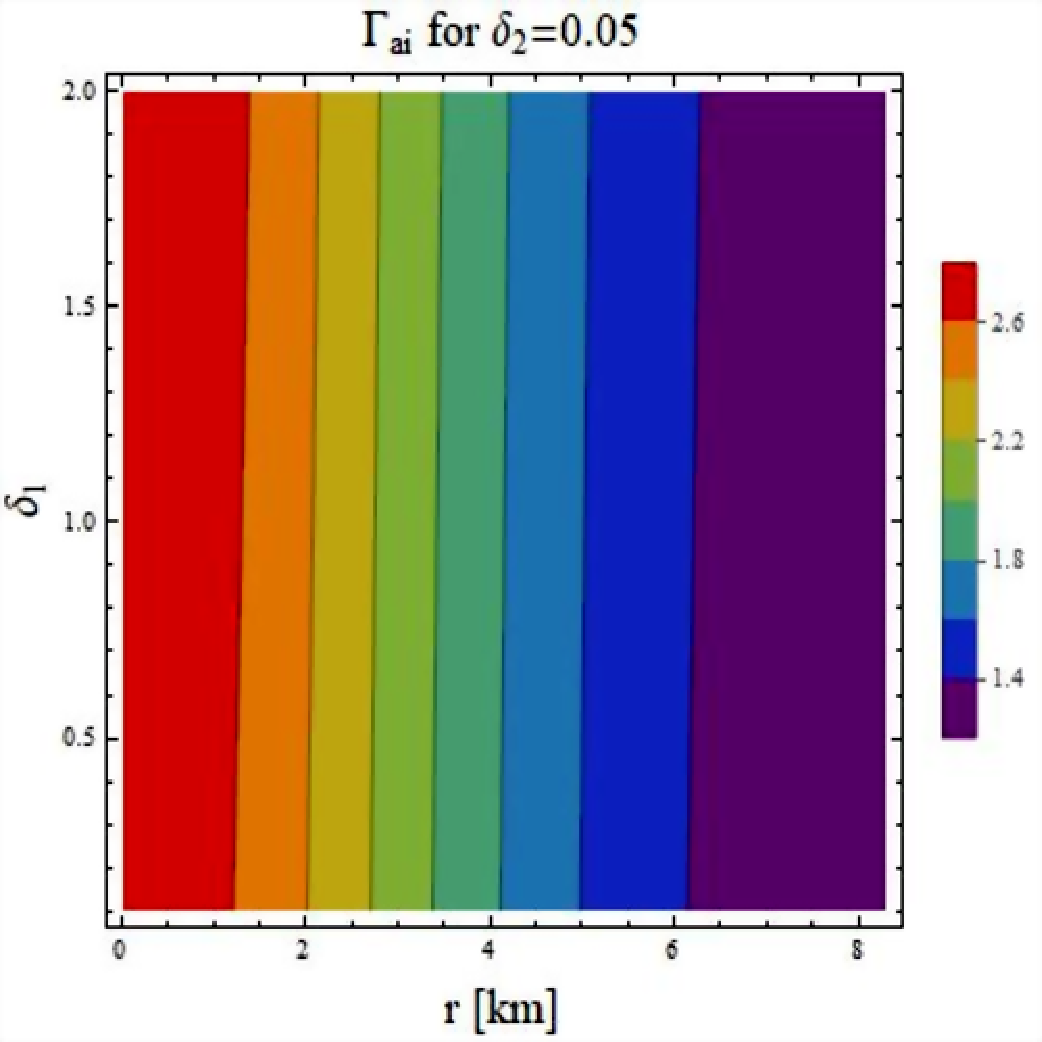,width=0.4\linewidth}\epsfig{file=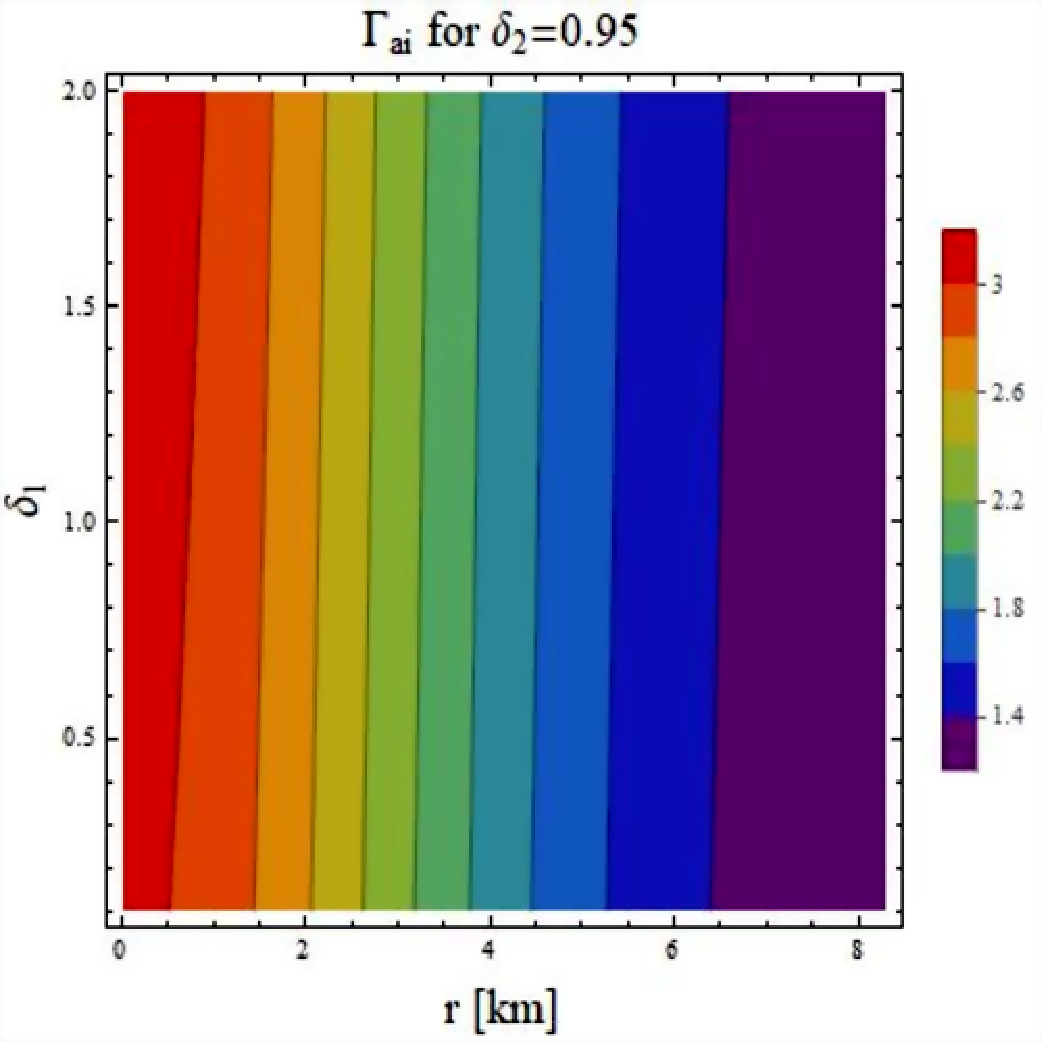,width=0.4\linewidth}
\caption{Stability analysis for model II with
$\mathcal{L}_{m}=-\rho$.}
\end{figure}

\section{Concluding Remarks}

This paper discusses multiple isotropic compact models which are
coupled with the electromagnetic field in the framework of
$f(\mathcal{R},\mathcal{L}_{m},\mathcal{T})$ gravitational theory.
For this purpose, we started off with the consideration of a static
geometry admitting spherical symmetry and induced an electric charge
through the addition of its corresponding Lagrangian in the modified
action defined in Eq.\eqref{g1}. We have then implemented the
least-action principle on this action and derived the field
equations possessing Lagrangian densities of both the fluid and
electromagnetic field. It was observed that there are three
independent components of equations of motion in the presence of
five unknowns, indicated the extra degrees of freedom and thus made
it impossible to find a unique solution. This only led to the
assumption of some constraints to deal with such issue. In this
regard, we have adopted the Karmarkar condition and a particular
$g_{tt}$ component form, resulted in computation of the $g_{rr}$
potential such that the ansatz becomes as follows
\begin{align}\nonumber
\delta_1(r)&=2b_2r^2+\ln b_3,\\\nonumber
\delta_2(r)&=\ln\big(1+b_2b_4r^2e^{2b_2r^2}\big).
\end{align}
There are actually three constants $(b_1,b_2,b_3)$ in the above
ansatz and the constant $b_4$ has been expressed in terms of this
triplet as $b_4=16b_1b_2b_3$. Therefore, we only needed three
conditions which have been provided by the matching conditions at
the interface, i.e., $\Sigma:~r=\mathrm{R}$ in terms of
$g_{tt},~g_{rr}$ and $g_{tt,r}$ components. The calculated values of
this quartet have been provided in Tables \textbf{I} and \textbf{II}
for five distinct compact objects from which we observed the impact
of charge on these constants.

We adopted two different (one minimal and one non-minimal) models in
this modified context, each of them has been discussed with two
different choices of the fluid Lagrangian. The model I contains two
parameters which are taken as $\beta_1 \in [0.1,2]$ and
$\beta_2=0.1,~0.8$. Further, the model II possesses one parameter
$\delta_1$ taken as same as $\beta_1$ along with an equation of
state parameter such as $\delta_2=0.05,~0.95$. The fluid doublet has
been observed acceptable because it fulfills the required behavior
of the energy density and pressure (Figures \textbf{1}, \textbf{2},
\textbf{8} and \textbf{9}). We also explored the mass function and
reached at the result that the model II possesses less massive
interior as compared to the first model for chosen parameters. A
necessary condition to be fulfilled is the validity of the energy
conditions which has been observed in Figures \textbf{4},
\textbf{5}, \textbf{10} and \textbf{11}, hence, our resulting
solutions are physically viable. Finally, the stability check has
been employed through two different techniques. We have found that
the minimal $f(\mathcal{R},\mathcal{L}_{m},\mathcal{T})$ theory
yields promising results in the context of astrophysical structures
for both $\mathcal{L}_{m}=P$ and $-\rho$. However, the non-minimal
modified model provides stable results only for former choice of the
Lagrangian density (Figures \textbf{6}, \textbf{7}, \textbf{12} and
\textbf{13}). It must be stressed here that disappearing the model
parameters reduces all these outcomes in GR.

\end{document}